%% file: main.tex
\documentclass[fleqn,11pt,table,anonymous]{wlscirep}
\usepackage[utf8]{inputenc}
\usepackage[T1]{fontenc}
\usepackage{soul}

\usepackage{booktabs}
\usepackage[counterclockwise]{rotating}
\usepackage[load-configurations=version-1]{siunitx} 

\usepackage{multirow}
\usepackage{soul}
\usepackage{enumitem}
\usepackage{array}
\usepackage{float}
\usepackage{caption}
\usepackage{subcaption}
\usepackage[most]{tcolorbox}

\usepackage[backend=biber, style=nature, maxbibnames=10, minbibnames=10]{biblatex}
\newcolumntype{L}[1]{>{\raggedright\let\newline\\\arraybackslash\hspace{0pt}}m{#1}}
\newcolumntype{C}[1]{>{\centering\let\newline\\\arraybackslash\hspace{0pt}}m{#1}}
\newcolumntype{R}[1]{>{\raggedleft\let\newline\\\arraybackslash\hspace{0pt}}m{#1}}

\title{EEG Foundation Models: A Critical Review of Current Progress and Future Directions}

\author[1,$\dagger$,*]{Gayal Kuruppu}
\author[2,$\dagger$]{Neeraj Wagh}
\author[3]{Vaclav Kremen}
\author[4]{Sandipan Pati}
\author[3]{Gregory Worrell}
\author[1,2]{Yogatheesan Varatharajah}

\affil[1]{Department of Computer Science \& Engineering, University of Minnesota Twin Cities, MN, USA}
\affil[2]{Department of Bioengineering, University of Illinois at Urbana-Champaign, Urbana, IL, USA}
\affil[3]{Department of Neurology, Mayo Clinic, Rochester, MN, USA}
\affil[4]{Department of Neurology, University of Minnesota Twin Cities, MN, USA}
\affil[$\dagger$]{Equal contributions}
\affil[*]{Correspondence: kurup016@umn.edu}

\begin{abstract}
\ul{\textit{Premise.}} Patterns of electrical brain activity recorded via electroencephalography (EEG) offer immense value for scientific and clinical investigations. The inability of supervised EEG encoders to learn robust EEG patterns and their over-reliance on expensive signal annotations have sparked a transition towards general-purpose self-supervised EEG encoders, i.e., EEG foundation models (EEG-FMs), for robust and scalable EEG feature extraction. However, the real-world readiness of early EEG-FMs and the rubrics for long-term research progress remain unclear.
\ul{\textit{Objective.}} In this work, \textcolor{black}{we conduct a review of ten early EEG-FMs} to capture common trends and identify key directions for future development of EEG-FMs.~\ul{\textit{Methods.}} We comparatively analyze each EEG-FM using three fundamental pillars of foundation modeling, namely the representation of input data, self-supervised modeling, and the evaluation strategy. Based on this analysis, we present a critical synthesis of EEG-FM methodology, empirical findings, and outstanding research gaps.
\ul{\textit{Results.}} We find that most EEG-FMs adopt a sequence-based modeling scheme that relies on transformer-based backbones and the reconstruction of masked temporal EEG sequences for self-supervision. However, model evaluations remain heterogeneous and largely limited, making it challenging to assess their practical off-the-shelf utility. In addition to adopting standardized and realistic evaluations, future work should demonstrate more substantial scaling effects and make principled and trustworthy choices throughout the EEG representation learning pipeline.
\ul{\textit{Significance.}} Our review indicates that the development of benchmarks, software tools, technical methodologies, and applications in collaboration with domain experts may advance the translational utility and real-world adoption of EEG-FMs.
\end{abstract}

\begin{document}

\flushbottom
\maketitle

\maketitle

\input{introduction}

\input{background}

\input{method}

\input{dim_comparisons}

\input{takeaways_gaps}

\input{future_directions}

\input{conclusion}

\input{acknowledgement}

\input{funding}

\input{data_availability}

\input{conflicts}

\clearpage
\printbibliography

\clearpage
\input{supplementary}

\end{document}

%% file: introduction.tex
\section{Introduction}
\label{sec:introduction}

Electroencephalography (EEG) is a widely used neurophysiological modality for measuring the brain’s electrical activity with high temporal resolution \cite{mushtaqOneHundredYears2024}. Multi-channel EEG recordings encode complex neural dynamics across spatial, temporal, and spectral dimensions, necessitating interpretation by highly trained experts. Despite extensive efforts to develop quantitative EEG representations using handcrafted features, advanced signal processing, and contemporary machine learning approaches, expert visual review remains the gold standard for clinical EEG interpretation and decision-making.
However, in the last decade, EEG research has seen significant advances in deep learning-based approaches for extracting application-specific features from raw EEG data (EEG-DL) \cite{royDeepLearningbasedElectroencephalography2019a, craikDeepLearningElectroencephalogram2019}. Although this line of research held significant potential to augment traditional visual EEG review, it met with limited success due to several reasons. \textcolor{black}{Much of the EEG-DL research comprised end-to-end learning methods that relied on supervised learning and were trained on a very narrow group of EEG tasks, i.e., use-cases, and datasets that resulted from laborious expert-driven annotation efforts \cite{obeid2016temple}.} These efforts were clearly unscalable to cover the wide range of EEG tasks that might be needed in clinical or scientific EEG review. Furthermore, such supervised EEG encoders were highly susceptible to overfitting on erroneous and noisy training instances, raising concerns about robustness and transferability to other tasks and datasets. \textcolor{black}{This lack of robustness was further exacerbated by the variability of EEG recordings across recording sites, acquisition systems, subjects, and sessions, leading to a lack of trust in supervised EEG encoders \cite{wagh2022evaluating, banville2021uncovering}. These limitations emphasized the need for EEG-DL solutions that rely less on expert EEG labels and yield robust, trustworthy, and explainable models with high translational value.}

The emerging paradigm of foundation models (FMs) \cite{bommasani2021opportunities}, based on label-free self-supervised learning (SSL) and efficient transfer learning, is a promising solution for these data-related challenges. Similar to the mainstream vision~\cite{yuan2021florence, kirillov2023segment, radford2021learning} and language~\cite{chowdhery2023palm, touvron2023llama, achiam2023gpt} FMs, EEG foundation models (EEG-FMs) are trained to identify salient EEG features from raw unlabeled EEG recordings by using various SSL pretext tasks, such as masked reconstruction or masked prediction, via a process known as \textit{pretraining}. EEG-FMs learn to represent EEG data as compressed embeddings in a latent space by leveraging the intrinsic properties found in the raw data. Pretrained EEG-FMs can then be adapted for various downstream applications using only very small amounts of labeled data, thereby alleviating the burden of expert EEG annotations -- the main bottleneck in supervised EEG-DL research. As such, EEG-FMs hold promise as powerful, off-the-shelf feature extractors (or \textit{encoders}) that can support scientific research, next-generation robust brain-computer interfaces, and augmented neurological decision support.

\begin{figure}[h]
    \centering
    \includegraphics[width=0.65\linewidth]{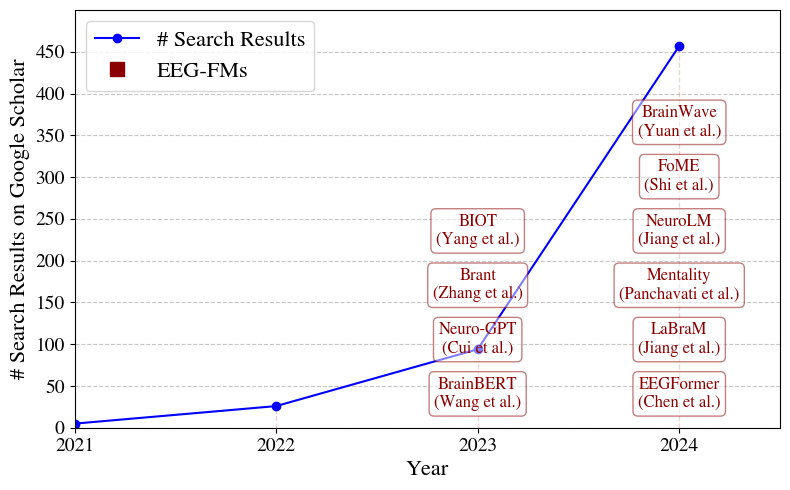}
    \caption{
    \textcolor{black}{\textbf{Electroencephalography foundation model (EEG-FM) publication trends}. The total annual Google Scholar search results for the term ``EEG Foundation Model'' between 2021 and 2024 (September) are shown in blue. The specific EEG-FMs reviewed in this study (i.e., according to the search criteria described in section \ref{sec:method}) are listed in red, in chronological order by their preprint publication dates.
    }}
    \label{fig:yearly_papers}
\end{figure}

Several EEG-FMs have been proposed over the last few years~\cite{wang2023brainbert, cui2024neuro, zhang2024brant, yang2024biot, chen2024eegformer, jiang_large_2024, panchavati2024mentality, jiang_neurolm_2024, shi_fome_2024, yuan2024brainwavebrainsignalfoundation}, whose chronological order along with total annual Google Scholar search results for the term ``EEG Foundation Model'' is shown in Figure \ref{fig:yearly_papers}. Despite the growing interest, many questions remain unanswered regarding the design choices in EEG-FMs, the learned representations, performance across various real-world applications, and the guarantees of robustness and trustworthiness. For example, the choice of EEG input representations, architectural components, and SSL pretext tasks can vary significantly across models, and the effects of those choices on the learned features are unclear. The complexity, quality, and flexibility of the representations learned by EEG-FMs and their relation to brain physiology have not been sufficiently studied, \textcolor{black}{raising questions about their explainability.} Furthermore, the performance and generalizability of these EEG-FMs beyond the common public datasets and benchmarks have not been adequately evaluated. These questions and concerns call for a \textcolor{black}{critical review of the early EEG-FMs}, focusing on architectural choices, pretraining approaches, evaluations, and trustworthiness, to identify rubrics for meaningful long-term progress in EEG-FM research and to advance their translational value.

\textcolor{black}{To that end, this topical review surveys ten early EEG-FMs (between January 2021 and September 2024) and critically analyzes their building blocks, identifies key takeaways and research gaps, and suggests directions for future EEG-FM research. Recent studies have begun to survey EEG-FMs \cite{yuxuan2025foundation,zhou2025brain,weng2025self}, but these reviews primarily emphasize technical components such as model architectures and self-supervised learning strategies. In contrast, critical factors related to data representation, evaluation scope, and rigor have received relatively little attention. Even more limited is the discussion of real-world translational requirements and the practical constraints under which EEG-FMs will ultimately need to operate. In this review, we offer a more holistic perspective on the development and assessment of EEG-FMs, adopting a data- and domain-centric viewpoint to evaluate their maturity for clinical translation and real-world use. We also highlight complementary research directions that can help position future EEG-FMs for meaningful practical impact.}
This review is organized as follows. Section \ref{sec:background} introduces the EEG recording procedures, the information content of EEGs, traditional quantitative EEG approaches based on feature engineering, the sources of data variability, and the different functional components of foundation models. Section \ref{sec:method} describes the search strategy we undertook to identify previously published EEG-FMs and provides brief summaries of those models. Section \ref{sec:dim_comparisons} provides a comprehensive analysis of each of the identified EEG-FMs along several dimensions, including training data, data preprocessing, input representations, model architectural components, and model evaluation strategies. Section \ref{sec:takeaways_gaps} discusses the key takeaways and research gaps, and Section \ref{sec:future_directions} lays out our view for future directions that would further advance the current state of the art.

%% file: background.tex
\section{Background}
\label{sec:background}

\textcolor{black}{EEG is a common neurophysiological technique for recording electrical activity generated by the brain using one or more electrodes. EEG emerged in the early nineteenth century and has since expanded into many areas of science and medicine.} It is widely used in clinical and research settings for diagnosing neurological disorders~\cite{tatum2018clinical}, studying brain function~\cite{da2013eeg}, and developing brain-computer interfaces~\cite{varbu2022past}. \textcolor{black}{EEG measures voltage potentials and their fluctuations resulting from ionic current flow mainly generated by firing populations of neurons~\cite{buzsaki2012origin}. Scalp EEG is typically captured using electrodes placed directly on the scalp. EEG can also be recorded invasively, with electrodes placed intracranially (within the brain) or subdurally (under the skin), commonly referred to as intracranial EEG (iEEG). While iEEG measures summated local field potentials generated by smaller neuronal populations, scalp EEG measures the electrical activity generated by large neuronal populations that pass through multiple layers of the brain and head (i.e., cerebrospinal fluid, meninges, skull, muscles, scalp). The conduction of neuronal activity through these layers causes significant attenuation and spatial mixing of the EEG signal~\cite{brunner2016volume}, including artifacts caused by movements of electrodes and muscle activity.} As a result, scalp EEG has considerably different spatio-temporal resolution, amplitude, frequency, and noise content compared to intracranial EEG~\cite{petroff2016comparison}.

\subsection{Contents and Sources of Variability in EEG Data}

The EEG signal is a noisy, high-dimensional, spatio-temporal measurement of electrical brain activity. Several factors influence the contents of EEG recordings, including physiological, pathological, and artifactual elements.
\textbf{Physiological elements}: \textcolor{black}{Most of the physiological elements present in EEG can be described using the various oscillatory components known as the delta, theta, alpha, beta, and gamma bands, each with distinct oscillatory frequencies and physiological mechanisms ~\cite{klimesch1996memory}.} The general attentional state of the subject during the recording (awake, drowsy, asleep) influences the spatio-temporal characteristics of EEG~\cite{oken2006vigilance}. \textcolor{black}{In addition, EEG also shows changes related} to natural aging~\cite{perinelli2022power} and other benign changes in brain structure and function.

\noindent\textbf{Pathological elements}: The primary diagnostic value of EEG is due to its ability to capture various pathological brain activity patterns. These electrographic patterns include seizures, interictal epileptiform discharges (e.g., spikes and sharp waves), triphasic waves, rhythmic discharges, lateralized periodic discharges, background slowing, and other diffuse or focal abnormal electroencephalogram patterns~\cite{tatum2018clinical}. 

\noindent\textbf{Artifactual and noise elements}: EEG artifacts can be introduced by biological phenomena (e.g., eye blinks, head movement, muscle activity, cardiac signals), recording conditions (e.g., signal discontinuities, transient filter effects), or external sources (e.g., electromagnetic interference)~\cite{amin2023normal}. \textcolor{black}{Eye- and muscle-related activities are the primary sources of artifacts in EEGs, introducing slow, high-amplitude changes and fast, low-amplitude changes, respectively.} In addition to these artifacts, consistent noise from the power line and its harmonics may also be present. Finally, certain medications, such as anti-depressants and anti-seizure medications, affect the EEG as well. For example, benzodiazepines and barbiturates are known to cause changes to the EEG~\cite{blume2006drug}.

The various practices and configurations used in clinical studies introduce variabilities across EEG datasets. \textcolor{black}{First, the hardware used to record EEGs can have preset configurations for sampling frequency, filtering, analog-to-digital quantization, amplification magnitude, and broadband noise levels~\cite{li2021review}.} Second, the EEG examination protocol may differ between clinical sites: the recording length could be different; recordings can have variable amounts of wake/sleep segments and eyes open/closed segments; and they may include cognitive tests and activation procedures such as sleep deprivation, photic stimulation, or hyperventilation. In addition, the EEG layout (i.e., the number of leads and their positions) and the reference leads may vary across sites. Finally, anatomical~\cite{cespedes2020influence}, genetic~\cite{smit2012individual}, and biochemical~\cite{gao2017inferring} differences introduce inter-individual variability. These factors make data ingestion and standardization non-trivial, especially in intracranial EEG datasets, where lead locations are subject-specific.

\subsection{Feature Extraction from EEG Signals}
Characteristics of clinical EEGs are assessed visually by experts trained to detect pathological events and adverse deviations from healthy brain activity~\cite{tatum2021handbook}. However, in non-clinical settings, the EEG is characterized by signal features derived computationally or analytically from the raw data. Quantitative EEG \textit{features} serve as lower-dimensional proxies of the raw signal, quantifying pertinent brain activity and providing objective EEG assessments for various applications. Classical EEG features are derived from signal processing, statistics, and information theory domains. Examples include relative band-limited power, power ratios, independent components, sample entropy, and kurtosis, among others~\cite{popa2020role}. In select instances, these EEG features isolate and quantify meaningful characteristics of underlying brain physiology, such as neuronal oscillations and functional connectivity, and are therefore considered robust and interpretable. In supervised EEG-DL models, EEG features are learned directly from raw EEG data to increase performance for a particular task~\cite{roy2019deep}. While EEG-DL models may push the boundaries of task performance, their black-box nature makes it non-trivial to relate them to brain physiology. Furthermore, EEG-DL features may overfit the noise in EEG signals, making them less likely to generalize out of the box. Hence, the robustness, reliability, and interpretability of EEG-DL feature extractors remain open research problems.

\subsection{Pillars of Foundation Modeling}

In this section, we deconstruct the FM monolith into a set of universal, domain-agnostic building blocks or pillars. These pillars are minimally required to systematically compare diverse existing EEG-FMs and to build new ones.

\noindent\textbf{Representation of input data:} 
EEG signals are acquired as a spatio-temporal data stream. Visual review of clinical EEGs is conducted primarily in this {\lq{native}\rq} view. However, the computational interpretation and analysis of EEGs can benefit from alternative, perhaps more informative, representations of raw data. For example, representing EEGs as a temporal stream of short spatio-spectral data segments (10s each) could make it easier for EEG-FMs to learn latent frequency patterns \textcolor{black}{and multi-scale information}. Alternatively, representing EEGs using spatial covariance matrices can make it easier to extract latent connectivity patterns. Broadly, \textcolor{black}{the choice of data representation at the input stage} may greatly influence the quality and semantics of learned latent representations~\cite{neumann2021smart}. In biosignal domains where data are sampled from underlying physiological processes, constructing informative input data views may itself require significant offline transformations or even domain-specific modeling, such as inverse modeling~\cite{michel2019eeg}. Without these offline efforts, data-driven learning of latent representations via FMs may be infeasible.

\noindent\textbf{Architectural and functional design of FMs:} 
At an architectural level, FMs are typically built from several non-linear transformer blocks that can be readily stacked to scale the model depth and size. Transformer blocks themselves contain multiple feed-forward layers followed by a multi-head attention module~\cite{vaswani2017attention}. Vision FMs may use convolutional blocks to exploit the desirable spatial inductive bias of convolutional operations~\cite {wu2021cvt}. Regardless of model size or low-level architectural choices, encoder-only FMs are understood to have only two high-level functional components: a backbone network and a final single fully connected layer. The backbone network serves as a feature extractor, providing low-dimensional embeddings (or representations) of the input data. The final layer, referred to as a \textit{task head}, then uses these embeddings for task-specific purposes such as classification or regression.

\noindent\textbf{Self-supervised learning (SSL):} 
In the SSL paradigm, supervision for training is derived strategically from unlabeled input data (pseudo-labels) rather than from human-curated labels, as in supervised learning~\cite{ericsson2022self}. SSL objectives (pretext tasks) are commonly either contrastive or generative. Contrastive pretext tasks use meaningfully augmented or noisy views of the input to learn robust data representations using encoder-only backbones~\cite{chen2020simple}. Generative pretext tasks mask portions of the input and reconstruct/generate the masked portions using encoder and decoder-based backbones~\cite{zhang2023survey}. In either case, SSL learns general intrinsic relationships within the data, thereby creating a task-agnostic backbone network that serves as a general feature extractor. The structure of representations learned by SSL tasks and general principles for effective SSL task design are under active study. Regardless, the value of SSL is particularly high in the EEG domain as it improves model generalizability~\cite{wagh2021domain} and partially alleviates the need for expert EEG annotations that are expensive to obtain, subjective, and error-prone~\cite{banville2021uncovering}.

\noindent\textbf{Transfer learning and model adaptation:} 
The backbone network enables the reuse, transfer, and adaptation of EEG features (i.e., data-driven EEG knowledge) across semantically overlapping tasks~\cite{thrun1998learning}. The proportion of the backbone adapted for a specific application is flexible. For example, the backbone may be entirely fixed/frozen (\textit{linear probing}), partially updated, or entirely updated (\textit{fine-tuning})~\cite{kumar2022fine}. This adaptation process may use slower or layer-specific learning rates to preserve pretrained knowledge within the backbone. In either case, any pretext task-specific layers at the top are discarded and replaced with a new trainable linear head that performs classification or regression for the `downstream' application. Using application-specific data splits and human-curated task-specific labels, the model is then trained further until convergence. The development of algorithms for robust and efficient adaptation is an active area of research. Broadly, transfer learning and model adaptation techniques obviate the need to train new EEG-DL models from scratch by leveraging knowledge from pre-existing models.

\noindent\textbf{Data and model scale:} 
The scaling up of pretraining data volume, \textcolor{black}{data diversity (i.e., the number of distinct data sources),} and model size has contributed towards improved pretrained latent features, increased downstream task performance, sample efficiency, and model generalizability, among other properties~\cite{sun2017revisiting, dehghani2023scaling, kaplan2020scaling, yao2024towards, isik2024scaling, edwards2024scaling}. Empirical scaling laws have established power-law relationships between FM error rates and three key modeling levers, namely the pretraining data size, model size, and compute~\cite{kaplan2020scaling, rosenfeld2019constructive}, which suggest that an exponential increase in data and model size can indeed improve FM performance, albeit with diminishing returns. \textcolor{black}{Overall, the balance between data and model size to maximize performance under a fixed compute budget remains a critical consideration in FM research. It is noteworthy that vision and language FMs are often multi-billion-parameter models pretrained on an internet-scale collection of diverse datasets~\cite{oquab2023dinov2, wenzek2019ccnet, dosovitskiy2020image}. However, the relative scarcity of large-scale EEG corpora may limit scaling efforts in EEG-FM research.}

%% file: method.tex
\section{Review Approach \& Summary of EEG-FMs}
\label{sec:method}

This section describes our search strategy and summarizes the EEG-FMs to date, highlighting the data used to train those models, model scale, architectural details, and evaluations.

\subsection{Search Strategy}
We conducted a comprehensive search across various web platforms, including Google Scholar, arXiv, DBLP, IEEE Xplore, PubMed, bioRxiv, and medRxiv, to identify relevant research in journals, conferences, workshops, and preprints. \textcolor{black}{\textcolor{black}{We limited our search query, (``EEG'' AND ``Foundation Model''}), to identify FMs that were developed primarily for scalp or intracranial EEG modalities. Our search starts from the year 2021 -- the year the term \textit{Foundation Model} was introduced \cite{bommasani2021opportunities} -- and includes studies that were published or archived until September 30th, 2024. We then removed duplicate instances and manually reviewed the title, abstract, and introduction sections to confirm relevant EEG-FMs by identifying phrases to the effect of \textit{"We developed an EEG foundation model…"} and \textit{"The proposed approach forms the basis for an EEG foundation model…"}. It is important to note that we excluded studies that, a) 
propose individual components that are building blocks of foundation models (e.g., pretraining strategies, spatio-temporal feature encoders), but not functional off-the-shelf foundation models, and b) finetune existing EEG-FMs to derive task-specific models. Furthermore, our review does not include multi-modal foundation models and sleep foundation models developed using polysomnography (PSG) data ~\cite{thapa2024sleepfm, ogg2024self}. Nonetheless, we briefly discuss two sleep foundation models published during the same time frame in the supplement for interested readers. Using those search criteria, we identified nine EEG-FMs, Neuro-GPT \cite{cui2024neuro}, Brant \cite{zhang2024brant}, BIOT \cite{yang2024biot}, EEGFormer \cite{chen2024eegformer}, LaBraM \cite{jiang_large_2024}, Mentality \cite{panchavati2024mentality}, NeuroLM \cite{jiang_neurolm_2024}, FoME \cite{shi_fome_2024}, and BrainWave \cite{yuan2024brainwavebrainsignalfoundation}. In addition, we also included BrainBERT because of its common presence as a baseline in other EEG-FM evaluations~\cite{zhang2024brant, chen2024eegformer, shi_fome_2024, yuan2024brainwavebrainsignalfoundation}.}

\subsection{EEG-FM Summaries}

Table \ref{tab:summary_table} summarizes all ten EEG-FMs. Although all the EEG-FMs share many commonalities, each FM is unique in its own right. In order to highlight the building blocks of each FM and their unique strengths, below we summarize each FM considering several factors, such as the amount of training data (in \textit{channel-hours}, calculated as the total recording duration multiplied by the number of EEG channels; see Supplement Table~\ref{tab:channelhours} for details), model size (in terms of the number of trainable parameters), the types of EEG data (scalp EEG and/or iEEG), the way inputs are configured (raw time series, power spectra, or time-frequency representation), architectural components (convolutional and/or transformer blocks), the SSL tasks used for pretraining (masked reconstruction, auto-regressive modeling, and/or contrastive learning), and the evaluations performed. 

\begin{table}[!h]
\centering
    \caption{\textcolor{black}{\textbf{Brief model summaries}. We provide brief summaries of the EEG foundation models (EEG-FMs) based on training data size, model size, input configurations, input data type, architectural components, and the self-supervised learning (SSL) tasks used for pretraining. Hyperlinks point to code and model weights, if available. \textcolor{black}{N/A represents cases where channel-hours or model size could not be determined.}}}
\fontsize{7pt}{7pt}\selectfont
\renewcommand{\arraystretch}{1.2}
\resizebox{\columnwidth}{!}{
    \begin{tabular}{|L{1.5cm}|L{1cm}|L{1cm}|L{1.5cm}|L{1.5cm}|L{2.5cm}|L{2cm}|}
        \toprule\toprule
        \textbf{Model} &
        \textbf{Training Data} (channel-hours) &
        \textbf{Number of Parameters} &
        \textbf{Input Configuration} & 
        \textbf{Data Type} &
        \textbf{Architectural Components} &
        \textbf{SSL Tasks}\\
        \hline
        BrainBERT [\href{https://github.com/czlwang/BrainBERT}{link}] &
        4.5k & 
        43.18M & 
        Single-channel spectrogram data & 
        Intracranial EEG &
        Transformer encoder and shallow decoder with two linear layers &
        Masked reconstruction \\
        \hline
        Neuro-GPT [\href{https://github.com/wenhui0206/NeuroGPT}{link}] &
        541k &
        79.53M &
        Fixed multi-channel time series data &
        Scalp EEG &
        Encoder with both convolution and transformer layers and GPT-2 as the decoder &
        Masked reconstruction (causally masked latent embeddings) \\
        \hline
        Brant [\href{https://huggingface.co/Daoze/Brant/tree/main}{link}] &
        281k &
        68M, 104M, 249M and 506M &

        Variable multi-channel time series data &
        Intracranial EEG &
        Two transformer encoders for time and space and a linear decoder &
        Masked reconstruction \\
        \hline
        BIOT [\href{https://github.com/ycq091044/BIOT}{link}] &
        312k &
        3.3M &
        Variable multi-channel spectral data &
        Scalp EEG &
        Linear transformer, encoder-only architecture &
        Contrastive learning\\
        \hline
        EEGFormer &
        541k &
        N/A &
        Multi-channel spectral data &
        Scalp EEG &

        A transformer encoder and a shallow transformer decoder & 
        Codebook-based reconstruction \\
        \hline
        LaBraM [\href{https://github.com/935963004/LaBraM}{link}]&
        80k &
        5.8M, 46M and 369M &
        Fixed multi-channel time series data &
        Scalp EEG &

        Convolutional temporal encoder and transformer encoder layers and a linear decoder. A separate decoder for tokenization. &
        Masked reconstruction (token-level) \\
        \hline
        Mentality &
        N/A &
        N/A &
        Fixed multi-channel time series data &
        Scalp EEG &

        Convolutional layers and Mamba blocks in both encoder and decoder &
        Masked reconstruction \\
        \hline
        NeuroLM [\href{https://github.com/935963004/NeuroLM}{link}]&
        546k &
        250M, 500M and 1.7B &
        Variable multi-channel time series data &
        Scalp EEG &
        Vector quantization for tokenization with convolutional temporal encoder and transformer spatial encoder &
        Autoregressive reconstruction (token-level) \\
        \hline
        FoME &
        N/A &
        476M and 745M &
        Variable multi-channel time series data & 
        Scalp EEG and intracranial EEG &
        Temporal and Spatial transformer encoder and a linear decoder &
        Masked signal reconstruction\\
        \hline
        BrainWave &
        878k &
        N/A & 
        Variable multi-channel spectrogram data &
        Scalp EEG and intracranial EEG &
        Transformer encoder with channel attention and a lightweight decoder & 
        Masked reconstruction (whole spectrogram)\\
        \bottomrule\bottomrule
    \end{tabular}
    }
    \label{tab:summary_table}
\end{table}

\noindent\textbf{BrainBERT}~\cite{wang2023brainbert}:
As the first released EEG-FM, BrainBERT is relatively smaller than others, with 43.18M parameters, and was pretrained using a modest dataset of 4.5k channel-hours of iEEG data. The inputs were represented as channel-wise spectrograms, and a \textcolor{black}{Bidirectional Encoder Representations from Transformers (BERT,~\cite{devlin2019bert})}-type model was pretrained to predict masked patches for different types of spectrograms based on the \textcolor{black}{Short-Time Fourier Transform (STFT,~\cite{gabor1946theory})} and Superlets~\cite{moca2021time}. The model comprised a transformer encoder and a shallow decoder with two linear layers. \textcolor{black}{Evaluations focused on evoked brain responses while watching movies~\cite{data:wang2024brain} and demonstrated generalizability to unseen subjects and electrode locations; however, the test data were drawn from the same distribution as the training data. The evaluations also showed that linear-probed BrainBERT performance matched that of randomly initialized supervised deep neural networks (DNNs) across most evaluation tasks. 
Additional evaluations showed that fine-tuned BrainBERT performance could match that of a randomly initialized DNN with as little as 15\% of the training data for one task. A task-agnostic \textcolor{black}{intrinsic dimensionality~\cite{ansuini2019intrinsic} (ID)-based} analysis showed that the BrainBERT embeddings of different brain regions had distinct IDs, whereas the distribution across electrodes was relatively constant under randomly initialized weights.}

\noindent\textbf{Neuro-GPT}~\cite{cui2024neuro}:
This mid-size EEG-FM with 79.53M parameters was pretrained entirely using scalp EEG data from the full \textcolor{black}{Temple University Hospital (TUH) EEG Corpus~\cite{data:obeid2016temple}}, which included 541k channel-hours of clinical scalp EEG data. The model takes raw time series of the 22 EEG channels in the standard 10-20 layout as input and learns EEG representations using a combination of convolutional and transformer layers.
Those representations are then used as input to a \textcolor{black}{Generative Pretrained Transformer (GPT-2,~\cite{radford2019language})} decoder, which autoregressively predicts the masked latents. It is noteworthy that the decoder has more parameters than the encoder in this setup, which is not common in other EEG-FMs. This model was evaluated on EEG data from a 4-class \textcolor{black}{Brain-Computer Interface (BCI)} motor imagery task with a different channel configuration than the pretraining data.
However, the downstream data were transformed to the original pretraining configuration using an inverse-forward approach~\cite{mosher2002eeg}. The results showed that fine-tuning or linear probing the pretrained model performed better than models trained from scratch, including some fully-supervised EEG-DL approaches.

\noindent\textbf{Brant}~\cite{zhang2024brant}:
Brant is a relatively larger FM with 500M parameters and was pretrained using 281k channel-hours of iEEG data. However, its pretraining data were limited to a single dataset comprising 9 subjects. \textcolor{black}{This model accepts raw EEG time series with different channel configurations as input.} The model consisted of a temporal encoder that learns long-term temporal dependencies, a spatial encoder that learns spatial correlations, and a simple linear decoder. The spatial encoder used in this model to capture spatial relationships is a novel contribution compared to previous EEG-FMs. \textcolor{black}{This model includes three scaled-down versions with 68M, 104M, and 249M parameters, respectively, all pretrained on the same data. These models were evaluated on short-term and long-term signal forecasting, frequency-phase forecasting, imputation, \textcolor{black}{seizure detection, and pathology detection} tasks. The results showed that Brant performs better than baselines in limited-label settings.}

\noindent\textbf{BIOT}~\cite{yang2024biot}:
This is the smallest of the ten reviewed EEG-FMs, with only 3.3M parameters, and was pretrained using a contrastive learning objective. The BIOT model introduced a novel approach to handling input data with variable lengths and variable numbers of channels: it tokenizes each channel into fixed-length segments representing frequency vectors, organizes them into `sentences', and uses channel and position embeddings to preserve spatio-temporal information. BIOT used linear transformers to reduce training time and was evaluated across multiple clinical tasks, including seizure detection and seizure type classification. BIOT showed superior performance even without pretraining and even better performance with pretraining, compared to previous supervised EEG-DL models.

\noindent\textbf{EEGFormer}~\cite{chen2024eegformer}:
This model was also pretrained on the full TUH corpus, which includes approximately 541k channel-hours of clinical scalp EEG data. It included a transformer-based encoder and shallow decoder, and was pretrained with a masked reconstruction objective. Latent features of input EEG patches generated by the encoder are used to train a vector quantizer to match neural codes generated by a neural codebook. The decoder then reconstructs the original EEG patches using these neural codes. Evaluations included several downstream tasks derived from the TUH corpus and an \textcolor{black}{out-of-distribution (OOD)} evaluation for neonatal seizure detection.
Additional experiments included an interpretability analysis using the learned codebook.

\noindent\textbf{LaBraM}~\cite{jiang_large_2024}:
\textcolor{black}{
This model used a relatively smaller dataset collected from multiple sources for pretraining, comprising 80k channel-hours of scalp EEG data. It was developed at three parameter scales: 5.8M, 46M, and 369M, respectively.}
\textcolor{black}{The LaBraM pretrained model consisted of two parts, the neural tokenizer and the neural transformer. The neural tokenizer is first trained to learn a codebook that accurately predicts the amplitude and phase of the raw time series data. Then, with the codebook frozen, the neural transformer is pretrained to predict the codebook-based codes for the masked input patches. Note that this differs from typical pretraining procedures, where the goal is to reconstruct masked patches in the time-domain itself. This transformer model consists of a convolutional temporal encoder and a set of transformer blocks, with temporal and spatial positional encodings used to learn relative relationships between input patches.}
Although the evaluations demonstrated performance gains in various subsets of the TUH corpus, it is unclear whether these results generalize to out-of-distribution data.

\noindent\textbf{Mentality}~\cite{panchavati2024mentality}: This model aims to capture the complex spatio-temporal dynamics of EEG signals using a Mamba~\cite{gu2023mamba}-based state-space model. The architecture of Mentality drew inspiration from other models such as SaShiMi~\cite{goel2022s}, U-Net~\cite{ronneberger2015u}, and EEGNet~\cite{lawhern2018eegnet} with the inclusion of Mamba blocks. However, the model was trained and \textcolor{black}{evaluated exclusively on a subset of the TUH corpus.}

\noindent\textbf{NeuroLM}~\cite{jiang_neurolm_2024}: This model was inspired by a previous EEG-FM, LaBraM. However, NeuroLM was trained on 7x more data and is one of the largest EEG-FMs, with 1.7B parameters. Smaller model variants include 250M- and 500M-parameter versions.
\textcolor{black}{The neural tokenizer is first trained to learn a codebook that is both aligned with a text embedding space and can accurately reconstruct the input time series data in both time and frequency domains. These codes are then fed into a large language model for multi-channel autoregressive pretraining. Finally, this pretrained model is adapted via joint multi-task instruction tuning, enabling a single model to perform diverse EEG tasks without task-specific fine-tuning. However, despite the novel EEG-text alignment and joint adaptation, NeuroLM performed worse on downstream tasks than LaBraM and other state-of-the-art models.}

\noindent\textbf{FoME}~\cite{shi_fome_2024}:
This model included two versions with 476M and 745M parameters, respectively. The size of the training data was not provided in the manuscript. The model takes masked time series patches and their power spectral densities as input, which are then transformed by a temporal encoder. The outputs of the temporal encoder are then reorganized by channel and fed to a spatial encoder to reconstruct masked time patches. FoME was evaluated across multiple downstream tasks, including classification, forecasting, and imputation; however, the evaluations were conducted on in-distribution data, limiting broader conclusions.

\noindent\textbf{BrainWave}~\cite{yuan2024brainwavebrainsignalfoundation}:
This model was pretrained using a large dataset of size 878k channel-hours, including both scalp EEG and iEEG. It includes a transformer encoder and a channel attention module that transform EEG spectrograms into latent representations, which are then decoded by a lightweight decoder. This encoder-decoder architecture was trained using a masked-reconstruction objective. BrainWave is one of the few EEG-FMs trained and evaluated on both scalp EEG and iEEG signals, demonstrating the benefits of joint pretraining over unimodal approaches. The model has been extensively evaluated under different settings, such as cross-subject, cross-hospital, cross-subtype and few-shot classification, showcasing the generalizability and robustness across various clinical tasks.

%% file: dim_comparisons.tex
\section{Comparative Analysis of EEG-FMs}
\label{sec:dim_comparisons}

\begin{figure}[!h]

    \centering
    \tcbox[size=small,on line]{\includegraphics[width=0.96\linewidth]{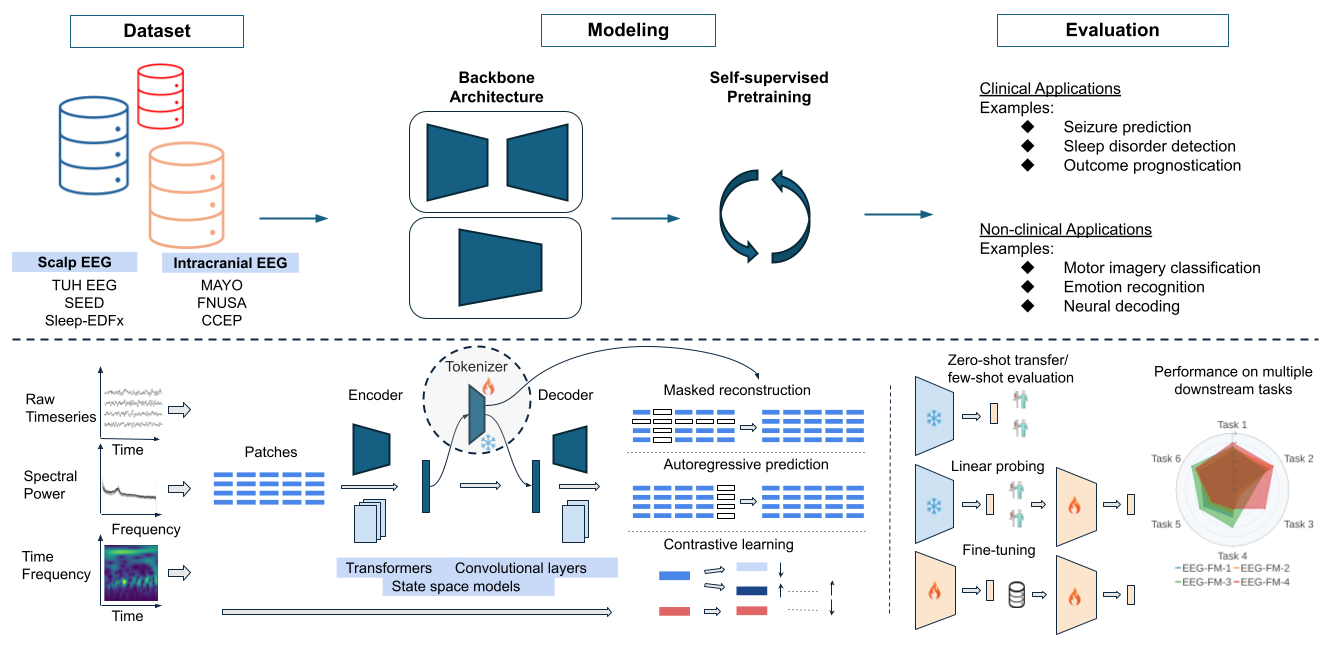}}
    \caption{\textcolor{black}{\textbf{Comparative analysis of EEG foundation models (EEG-FMs)}. Our review analyzes EEG-FMs along three major dimensions; input data configuration, modeling, and evaluation (top figure). A summary of the various approaches undertaken by the EEG-FMs to address those components is shown in the bottom figure. EEG data is represented in one of three forms: raw time series, magnitude power spectrum, and time-frequency representation. Model architecture may include convolutional blocks to learn low-level patterns and/or transformer blocks to learn higher-level relationships. Models are pretrained primarily using self-supervised learning (SSL) approaches; the common SSL approaches used are masked reconstruction, auto-regressive modeling, and contrastive learning. The pretrained models are then evaluated on various downstream tasks, including clinical and non-clinical tasks. 
    }}
    \label{fig:overview}
\end{figure}

This section surveys the ten EEG-FMs and compares their design and construction along three major axes covering the pillars of foundation modeling described in section \ref{sec:background}: a) preparation and representation of input data, b) model architecture and self-supervised pretraining, and c) model evaluations. The various considerations along these axes are illustrated in Figure \ref{fig:overview}.

\subsection{Preparation and Representation of Input Data}
\label{subsec:input}
\noindent\textbf{Datasets}: Pretraining data functions as the knowledge base of an FM, making it a highly crucial component. Key factors, such as diversity, data volume, relevance, and, most importantly, data quality, can influence the generalizability of the patterns learned by an EEG-FM. Table~\ref{tab:datasets} shows the datasets that were used in the pretraining and evaluation phases of the ten EEG-FMs reviewed in this article.

Most EEG-FMs (e.g., LaBraM, EEGFormer, Neuro-GPT, Mentality, BIOT, and NeuroLM) were trained only on scalp EEG data, although some (e.g., Brant and BrainBERT) were trained exclusively on iEEG data. Two EEG-FMs, FoME and BrainWave, were trained on both scalp and intracranial EEG data. 
\textcolor{black}{
A large portion of the pretraining data in the models used scalp EEGs from the Temple University Hospital (TUH) EEG corpus~\cite{obeid2016temple}, which comprises 541k channel-hours of scalp EEG data. This corpus also included smaller derived subsets containing expert annotations for EEG abnormalities (TUAB), seizures (TUSZ), and events (TUEV).
}
Other scalp EEG datasets used for pretraining include CHB-MIT~\cite{data:guttag2010chb} and small research datasets. Some FMs utilized EEG data available within PSG datasets such as SHHS~\cite{data:zhang2018national}, Sleep-EDF \cite{data:kemp2000analysis}, and CAP Sleep \cite{data:terzano2001atlas}. Additionally, public or private iEEG data accounted for a large portion of the training data in some models, such as Brant and BrainWave. The common publicly available iEEG datasets used for pretraining were MAYO ~\cite{nejedly_multicenter_2020}, FNUSA ~\cite{nejedly_multicenter_2020}, Brain TreeBank~\cite{data:wang2024brain}, and CCEP ~\cite{data:ds004080:1.2.4}. Training data volume ranged from 4.5k channel-hours in BrainBERT to 878k channel-hours in BrainWave. BrainWave utilized the most diverse pretraining dataset, including clinical scalp EEG datasets, sleep EEG datasets, various smaller-scale scalp EEG repositories, and several iEEG datasets. LaBraM and NeuroLM also used moderately diverse pretraining datasets, but were limited to the scalp EEG modality.

\noindent\textbf{Preprocessing and normalization}:
The content of EEG data is highly sensitive to the preprocessing and normalization steps performed \textcolor{black}{before model training or evaluation.}
The choices made for resampling, bandpass and notch filtering, artifact removal, \textcolor{black}{and temporal data segmentation into patches or epochs can directly impact the representations learned by an EEG-FM.} Although some models did not comprehensively describe the preprocessing steps, most applied standard procedures such as bandpass filtering (0.5Hz to a higher cutoff frequency), powerline interference removal via notch filters (50 Hz or 60 Hz, including harmonics), and downsampling to 200 Hz or 250 Hz to standardize sampling rates across datasets. EEG signals are then segmented into smaller epochs, typically 1-10 seconds long. However, none of the EEG-FMs performed explicit artifact removal or outlier exclusion, except in cases of expertly labeled bad data or missing channels. Other, less common steps included DC offset removal and linear trend removal by FoME and Neuro-GPT. Data normalization practices after preprocessing were not described in several manuscripts. Neuro-GPT applied the commonly used z-transform along the time dimension to normalize EEG signals, whereas NeuroLM, BIOT, and LaBraM used a constant scaling factor based on the input range.

\noindent\textbf{Input representation}:
EEG is naturally a spatio-temporal data modality recorded using multiple channels. 
In addition to spatial and temporal information, spectral information is useful for interpreting EEG data, which is most often used as expert-derived features in statistical machine learning (ML) models. Furthermore, time-frequency representations, such as spectrograms and wavelet~\cite{grossmann1984decomposition} transforms, are also used to preserve information in all three domains. All EEG-FMs used one or a combination of these representations as input.

The most widely used input representation is the multivariate time-series format, adopted by six of the ten EEG-FMs. With the adoption of a transformer architecture, the windows segmented from the original recording require further segmentation into smaller patches, also known as tokens. These patches or tokens can then be augmented with temporal and spatial information via positional encodings, allowing the models to be trained on EEG data with different channel configurations. This approach was adopted by models such as Brant, NeuroLM, and FoME to pretrain on datasets with varying channel counts. Two models, BIOT and EEGFormer, used power-spectral data as input instead of the original time-series patches, while two other models, Brant and FoME, added power-spectral data to complement the raw time-series inputs. Some EEG-FMs, such as LaBraM, Neuro-GPT, and Mentality, standardized different EEG datasets with different channel configurations to a fixed set of input channels. Two other models, BrainBERT and BrainWave, utilized channel-wise time-frequency representations of 1-5 seconds of EEG data as input. BrainBERT evaluations utilized spectrograms generated using classical methods like STFT, as well as scalograms~\cite{grossmann1984decomposition} generated using modern methods such as Superlets.

\begin{table}[]
\definecolor{clinical}{RGB}{255, 182, 193}  
\definecolor{nonclinical}{RGB}{173, 216, 230}  
 \caption{\textcolor{black}{\textbf{Datasets used for pretraining and evaluation.} Here we list the diverse set of datasets and tasks that were used to pretrain and evaluate EEG foundation models (EEG-FMs). We highlight datasets that are not publicly available in bold, clinical and non-clinical datasets in pink and blue backgrounds, respectively, and use superscripts $*$ and \# to denote in-distribution and out-of-distribution evaluations, respectively.}}
  \centering\scriptsize
  \resizebox{\columnwidth}{!}{
\begin{tabular}{|L{7cm}|c|c|c|c|c|c|c|c|c|c|}
\toprule
 \textbf{Dataset/EEG-FM} & \textbf{BrainBERT} & \textbf{Neuro-GPT} & \textbf{Brant} & \textbf{BIOT} & \textbf{EEGFormer} & \textbf{LaBraM} & \textbf{Mentality} & \textbf{NeuroLM} & \textbf{FoME} & \textbf{BrainWave} \\ \midrule

 \centering\textbf{Scalp EEG} &  &  &  &  &  &  &  &  &  &  \\ \midrule

 \cellcolor{clinical}TUEG~\cite{data:obeid2016temple} &  & $train$ &  &  & $train$ &  &  & $train$ & $train$ & $train$ \\ \midrule
 \cellcolor{clinical}\quad TUAB (Abnormal/normal classification - binary) &  &  &  & $eval^\#$ & $eval^*$ & $eval^*$ &  & $eval^*$ &  &  \\ \midrule
 \cellcolor{clinical}\quad TUAR (EEG artifact classification - multiclass)  &  &  &  &  & $eval^*$ & $train$ &  &  &  &  \\ \midrule
 \cellcolor{clinical}\quad TUEP (Epilepsy classification - binary) &  &  &  &  &  & $train$ &  &  &  &  \\ \midrule
 \cellcolor{clinical}\quad TUEV (EEG Events classification - multiclass) &  &  &  & $eval^\#$ &  & $eval^*$ &  & $eval^*$ & $eval^*$ &  \\ \midrule
 \cellcolor{clinical}\quad TUSL (EEG slowing classification - multiclass) &  &  &  &  & $eval^*$ & $train$ &  & $eval^*$ &  &  \\ \midrule
 \cellcolor{clinical}\quad TUSZ (Seizure type classification - multiclass) &  &  &  &  & $eval^*$ & $train$ & $train/eval^*$ &  &  &  \\ \midrule
 \cellcolor{clinical}CHB-MIT~\cite{data:guttag2010chb} (Peadiatric seizure detection - binary) &  &  &  & $eval^\#$ &  &  &  &  & $train$ & $eval^\#$ \\ \midrule
 \cellcolor{clinical}Sleep-EDFx~\cite{data:kemp2000analysis} (Sleep stage classification - multiclass) &  &  &  &  &  &  &  &  & $train/eval^*$ & $train$ \\ \midrule
 \cellcolor{clinical}Siena Scalp EEG~\cite{data:detti2020eeg} (Seizure classification - binary) &  &  &  &  &  & $train$ &  & $train$ &  & $train$ \\ \midrule
\cellcolor{clinical}SHHS~\cite{data:quan1997sleep}\cite{data:zhang2018national} (Sleep stage classification - multiclass) &  &  &  & $train$ &  &  &  &  &  &  \\ \midrule
 \cellcolor{clinical} \textbf{PREST} (Abnormal event detection - binary) &  &  &  & $train$ &  &  &  &  &  &  \\ \midrule
 \cellcolor{clinical}CAP Sleep~\cite{data:terzano2001atlas} (Sleep stage classification - multiclass) &  &  &  &  &  &  &  &  &  & $train$ \\ \midrule
 \cellcolor{clinical}HMC~\cite{data:alvarez2021inter} (Sleep stage classification - multiclass) &  &  &  &  &  &  &  & $eval^\#$ &  & $train$ \\ \midrule
 \cellcolor{clinical}SRM~\cite{data:hatlestad2022bids} (Resting state EEG data) &  &  &  &  &  &  &  &  &  & $train$ \\ \midrule
 \cellcolor{clinical}Schizophrenia-81 (Schizophrenia classification - binary) &  &  &  &  &  &  &  &  &  & $train$ \\ \midrule
 \cellcolor{clinical}Stroke-50~\cite{data:liu2024eeg} (Hand movement classification - binary) &  &  &  &  &  &  &  &  &  & $train$ \\ \midrule
 \cellcolor{clinical}PD-31~\cite{data:ds002778:1.0.5} (Parkinson disease classification - binary) &  &  &  &  &  &  &  &  &  & $train$ \\ \midrule
 \cellcolor{clinical}IowaDataset~\cite{data:anjum2020linear} (Unknown task/s) &  &  &  &  &  &  &  &  &  & $train$ \\ \midrule
\cellcolor{clinical}UNMDataset~\cite{data:cavanagh2018diminished} (Parkinson disease classification - binary) &  &  &  &  &  &  &  &  &  & $train$ \\ \midrule
 \cellcolor{clinical}AD-184~\cite{data:vicchietti2023computational} (Alzheimer’s disease classification - binary) &  &  &  &  &  &  &  &  &  & $train$ \\ \midrule
 \cellcolor{clinical}Neonate dataset~\cite{data:stevenson2019dataset} (Seizure detection - binary) &  &  &  &  & $eval^\#$ &  &  &  &  &  \\ \midrule
 \cellcolor{clinical}IIIC Seizure~\cite{data:ge2021deep} (IIIC pattern classification) - multiclass &  &  &  & $eval^\#$ &  &  &  &  &  &  \\ \midrule
 \cellcolor{clinical} \textbf{Absence-16} (Seizure type classification - multiclass) &  &  &  &  &  &  &  &  &  & $eval^\#$ \\ \midrule
 \cellcolor{clinical} \textbf{Clonic-6} (Seizure type classification - multiclass) &  &  &  &  &  &  &  &  &  & $eval^\#$ \\ \midrule
 \cellcolor{clinical}\textbf{ Atonic-5} (Seizure type classification - multiclass) &  &  &  &  &  &  &  &  &  & $eval^\#$ \\ \midrule
 \cellcolor{clinical} \textbf{DRE-Clinical} (Seizure detection and localization - binary) &  &  &  &  &  &  &  &  &  & $eval^\#$ \\ \midrule
 \cellcolor{clinical}SD-71~\cite{data:xiang2024resting} (Sleep deprivation detection - binary) &  &  &  &  &  &  &  &  &  & $eval^\#$ \\ \midrule
 \cellcolor{clinical}ADHD-Adult~\cite{data:trinh2023task} (ADHD classification - binary) &  &  &  &  &  &  &  &  &  & $eval^\#$ \\ \midrule
 \cellcolor{clinical}ADHD-Child~\cite{data:motie2020eeg} (ADHD classification - binary) &  &  &  &  &  &  &  &  &  & $eval^\#$ \\ \midrule
 \cellcolor{clinical}Schizophrenia-28~\cite{data:olejarczyk2017graph} (Schizophrenia classification - binary) &  &  &  &  &  &  &  &  &  & $eval^\#$ \\ \midrule
 \cellcolor{clinical}Depression-122~\cite{data:ds003478:1.1.0} (Depression classification - binary) &  &  &  &  &  &  &  &  &  & $eval^\#$ \\ \midrule
 \cellcolor{clinical}MDD-64~\cite{data:mumtaz2016mdd} (Major depression detection - binary) &  &  &  &  &  &  &  &  &  & $eval^\#$ \\ \midrule
 \cellcolor{clinical}AD-65~\cite{data:miltiadous2023dataset} (Alzheimer’s disease classification - binary) &  &  &  &  &  &  &  &  &  & $eval^\#$ \\ \midrule

 \cellcolor{nonclinical}BCI Competition IV-1~\cite{data:blankertz2007non} (Movement classification - binary) &  &  &  &  &  & $train$ &  & $train$ &  &  \\ \midrule
\cellcolor{nonclinical}Emobrain~\cite{data:savran06_einterface} (Emotion classification - multiclass) &  &  &  &  &  & $train$ &  & $train$ &  &  \\ \midrule
 \cellcolor{nonclinical}Grasp/Lift Challenge~\cite{data:luciw2014multi} (Grasp classification/regression) &  &  &  &  &  & $train$ &  & $train$ &  &  \\ \midrule
 \cellcolor{nonclinical}Inria BCI Challenge~\cite{data:margaux2012objective} (Next letter prediction - multiclass) &  &  &  &  &  & $train$ &  & $train$ &  &  \\ \midrule
 \cellcolor{nonclinical}EEG Motor Imagery ~\cite{data:schalk2004bci2000} (Classification - multiclass) &  &  &  &  &  & $train$ &  & $train$ & $train$ &  \\ \midrule
 \cellcolor{nonclinical}Visual Categorization EEG~\cite{data:trujillo2019mental} (Classification - multiclass) &  &  &  &  &  & $train$ &  & $train$ &  &  \\ \midrule
 \cellcolor{nonclinical}Resting-state EEG ~\cite{data:trujillo2017effect} (Eye state classification - binary) &  &  &  &  &  & $train$ &  & $train$ &  &  \\ \midrule
 \cellcolor{nonclinical}SEED Series~\cite{data:zheng2015investigating, data:zheng2018emotionmeter, data:liu2022identifying} (Emotion classification - multiclass) &  &  &  &  &  & $train/eval^*$ &  & $train/eval^*$ & $train/eval^*$ &  \\ \midrule
 \cellcolor{nonclinical}SPIS Resting State EEG~\cite{data:torkamani2020prediction} (CVS and HRT regression) &  &  &  &  &  & $train$ &  & $train$ &  &  \\ \midrule
 \cellcolor{nonclinical}Brain-Invaders~\cite{data:korczowski2019brain} (Target classification - binary) &  &  &  &  &  & $train$ &  & $train$ &  &  \\ \midrule

 \cellcolor{nonclinical}Multiple data sources~\cite{data:jiang2021discriminating, data:jiang2023multimodal, data:luo2022multimodal, data:li2021discrimination, data:tao2020emotion} (Unknown tasks) &  &  &  &  &  & $train$ &  & $train$ &  &  \\ \midrule

 \cellcolor{nonclinical}MoBI~\cite{data:he2018mobile} (Gait angle regression) &  &  &  &  &  & $eval^\#$ &  &  &  &  \\ \midrule
 
 \cellcolor{nonclinical}BCI Competition IV Dataset 2a~\cite{data:brunner2008bci} &  & $eval^\#$ &  &  &  &  &  &  &  &  \\ \midrule
 \cellcolor{nonclinical}Workload~\cite{data:zyma2019electroencephalograms} (Workload classification - binary) &  &  &  &  &  &  &  & $eval^\#$ &  &  \\ \midrule

 \centering\textbf{Intracranial EEG} 
 &  &  &  &  &  &  &  &  &  &  \\ \midrule

  \cellcolor{clinical}Brain TreeBank~\cite{data:wang2024brain} (EEG classification - multiclass) & $train/eval^*$ &  &  &  &  &  &  &  &  &  \\ \midrule
 \cellcolor{clinical} \textbf{Private single dataset without source} (Unknown) &  &  & $train/eval^*$ &  &  &  &  &  &  &  \\ \midrule
 \cellcolor{clinical} \textbf{Private dataset collection without source} (Unknown) &  &  &  &  &  &  &  &  &  & $train$ \\ \midrule
 \cellcolor{clinical}MAYO~\cite{data:nejedly2020multicenter} (Pathology detection \& events classification) &  &  & $eval^\#$ &  &  &  &  &  & $train/eval^*$ & $eval^\#$ \\ \midrule
 \cellcolor{clinical}FNUSA~\cite{data:nejedly2020multicenter} (Pathology detection \& events classification) &  &  & $eval^\#$ &  &  &  &  &  & $train/eval^*$ & $eval^\#$ \\ \midrule
 \cellcolor{clinical}CCEP~\cite{data:ds004080:1.2.4} (Seizure onset zone localization - binary) &  &  &  &  &  &  &  &  &  & $train$ \\ \bottomrule

\end{tabular}
 \label{tab:datasets}
 }
\end{table}

\subsection{Modeling and Pretraining}
\label{subsec:ssl_modeling}

\noindent \textbf{Patching and context length}: The temporal resolution of tokens or patches, i.e., how much EEG data each patch contains, varies between models, reflecting different design choices in model architecture and target tasks. Brant and FoME used patches corresponding to six seconds of EEG, while BrainBERT used 5-second STFT/Superlet windows. Models such as LaBraM, NeuroLM, BIOT, and BrainWave adopted a finer resolution, using 1-second segments per patch, whereas Neuro-GPT utilized a patch size of two seconds. Similarly, context length also varied among EEG-FMs. FoME was trained on 90-second segments, each represented by a sequence of 15 tokens. LaBraM could take up to 256 tokens as input, with the number of temporal tokens depending on the number of channels used. Neuro-GPT utilized a 32-token context window with overlapping segments, covering $\sim$57.8 seconds of EEG. BIOT utilized 19 tokens per channel and EEGFormer utilized 12 seconds of EEG context as input.

\noindent\textbf{Architectural components}: The model architecture and training objectives play a central role in determining the effectiveness of a foundation model in extracting information from the input data~\cite{bommasani2021opportunities}. EEG-FMs utilized various components at different stages of the pipeline, including tokenizers, local representation learning modules, spatial and/or temporal attention modules to learn dependencies, and task heads. Almost all the EEG-FMs we review in this article are transformer-based, except Mentality, which uses a combination of convolutional layers and a mamba-based architecture. In the transformer-based models, the input signal is divided into a fixed number of small patches/tokens to construct the input sequence. Some models that used raw time-series inputs leveraged convolutional layers to learn morphological features (e.g., Neuro-GPT, LaBraM, Mentality, NeuroLM).

\noindent\textbf{Neural tokenizers}: Some models, such as LaBraM, NeuroLM, and EEGFormer, integrated separate tokenizers to transform the input signals into discrete tokens or codes. These EEG-FMs adapted a VQ-VAE~\cite{van2017neural}-like architecture to train a neural tokenizer, which mapped input patches into a fixed set of discrete embeddings called the \textit{codebook}. During tokenizer training, latent features are mapped to nearest-neighbor embeddings from the codebook to reconstruct known features of the input, such as power and phase in LaBraM, and temporal and spectral features in NeuroLM. The models used neural codebooks to reconstruct raw time-series inputs using a transformer-like architecture. While NeuroLM and EEGFormer used codebook embeddings to map encoder outputs to discrete codes prior to decoding, LaBraM used codebook embeddings of masked input patches as targets during model pretraining.

\noindent\textbf{Spatial encoding}: EEG-FMs implemented different spatial encoding strategies, although they differ in how explicitly they handle spatial information. Brant and FoME, for example, used transformer encoders with spatial attention but without dedicated spatial positional encodings. In contrast, LaBraM, BIOT, and NeuroLM injected spatial information through explicit spatial encodings to enhance positional encoding and capture the topographic layout of EEG channels. Among them, only BIOT supported multiple electrode configurations, while LaBraM and NeuroLM are limited to a fixed 10–20 layout for spatial encoding. Models such as Neuro-GPT and Mentality learned spatial dependencies implicitly using convolutional layers, without dedicated spatial embeddings. BrainWave incorporated spatial attention mechanisms to model inter-channel dependencies directly. In contrast, BrainBERT and EEGFormer omit spatial modeling altogether, using only transformer layers without convolution or spatial priors.

\noindent\textbf{Self-supervision}: SSL approaches are key components that enable the development of foundation models using massive unlabeled datasets~\cite{bommasani2021opportunities}. Most EEG-FMs (seven out of ten) employed reconstruction of masked data as the primary SSL approach. EEGFormer model was pretrained by minimizing the reconstruction loss between input signals and signals decoded from codebook mappings of transformer-encoded input patches. NeuroLM employed an autoregressive approach to predict future patches from past patches, including an EEG-text alignment objective, and BIOT adopted a contrastive learning approach where the embeddings of masked inputs and the same inputs with different augmentations are minimized (i.e., the SimCLR~\cite{chen2020simple} pretraining approach). Only one model — Neuro-GPT — employed masked data reconstruction in the latent space. Additionally, the LaBraM model employed a masked-signal reconstruction approach; however, the codebook embeddings of masked input patches from a previously trained neural tokenizer were used as reconstruction targets.

\noindent \textbf{Masking}: The granularity of the learned representations in the EEG-FMs is influenced by the type and extent of masking used during self-supervised pretraining~\cite{he2022masked}. In models based on masked-signal reconstruction, masking is typically applied to random patches across temporal and spatial dimensions, as opposed to masking entire channels or full time segments. For example, Brant and FoME applied a $\sim$40\% uniform random masking, zeroing out the masked input signal patches across time and channel dimensions. LaBraM similarly utilized a 50\% patch-level masking, while BrainWave, despite operating in the spectrogram domain, performed masking entire spectrograms of several channels. On the other hand, BrainBERT applied a more fine-grained masking strategy, masking 5\% from the time domain and 5\% from the frequency domain, effectively masking $\sim$10\% of the input spectrogram. Mentality applied random channel masking, targeting specific channels rather than patches. NeuroLM introduced a stair-step masking scheme, in which next token is auto-regressively predicted in a channel-conditioned manner, whereas Neuro-GPT utilized a causal masking approach in the latent space, consistent with autoregressive generation.

\vspace{-0.5mm}
\noindent\textbf{Model scale}:
The number of trainable parameters (or weights) varied significantly between the ten EEG-FMs, ranging from 3.3M in BIOT to 1.7B in NeuroLM, although most models fall within the 200-500M parameter range. Some EEG-FMs, such as FoME, LaBraM, and NeuroLM, were developed at multiple scales with the same architecture, allowing users to choose a model scale according to their specific resource constraints.

\vspace{-0.5mm}
\subsection{Model Evaluations}
\label{subsec:model_evaluations}

\noindent\textbf{Downstream task evaluations}: Performance on downstream tasks post-adaptation was the primary metric used for evaluation in all the EEG-FMs. However, the clinical and non-clinical downstream tasks used for evaluation and the adaptation techniques varied significantly between them. The various datasets used for evaluation are shown in Table \ref{tab:datasets}. While most EEG-FM evaluations were restricted to either scalp or intracranial EEG, two models, namely FoME and BrainWave, were evaluated on both modalities. Furthermore, most FMs used common adaptation techniques, namely linear probing and fine-tuning, while some used prototype-based few-shot evaluation and instruction tuning.

\textcolor{black}{Almost all EEG-FMs, except Mentality, investigated fine-tuning for downstream task evaluations, in which the pretext task decoder is discarded but the encoder is trained further using task-specific labels.} On the other hand, five EEG-FMs (i.e., Brant, BrainBERT, LaBraM, EEGFormer, and Neuro-GPT) evaluated linear probing, where the encoder is typically frozen, and additional task-specific heads on top of the encoder are trained using labeled downstream data. The BrainWave model was evaluated using a prototype-based few-shot learning approach, where class prototypes were created by averaging the latent representations of a few examples per class, and downstream classification was performed by measuring similarity to those prototypes in the learned latent space. Furthermore, NeuroLM, which was pretrained with an EEG-text alignment objective, was evaluated using a text-prompting approach. The model was fine-tuned (instruction-tuned) using an auto-regressive prediction task utilizing EEG-text combined tokens, where the text tokens contained both class prompts and labels. Subsequently, the model was evaluated on various tasks by providing queries containing EEG tokens and prompts, for which the model predicted the class labels. In the following, we briefly summarize the evaluations performed by each EEG-FM and the results.

\noindent \ul{BrainBERT}: 
This model was trained and evaluated on the same intracranial EEG dataset, Brain TreeBank. The downstream task focused on predicting the features of the movies that the subjects were watching during the EEG recording.
Fine-tuning the pretrained BrainBERT model on this task substantially outperformed fully-supervised linear and multi-layer neural network models. Linear probing also achieved Area Under the Receiver Operating Characteristic Curve (AUC) values similar to those of the fully supervised models.

\noindent \ul{Neuro-GPT}:
This model was evaluated on a single scalp EEG dataset that was not used during pretraining, with the downstream task being 4-class motor-imagery classification. The fully-supervised baselines considered in this study include BENDR~\cite{kostas2021bendr}, SVM~\cite{oikonomou2017comparison}, EEGNet~\cite{lawhern2018eegnet}, CTCNN~\cite{schirrmeister2017deep}, CCNN~\cite{amin2019deep}, and NG-CRAM~\cite{zhang2020motor}. When used in a supervised setting without any pretraining, this model performed similarly to other supervised baselines. However, when the pretrained model was fine-tuned for the downstream task, it provided significant improvements in certain configurations. In contrast, linear probing performed worse than fully-supervised baselines.

\noindent \ul{Brant}:
\textcolor{black}{This model was evaluated on three unseen intracranial EEG datasets, including one private and two public datasets (MAYO and FNUSA).} The downstream tasks evaluated include signal forecasting, imputation, \textcolor{black}{seizure detection, and pathology detection}. For all these tasks, the pretrained model was fine-tuned or linear-probed using task labels. The self-supervised pretrained baselines considered in this study included (RP, TS, CPC)~\cite{banville2021uncovering}, BENDR~\cite{kostas2021bendr}, MVTS~\cite{potter2022unsupervised}, and BrainBERT that are designed for brain signals, and CoST~\cite{woo2022cost}, TF-C~\cite{zhang2022self}, PatchTST~\cite{nieTimeSeriesWorth2023}, TS-TCC~\cite{eldele2021time} that are designed for general time series. Fully-supervised baselines without any pretraining included handcrafted features-based methods such as spectral power~\cite{zhang2015low}, rhythmicity spectrogram~\cite{handa2021epileptic}, and amplitude-integrated EEG~\cite{kadivar2019comparison} along with SEEG-Net~\cite{wang2022seeg}. In signal forecasting and imputation tasks, fine-tuned Brant consistently outperformed several time series baselines and its own linear probing version. In the \textcolor{black}{seizure detection and pathology detection tasks}, both the fine-tuned and linear probed versions of Brant outperformed other supervised and EEG-FM baselines, including BrainBERT. These observations held true in low-labeled settings as well.

\noindent \ul{BIOT}:
This model was evaluated entirely on scalp EEG datasets on various clinical tasks. The supervised baselines considered in this study included SPaRCNet~\cite{jing2023development}, ContraWR~\cite{yang2021self}, CNN-Transformer~\cite{peh2022transformer}, FFCL~\cite{li2022motor}, ST-Transformer~\cite{song2021transformer}. Evaluations showed that the finetuned BIOT encoder outperformed several fully-supervised baselines in CHB-MIT seizure detection and TUAB normal/abnormal classification. In two other tasks, namely IIIC seizure classification and TUEV event classification, both the vanilla untrained BIOT in the fully supervised setting and the pretrained BIOT model in the fine-tuning setting outperformed all supervised baselines.

\noindent \ul{EEGFormer}: 
\textcolor{black}{
The three variants of EEGFormer were evaluated against four fully-supervised models (i.e., EEGNet~\cite{lawhern2018eegnet}, TCN~\cite{bai2018empirical}, EEG-GNN~\cite{tang2021self}, and GraphS4mer~\cite{tang2023modeling}) and BrainBERT as baselines, using five different scalp EEG datasets/tasks.
}
Fine-tuned EEGFormer outperformed all baselines in all the tasks except one, where its performance was marginally lower than EEG-GNN. Their evaluations also showed that a smaller version of EEGFormer demonstrated less variability in some downstream evaluations.

\noindent \ul{LaBraM}:
This model was evaluated on four scalp EEG datasets: two clinical (TUAB and TUEV) and two non-clinical (SEED-V and MoBI), with MoBI as the only external dataset. The evaluations considered several supervised models, SPaRCNet~\cite{jing2023development}, ContraWR~\cite{yang2021self}, CNN-Transformer~\cite{peh2022transformer}, FFCL~\cite{li2022motor}, ST-Transformer~\cite{song2021transformer}, and BIOT as baselines. \textcolor{black}{All variations of LaBraM (base, large, and huge) with fine-tuning outperformed baselines in all tasks.}

\noindent \ul{Mentality}:
This model was evaluated on the TUSZ seizure detection dataset, which was also used for pretraining. The evaluation compared an untrained Mentality model trained in a fully supervised setting with a pretrained Mentality model linearly probed with two linear layers, showing that pretraining is beneficial for this task.

\noindent \ul{NeuroLM}:
\textcolor{black}{This model was fine-tuned jointly on six tasks and was compared against seven baselines fine-tuned individually per task.}
The evaluations considered supervised models SPaRCNet~\cite{jing2023development}, ContraWR~\cite{yang2021self}, CNN-Transformer~\cite{peh2022transformer}, FFCL~\cite{li2022motor}, ST-Transformer~\cite{song2021transformer}, BIOT, and LaBraM as baselines. Evaluations showed that joint fine-tuning with prompting is feasible and achieves reasonable performance across all tasks. However, NeuroLM did not perform as well as the LaBraM model on any of the tasks, where LaBraM was fine-tuned separately for each task. Evaluations showed that the TUAB and TUEV performance of NeuroLM variants was not sensitive to model size, whereas the performance on HMC and Workload tasks decreased with larger variants.

\noindent \ul{FoME}:
This model was evaluated on 7 tasks across 4 datasets (MAYO, FNUSA, SEED, SleepEDFx), and evaluations considered several fully-supervised models (LSTM~\cite{hochreiter1997long}, ConvNeXt~\cite{liu2022convnet}), general time series models (PatchTST, TimesNet), and some EEG-FMs (BrainBERT, Neuro-GPT, LaBraM) as baselines. Evaluations showed that, in emotion classification, sleep staging, \textcolor{black}{events classification, }and pathology detection tasks, FoME performed competitively or better than all baselines, whereas its performance was significantly better in signal forecasting tasks. Interestingly, the evaluations also demonstrated that general time-series models, such as PatchTST and TimesNet, perform reasonably well on pathology detection after fine-tuning, despite not being pretrained on brain signals.

\noindent\ul{BrainWave}:
This model was evaluated using multiple scalp and intracranial EEG datasets and was compared against LaBraM -- a scalp-EEG-FM, BrainBERT -- an intracranial EEG-FM, and MOMENT~\cite{goswami2024moment} -- a general time series FM. Evaluations considered different criteria, including class prototype-based few-shot classification and cross-domain transfer learning across subjects, sites, and diagnostic subtypes. Results showed that BrainWave significantly outperformed all the baselines in all settings, with an average improvement of 0.21 AUC points across 12 tasks, establishing its superiority in challenging out-of-distribution evaluations.

\begin{figure}[!h]
     \centering
     \begin{subfigure}[b]{0.495\textwidth}
         \centering
         \includegraphics[width=\textwidth]{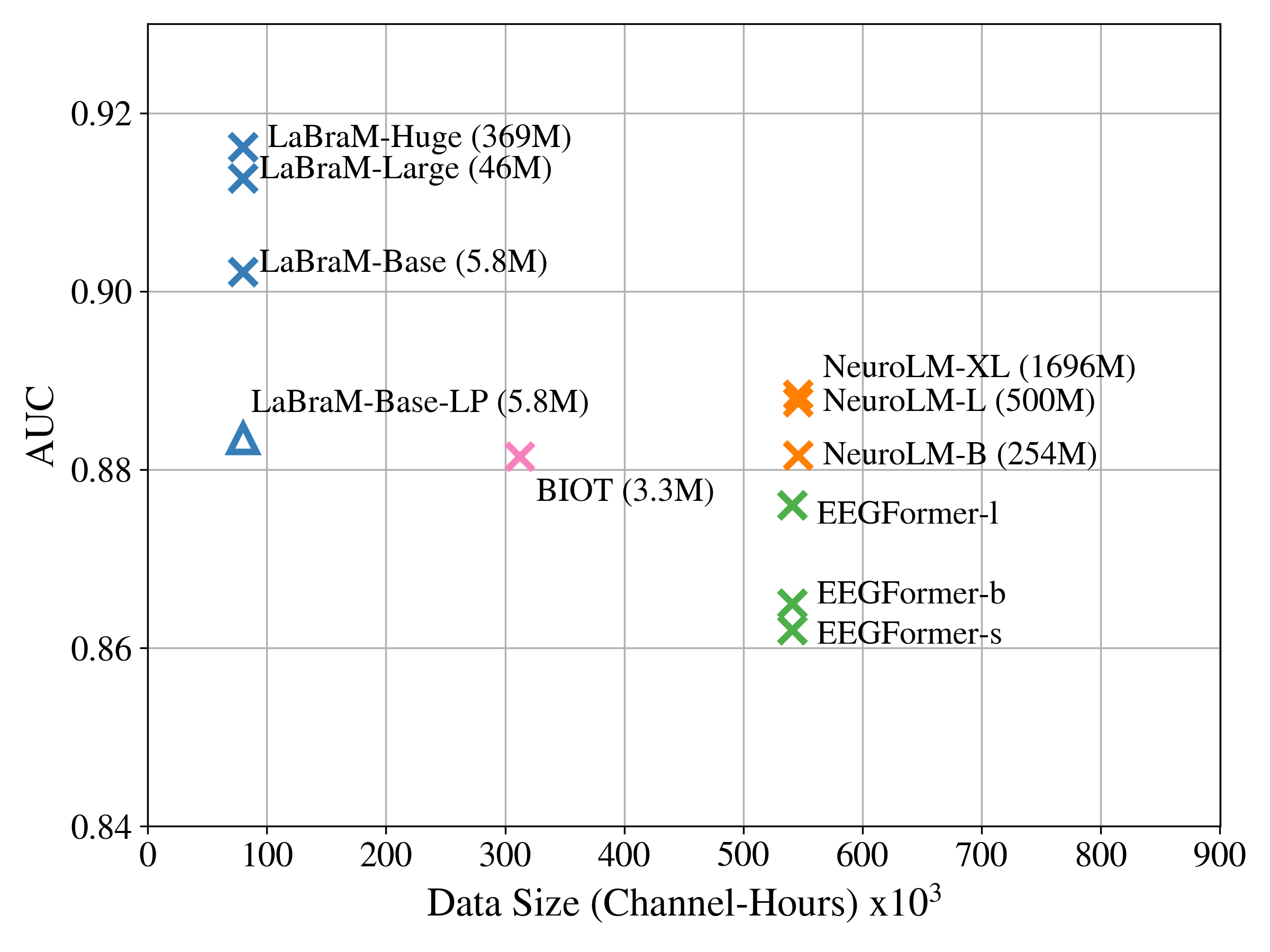}
         \centering
         \caption{
         TUAB: normal vs. abnormal classification.
         }
         \label{fig:tuab}
     \end{subfigure}
     \hfill
     \begin{subfigure}[b]{0.495\textwidth}
         \centering
         \includegraphics[width=\textwidth]{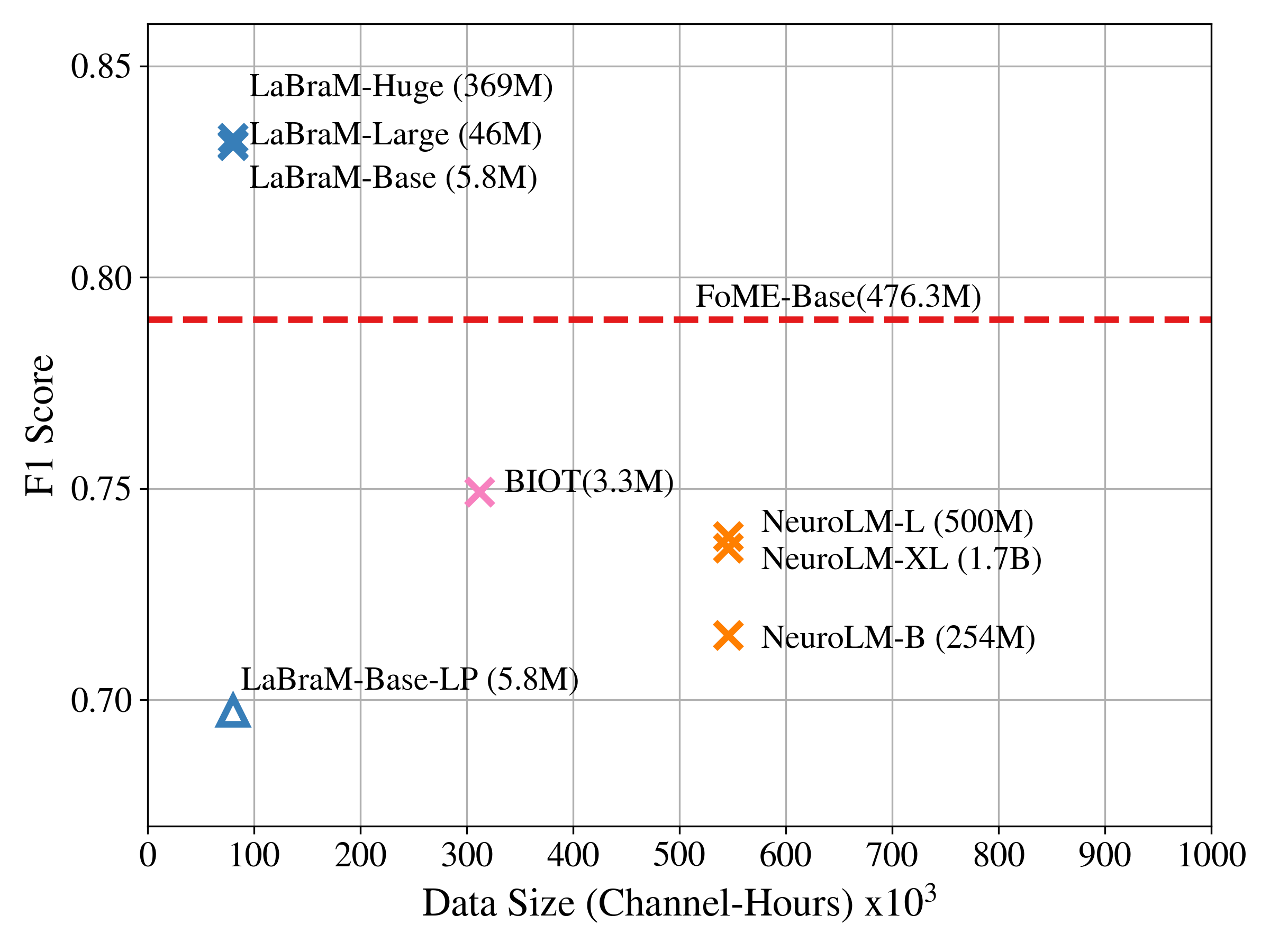}
         \caption{
         TUEV: six-class event classification.
         }
         \label{fig:tuev}
     \end{subfigure}
     
    \caption{\textcolor{black}{\textbf{Performances on common tasks}. Here we compare EEG foundation models based on their performance on the common Temple University EEG Corpus tasks -- TUAB (abnormal EEG classification) and TUEV (event classification) -- along with the size of the dataset used for pretraining. The performance of FoME in \ref{fig:tuev} is shown as a line because the pretraining data size was unavailable. Channel-hours are on a $10^{3}$ scale. All scores represent fine-tuned model performance, except for the triangular markers, which represent linear-probed model performance.\vspace{-1em}}}
    \label{fig:comparison}
\end{figure}

\noindent\textbf{Evaluations on common tasks}: We observed that two datasets, TUAB for abnormal EEG classification (two classes) and TUEV for EEG event detection (six classes), have been widely utilized for downstream evaluation. Note that both TUAB and TUEV are scalp EEG datasets derived from the larger TUH EEG corpus. In Figure~\ref{fig:tuab}-\ref{fig:tuev}, we compare the performances reported on these common tasks. We also analyzed the impact of the pretraining data size on the x-axis. For comparison, we use AUC and F1 scores as performance metrics, as they are the most commonly reported metrics for these two tasks. Linear probing results were available only in LaBraM and are indicated by a triangle in the figure. The remaining results are based on fine-tuning experiments conducted by LaBraM, NeuroLM, EEGFormer, and BIOT. We find that AUC values ranged $0.86-0.92$ for TUAB abnormal EEG classification and F1-scores ranged $0.70-0.83$ for TUEV event classification, highlighting significant performance variation across models. \textcolor{black}{These comparisons also suggested that increasing the size of the pretraining dataset did not necessarily improve model performance on these two tasks.}

Apart from downstream task evaluations, some models performed additional evaluations focused on model interpretability, pretraining quality, and ablations. We briefly summarize those evaluations below.

\noindent\textbf{Interpretability analyses}:
An analysis performed in BrainBERT used the intrinsic dimensionality (ID) measure to assess task-specific embeddings post-finetuning. ID is a geometric measure that quantifies the minimum number of parameters required to represent embeddings \cite{ansuini2019intrinsic}. 
\textcolor{black}{
They showed that, in their model, the electrode ID distribution lies in a lower-dimensional space compared to a randomly initialized model. Furthermore, they showed that electrodes in the top 10th percentile of IDs mainly fall in brain regions associated with the task.}
Similarly, EEGFormer, one of the three models that utilized a vector quantizer to train a neural codebook, analyzed the learned codebook to attribute parts of the input EEG data to downstream task predictions. \textcolor{black}{
In their analysis, codes most likely to predict seizures were mapped back to the EEG traces to visualize where the model focused. These highlighted regions corresponded to known epileptiform patterns, indicating that the model’s decisions aligned with meaningful clinical markers.}

\noindent\textbf{Pretraining quality}: 
Some models evaluated pretraining quality through qualitative and quantitative assessments. Most qualitative assessments were based on visualizations of data reconstructed by the specific models, which were run in Mentality, LaBraM, and BrainBERT. Quantitative assessments focused on reconstruction error (e.g., Mentality) and performance on forecasting and imputation tasks (e.g., Brant, FoME).

\noindent\textbf{Ablations}: Few models evaluated the contributions of various model components or datasets to task performance via ablations. An experiment in Neuro-GPT, which combined a convolutional transformer and a GPT decoder, showed that removing the GPT decoder significantly improved downstream performance. Experiments in Brant evaluated the contributions of temporal, spatial, and frequency encoders and demonstrated that, while all components were valuable for learning EEG representations, the temporal encoder made the greatest contribution. 
Additionally, LaBraM showed that spatial embeddings are crucial for pretraining to converge and for downstream performance. 
Furthermore, LaBraM studied the effect of the codebook during pretraining by replacing the neural tokenizer with traditional time- and frequency-domain masked reconstruction and showed that this degraded performance for complex downstream tasks such as TUEV.
NeuroLM showed that different neural tokenizer training tasks (frequency vs. temporal vs. frequency and temporal reconstruction) can be beneficial for different downstream tasks.
Additionally, NeuroLM examined the impact of model pretraining duration by fine-tuning different checkpoints, but the results were inconclusive.
An experiment in BrainWave performed data ablation by training separate models with scalp and intracranial EEG and compared their performance with joint scalp-iEEG training. Their results indicated that joint training generally improved downstream performance in all tasks, except one.

%% file: takeaways_gaps.tex
\section{Key Takeaways and Research Gaps}
\label{sec:takeaways_gaps}
This section presents several key observations made across the preceding comparative analyses (Section \ref{sec:dim_comparisons}) and highlights the research gaps that remain. We believe that addressing these gaps would make future EEG-FMs more principled in their methodology and trustworthy for scientific and clinical investigations.

\begin{figure}[!h]
    \centering
    \begin{subfigure}[b]{0.95\textwidth}
         \centering
         \includegraphics[width=\textwidth]{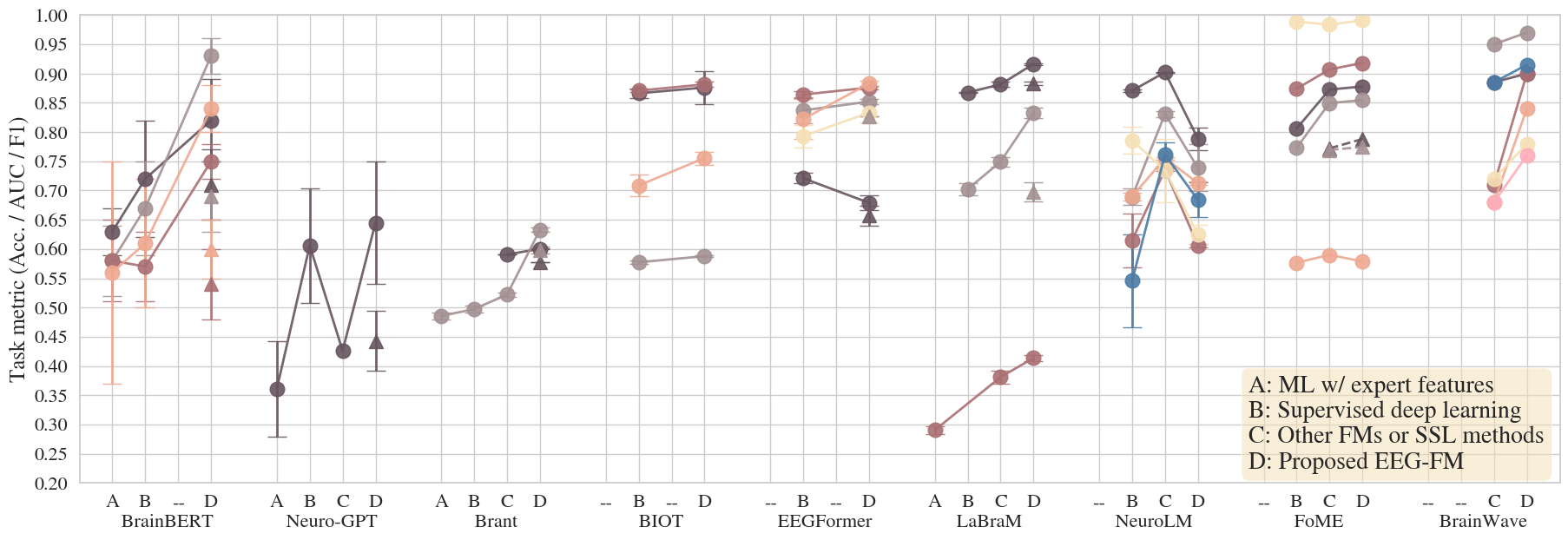}
         \caption{Impact of progressive learning paradigms (A \textrightarrow{} D) on downstream EEG classifications.}
         \label{fig:feature_paradigm}
     \end{subfigure}
    \begin{subfigure}[b]{0.8\textwidth}
         \centering
         \includegraphics[width=\textwidth]{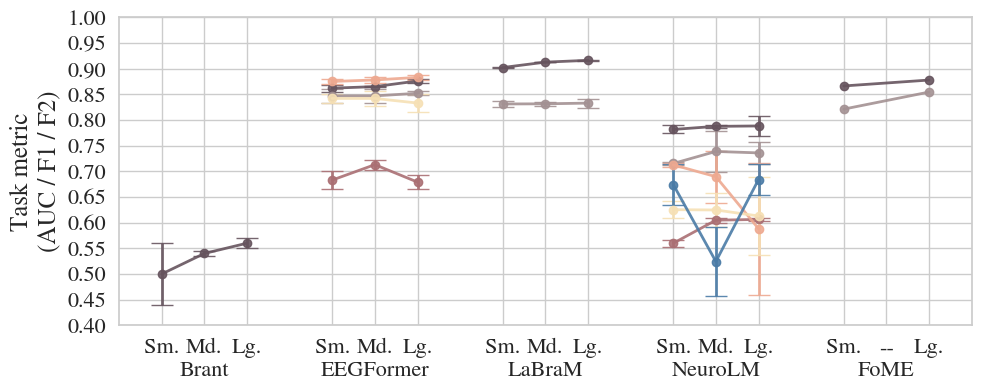}
         \caption{Model scaling and performance gains.}
         \label{fig:model_scaling}
     \end{subfigure}
    \caption{\textcolor{black}{\textbf{The impact of learning paradigm and model scaling on task performance.} In \ref{fig:feature_paradigm}, we compare the impact of learning paradigms -- feature-based statistical machine learning (ML), supervised deep learning, and self-supervised pretraining (proposed EEG-FMs and other baselines) -- on task performance. For each model, a specific task is represented using a unique color, and finetuning and linear-probing evaluations are represented using $\bigcirc$ and $\triangle$, respectively. Note that downstream tasks and the metrics differ across models. In \ref{fig:model_scaling}, we analyze the impact of model scaling on task performance. Although model sizes are specific to each study, we use \textit{`Sm.'}, \textit{`Md.'}, \textit{`Lg.'}, to represent the smallest, intermediate, and the largest variants, respectively. Within each model, a specific task is represented using a unique color. \vspace{-1em}}}
    \label{fig:takeaway}
\end{figure}

\subsection{Key Takeaways}

\noindent \textbf{Diversity of pretraining data:} Several EEG-FMs (LaBraM, NeuroLM, FoME, and BrainWave) leveraged a diverse set of EEG domains spanning clinical, sleep, and task-based BCI. Notwithstanding potential confounders such as preprocessing, data splits, and model adaptation strategies, LaBraM's lead in the TUAB and TUEV evaluations (Figure \ref{fig:comparison}) may have emerged from its higher diversity in pretraining data, i.e., the number of distinct data sources, compared to EEGFormer and BIOT. LaBraM also performed competitively in \textcolor{black}{seizure detection and events} classification tasks on the MAYO and FNUSA datasets as reported in FoME's evaluations. Note that LaBraM was pretrained on significantly fewer channel-hours of scalp EEG data and while TUAB and TUEV evaluations are both in-distribution assessments, MAYO-FNUSA evaluations are both iEEG datasets. The mixed use of scalp and iEEG, as done in FoME and BrainWave, can be considered as another form of data diversity. Using data ablations (scalp EEG vs. iEEG), BrainWave demonstrated that joint pretraining (scalp EEG + iEEG) boosts downstream task performance and transfer to unseen data types (electrocardiograms). Overall, the notion of data diversity, in conjunction with data volume, may influence the performance, generalizability, and transferability of EEG-FMs.

\noindent \textbf{Minimal data preprocessing:} Most EEG-FMs perform minimal and simple data preprocessing steps, namely filtering and resampling, to standardize EEGs from various sources. Notably, the removal of noise-related outliers, suppression of EEG artifacts, and site-related harmonization were not explicitly pursued. Moreover, data normalization strategies that produce training-ready samples were not sufficiently described in most studies. It remains unclear whether or how various offline data handling strategies impact EEG-FM pretraining and downstream task performance, particularly with out-of-distribution test data.

\noindent \textbf{Multivariate time series EEG representation:} All but four EEG-FMs utilized the native multivariate time series representation of EEG, while two models utilized the spectral representations. Two other models combined time-series and spectral representations as input, while the remaining two models exclusively adopted time-frequency representations.
Subsequently, the respective EEG inputs were positionally encoded with their spatial and temporal order. Due to a lack of shared downstream evaluations and ablation studies, it is difficult to assess the relative contributions of different EEG input representations and spatial positional encoding schemes. Furthermore, the context length of the EEG-FMs did not exceed 90 seconds (FoME), and as such, they may struggle to capture long-range EEG patterns, relationships, or dependencies.

\noindent \textbf{Temporal sequence modeling:} Sequence-based transformer blocks were the primary workhorse of representation learning in most EEG-FMs, with the exception of Mentality, which was based on a Mamba-based architecture. A few models (NeuroLM, Neuro-GPT) utilized convolutions to capture low-level morphological features of time-domain EEG. However, the modeling of spatial EEG relationships was either ignored or limited to the positional encoding step. BrainWave, a notable exception in this trend, integrated a spatial attention mechanism. Supporting such emphasis on temporal modeling, the ablation experiments in Brant showed that their temporal encoder provided the largest contribution to downstream task performance, compared to the spatial encoder and frequency encoding.

\noindent \textbf{Pretraining using masked reconstruction:} Reconstruction of masked temporal EEG sequences was the predominant EEG-FM pretraining paradigm, albeit with varying strategies for masking of sequence tokens/patches. Despite its origins in vision and language domains, this SSL strategy seemingly holds merit in the EEG domain. However, the generalizability of representations learned during pretraining is generally unclear since most evaluations were performed after fine-tuning on downstream data. 
In addition, several studies (LaBraM, NeuroLM, EEGFormer) employed a learned discrete neural codebook to further facilitate the pretraining process. In auxiliary efforts, this codebook can support interpretability (EEGFormer) and interface with discrete language vocabularies (NeuroLM).

\noindent \textbf{Limited model evaluations:} Task performance after fine-tuning was the primary paradigm of EEG-FM evaluation. However, in four of the ten reviewed studies (BrainBERT, Mentality, NeuroLM, and FoME), \textcolor{black}{the downstream evaluation datasets were already used for pretraining, i.e., the evaluations were in-distribution.} However, notably, BrainWave and Brant are the only studies to have performed out-of-distribution evaluations using unseen datasets (see Supplementary Table~\ref{tab:eval_table_2}).
Direct model rankings beyond the TUAB and TUEV tasks are difficult to determine due to heterogeneous selections of downstream tasks across most EEG-FMs (see Table \ref{tab:datasets}). Even when considering the TUAB and TUEV tasks, only four out of the ten EEG-FMs can be ranked, as depicted on the y-axis in Figures \ref{fig:tuab} and \ref{fig:tuev}, respectively. In addition, linear probing and few-shot evaluations were rarely reported (see Fig. \ref{fig:feature_paradigm} and Supplementary Table~\ref{tab:eval_table_2}, respectively). Overall, the number of meaningful EEG classification tasks evaluated in each study ranged from 1 (Mentality) to 12 (BrainWave), with most studies evaluating proposed EEG-FMs on at most 5 tasks. Overall, the universality and robustness of EEG-FMs have not been convincingly demonstrated in most studies, with BrainWave being a notable exception.

\noindent\textbf{Model scaling and task performance}: In Figure \ref{fig:model_scaling}, we analyze the impact of model scaling on task performance using studies with at least two model variants. Each line plot indicates a specific downstream task and the x-axis shows model variants, which were typically classified as small, intermediate, and large. Some marginal improvements can be observed for certain tasks and models, although these variants were developed with a fixed amount of pretraining data and were evaluated within study-specific experimental and methodological contexts. Notably, LaBraM investigated the combined effects of pretraining data and model scale on downstream TUAB/TUEV classifications. Overall, it is unclear whether a clear and strong trend exists with model scaling, especially within the current EEG-FM parameter regime ranging from 3.3M (BIOT) to 1.7B (NeuroLM, largest variant).

\noindent\textbf{Performance of general-purpose time series models}:
Interestingly, we observed that general time series foundation models, such as TimesNet~\cite{wu2022timesnet}, performed reasonably well on several EEG tasks after fine-tuning and sometimes outperformed EEG-FMs (e.g., sleep-stage classification in FoME). Additionally, experiments in BrainWave show that a time series FM -- MOMENT~\cite{goswami2024moment} -- outperformed EEG-FMs in specific tasks, such as \textcolor{black}{seizure detection and pathology detection}. Experiments in Brant show that general time series architectures, such as PatchTST~\cite{nie2022time} and CoST~\cite{woo2022cost}, perform relatively better on some tasks, such as short- and long-term signal forecasting and imputation, respectively, than some EEG-specific architectures. 
\textcolor{black}{However, other results reported in BrainWave and FoME suggest that EEG-specific inductive biases and modeling choices (e.g., spatial modeling, multi-scale modeling of intracranial and scalp data, time-frequency representations, diverse clinical characteristics) can indeed help EEG-FMs outperform general time-series FMs (TS-FMs) in EEG-specific tasks. Overall, these findings indicate that TS-FMs may provide moderate value in certain EEG tasks due to efficient model adaptation via fine-tuning or larger-scale general pretraining on time series containing richer and more diverse semantics than EEG.}

\noindent \textbf{Data scaling and task performance:} 
The trend along the x-axis of Figures \ref{fig:tuab} and \ref{fig:tuev} suggests that scaling up pretraining data may not necessarily increase downstream task performance, even with significant model scale-up (e.g., LaBraM vs. EEGFormer/NeuroLM). Notably, LaBraM demonstrated that the effects of pretraining data scaling on TUAB and TUEV classifications are sharpest under $\sim$1000 hours of data and begin to plateau thereafter. Overall, the evidence for data scaling is weak, if any, based on the limited shared tasks and models evaluated thus far.

\noindent \textbf{Advance over other feature paradigms:} 
\textcolor{black}{
In Figure \ref{fig:feature_paradigm}, we compare progressive feature paradigms (i.e., expert features, supervised EEG-DL features, SSL baselines, and proposed EEG-FMs) on various tasks. In a majority of tasks, proposed EEG-FMs (`D') provided at least some improvement after fine-tuning, if not drastic, over previous DL (`B') and self-supervised baselines, including previous EEG-FMs (`C'). Linear probing results, however, were relatively worse. Fine-tuned EEG-FMs showed substantial improvements over classical machine learning (ML) models with expert features (`A'), although such assessments were reported in only four studies.
}

\subsection{Current Research Gaps}

\noindent\textbf{Data and model scaling:} 
The scaling up of data and models is a defining principle of foundation modeling.
However, empirical evidence of scaling in EEG-FMs has been either weak, limited, or inconclusive.
Investigations that scale up data volume (channel-hours), model size (trainable parameters), and evaluate on an expansive set of downstream tasks are lacking in current EEG-FMs, particularly at sufficiently large scales where effects are clear and discernible.

\noindent\textbf{Preprocessing and normalization effects:} 
EEG datasets can require significant offline preprocessing to manage data quality and suppress artifacts. Since current studies perform minimal processing, it is unclear whether or to what extent data outliers and noise impact EEG-FM pretraining and task performance. Even clean EEG datasets require careful consideration of normalization strategies as EEG can vary across channels, subjects, and acquisition sites \cite{wagh2022evaluating}. The choice of input representation can further complicate decisions related to preprocessing and normalization. As such, there is a need to better understand the impact of these choices on downstream EEG-FM modeling.

\noindent\textbf{Model ablations:}
EEG-FM design involves two significant choices: the input representations and the internal architectural components. However, the lack of systematic exploration of the effects of these choices in existing EEG-FM literature prevents principled choices in EEG-FM design. The relative merits of time-domain and time-frequency domain inputs remain unclear, although both appear to be effective. The contributions of positional encoding schemes, both temporal and spatial, can be better understood. Similarly, the effects of time-domain and channel-domain spatial attention mechanisms on learned patterns remain poorly understood.

\noindent\textbf{Long temporal context and spatial modeling:} 
\textcolor{black}{
Slow variations over long timescales can be observed in multi-day intracranial and scalp EEGs \cite{mivalt2023impedance}. Additionally, there is substantial spatial and temporal variation in EEG across behavioral states (e.g., deep sleep versus wake, eyes open versus eyes closed).} However, current EEG-FMs can only process patterns within EEG sequences spanning 90 seconds or less. There is a need for solutions that expand the effective context length of EEG-FMs. Moreover, the explicit modeling of spatial or inter-channel relationships and their contributions relative to temporal modeling remains to be investigated.

\noindent\textbf{Quality of pretraining strategy:} 
EEG-FM linear probing evaluations reported by some studies have performed significantly worse than fine-tuned versions and other baselines in several instances (see Figure \ref{fig:feature_paradigm}). However, some gains can be observed when FMs are fine-tuned on task data compared to fully supervised training on the same data. This contrasting observation casts doubt on the quality of the representations learned via self-supervision in EEG-FMs. Further investigations into the effects of SSL pretraining on downstream evaluations are needed to fully understand the extent of transferability achieved through SSL. 
Beyond the SSL strategy itself, the effects of data diversity, data volume, and model scale on the quality of EEG-FM pretraining remain unknown.

\noindent\textbf{Practically relevant evaluations and metrics:}
Current fine-tuning evaluations are limited in their ability to assess the practical utility of EEG-FMs in real-world settings. There is a need to adopt evaluation schemes and task metrics that capture the reality of EEG research and clinical use. \textcolor{black}{Evaluations on out-of-distribution data and novel tasks are required to assess the off-the-shelf value of EEG-FMs. Few-shot or low-label performance, when measured in absolute terms rather than percentages, can capture how efficiently EEG-FMs leverage labels that are typically expensive and laborious to collect.} Out-of-distribution performance on a known task without fine-tuning can help understand EEG-FM robustness to the idiosyncratic cross-subject and cross-site variability of EEG. Furthermore, application-specific, practically relevant metrics, such as false positives per hour for seizure detection, and comparisons with expert features can clarify the real-world utility of EEG-FMs.
\textcolor{black}{
Finally, evaluations must account for the unbalanced nature of physiological data and its influence on metrics and model outcomes. For example, in seizure detection settings, patient data consists mostly of interictal segments, with seizure events occurring rarely. Evaluations must reflect this natural imbalance to assess the model's real-world utility faithfully.
}

\noindent\textbf{Standardized benchmarking tasks:}
The comparative analysis revealed significant heterogeneity in the tasks used for EEG-FM evaluation (see Table \ref{tab:datasets}). The lack of common tasks across EEG-FM evaluations makes it challenging to understand the state of the art and highlights the need to identify a common core set of evaluations for future EEG-FM development. This set must cover multiple task types and include both classification and regression tasks, with dense (one label per EEG recording) and sparse (one label per EEG segment) labels. Additionally, the tasks must be challenging enough for previous generations of EEG-DL models, with ample room for improvement, unlike TUAB, where performance may already have saturated (85-87\% accuracy) with traditional approaches \cite{kiessner2024reaching}.
\textcolor{black}{
Finally, the tasks must reflect the data heterogeneity encountered in clinical settings, including wake and sleep patient states.
}

\noindent\textbf{Trustworthy modeling:} 
\textcolor{black}{Despite the importance of explainability and uncertainty handling in the high-risk, expert-centric domain of medicine, experiments probing trust-related aspects of EEG-FMs have been limited. 
Notably, preliminary interpretability analyses based on intrinsic dimensionality and codebooks were conducted for BrainBERT and EEGFormer, respectively.}
However, studies that further demystify the EEG-FM black box are needed to gain insight into the knowledge learned by EEG-FMs (EEG patterns, dependencies, relationships) and the practical robustness of their decision-making process for downstream applications. Connections to known patterns of brain physiology or pathology may be necessary to make EEG-FMs trustworthy in the eyes of expert and clinical users.

%% file: future_directions.tex
\section{Proposed Future Directions}
\label{sec:future_directions}

\begin{figure}[!h]
  \begin{center}
    \includegraphics[width=\linewidth]{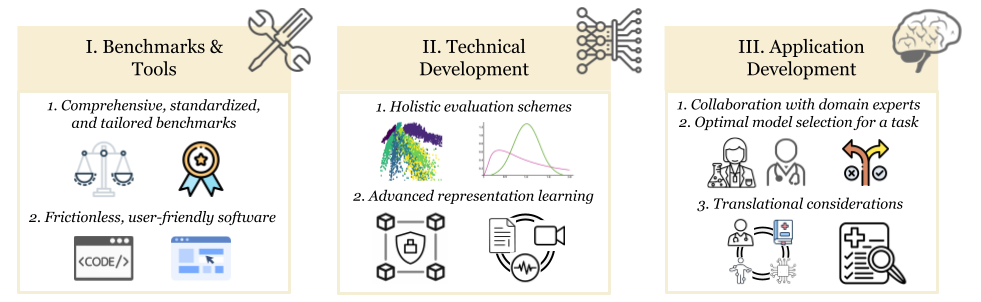}
    \caption{
    \textcolor{black}{\textbf{Suggested future directions.} (I) \ul{Benchmarks and tools:} future EEG foundation models can be compared using standardized benchmarks against prevailing feature paradigms and participate in community-specific EEG challenges to establish their real-world utility. Frictionless and user-friendly software tools are needed to quickly adopt and experiment with off-the-shelf models. (II) \ul{Technical modeling:} holistic evaluation frameworks that test embedding space semantics, robustness, and transfer efficiency can meaningfully track the state of the art. Advanced representation learning techniques, such as federated or multi-modal learning, can enhance large-scale pretraining. (III) \ul{Applications:} collaborations with domain experts can inspire novel applications. Strategies that help identify suitable off-the-shelf models for a particular task and address translational hurdles, such as clinical interpretability, prospective validation, and operational feasibility, can increase adoption and impact.
    }}
    \vspace{-2.5em}
    \label{fig:future_directions}
    \end{center}
\end{figure}

From a time series perspective, EEG signals are distinct from general signals or sequences due to their non-linearity \cite{klonowski2009everything}, non-stationarity \cite{klonowski2009everything}, complex spatial relationships, and 1/f spectral characteristics \cite{donoghue2020parameterizing}. These peculiarities would suggest that foundation modeling for EEG signals requires domain-specific inductive biases. Indeed, we have cultivated an EEG data-centric and domain-sensitive perspective throughout this review. However, counter-intuitively, EEG-FMs have enjoyed some success in drawing from vision and language domains in their adoption of patch-based sequences and transformer architecture design (transformer~\cite{vaswani2017attention}, vision transformer~\cite{dosovitskiy2020image}), masked reconstruction-based SSL (masked autoencoders~\cite{he2022masked}), and discrete neural codebooks (VQ-VAE~\cite{van2017neural}, BEiT v2~\cite{peng2022beit}). Moreover, time series FMs pretrained on generic, non-neural time series have performed similarly to EEG-FMs on certain EEG tasks. Therefore, it is likely that future EEG-FMs could benefit from embracing both domain-specific insights and innovations from external data domains. These domains may be semantically distant to EEG, such as vision and natural language, or loosely related, such as non-neural time series, audio, speech, or other biosignals.
Below, we outline the future research directions that we believe could support meaningful and sustained progress in the EEG-FM research domain. As illustrated in Figure \ref{fig:future_directions}, these recommendations are organized under three broad themes of development spanning benchmarks and software tools, technical advances, and real-world applications. These suggestions build upon the research gaps identified earlier and provide additional guidance for advancing and accelerating EEG-FM research and adoption.

\subsection{Benchmarks and Tools}

\noindent \textbf{Benchmarking against prevailing feature paradigms:}
Quantitative EEG research currently employs two ML feature paradigms: expert-crafted EEG features and task-specific supervised features. As such, data-driven self-supervised features, such as those provided by EEG-FMs, remain in their infancy. Therefore, to establish a compelling case for EEG-FM adoption, studies proposing novel EEG-FMs should evaluate against the prevailing feature paradigms. In doing so, statistical tests of superior performance must be presented when necessary.
\textcolor{black}{
Additionally, identical subject-level splits must be used to control for inter-subject variability, and temporal past/future splits must be used to account for the temporal order of events in longer-term EEG data.
}

\noindent \textbf{Standardized community-specific benchmarks:}
\textcolor{black}{
The primary failure mode of EEG encoders remains their inability to learn robust latent features that generalize across realistic EEG distribution shifts, including those seen across subjects, age groups, acquisition setups, and experimental or clinical conditions, among other factors. However, each EEG user community (e.g., clinicians, brain computer interface users, neuroscientists) may have different practical requirements or definitions of model robustness and generalization. Therefore, future work can design community-specific EEG-FM benchmarks with clear task definitions, realistic performance criteria beyond traditional metrics, real-world datasets, and expert baselines. In doing so, EEG researchers can draw inspiration from existing FM benchmarks in other data domains. Standardized benchmarks tailored to each community’s needs can significantly boost the chances of real-world translation and adoption of EEG-FMs.
}

\noindent \textbf{Ranking EEG-FMs in global EEG competitions:}
Novel EEG-FM methodology can leverage international predictive modeling competitions to establish real-world performance and utility, as those competitions can highlight the most significant empirical challenges or milestones within the field. Such competitions have already been developed for time series~\cite{wang2024deep}, electrocardiograms~\cite{wan2025openecg}, and speech~\cite{arora2024evaluation} domains. Notably, within the EEG domain, previous competitions have highlighted issues with transfer learning under distribution shifts \cite{wei20222021}, abnormality detection \cite{hms-harmful-brain-activity-classification}, coma prognosis \cite{reyna2023predicting}, \textcolor{black}{seizure detection \cite{dan2024szcore}, and seizure forecasting \cite{brinkmann2016crowdsourcing}.}

\noindent \textbf{Frictionless software:}
Several software modifications and code debugging are required to run the currently available EEG-FMs. The deep technical nature of EEG-FMs and the computing skills required to run these models pose a significant entry barrier for non-technical or non-computational researchers. Future studies will need to ensure that EEG-FMs can be used as off-the-shelf feature-extraction tools, with no technical modifications required. Graphical software for plug-and-play EEG-FM analytics, such as simple binary classifications, can greatly increase real-world EEG-FM adoption, testing, and experimentation. Encouragingly, a mature ecosystem for large-scale EEG analytics already exists in Python \cite{gramfort2013meg, HBM:HBM23730, schiratti2018ensemble} and can support the development of frictionless EEG-FM software.

\subsection{Technical Development}

\noindent \textbf{Holistic evaluation schemes:}
\textcolor{black}{Future studies may benefit from focusing specifically on out-of-distribution evaluations, as they can directly assess the effectiveness of EEG-FM pretraining and performance in realistic, off-the-shelf, and low-label scenarios.} Furthermore, evaluating EEG-FMs based on their task performance on a narrow set of commonly available EEG datasets/tasks is insufficient to measure their real-world readiness.
Novel evaluation and model ranking schemes can be designed to assess the computational costs, semantic EEG-FM embedding quality, transfer efficiency, and robustness to real-world EEG variability and noise sources.

\noindent \textbf{Privacy-preserving learning:}
EEG datasets required for large-scale EEG-FM pretraining can be siloed due to patient privacy and legal concerns. Research collaborations involving multiple clinical sites can utilize federated learning \cite{mcmahan2017communication} techniques to train EEG-FMs without requiring centralized data access or sharing. Such techniques may also support the derivation of personalized, i.e., patient-specific, models from EEG-FMs.

\noindent \textbf{Multi-modal learning:}
Latent representations learned during pretraining may benefit from cross-modal supervision from text reports, video recordings, or other biosignals, such as electrocardiograms. Some of these modalities are jointly recorded with EEGs in many clinical and scientific recording settings. Notably, downstream applications can reap the performance benefits of multi-modal pretraining without requiring those modalities during evaluation.

\noindent \textbf{Reasoning and agentic capabilities:}
Future EEG-FMs could be fine-tuned to mimic expert reasoning processes, adhere to expert grading criteria \cite{tatum2016american} or instructions, and \textcolor{black}{autonomously retrieve relevant evidence from verified medical literature repositories and knowledge bases.} Augmented healthcare workflows could leverage insights from multiple specialized modality-specific agents to inform clinical decisions.

\noindent \textbf{Scaling down task-specific EEG-FMs:}
FMs may require model parameter scaling to obtain superior pretrained features. However, large EEG-FMs may not be suitable for applications involving streaming EEG data, real-time EEG processing, or deployment in resource-constrained settings, such as brain-computer interfaces, medical devices, and clinical environments. In such contexts, knowledge distillation \cite{hinton2015distilling}, pruning \cite{han2015deep}, or quantization \cite{gong2014compressing} techniques can deliver smaller, computationally inexpensive models that maintain high task performance.

\noindent \textbf{Timeseries FMs, task semantics, and inductive biases:}
\textcolor{black}{
The relative maturity of general time-series FMs (TS-FMs) can serve as a source of inspiration for EEG-FM development. For example, an understanding of how TS-FMs learn effectively from large, diverse datasets across economic, energy, traffic, climate, and industrial domains may improve EEG-FM scaling. TS-FM performance may be particularly instructive when an EEG task shares underlying semantics with more general time-series tasks, such as forecasting or anomaly detection. There may exist opportunities to `specialize' TS-FMs by injecting EEG-specific biases into data representation, model design, and the self-supervised learning objective. Future EEG-FM studies can assess the utility of such biases in adapting TS-FMs for EEG tasks and the relative merits and weaknesses of TS-FMs.
}

\subsection{Application Development}

\noindent \textbf{Cross- and inter-disciplinary collaboration:}
The technical development of EEG-FMs should involve close collaboration with scientific and clinical experts. Such collaborations can productively constrain and inspire novel modeling decisions, identify novel research gaps, establish clear criteria for real-world success or progress, and facilitate EEG-FM experimentation among niche application-specific audiences.

\noindent \textbf{Model cards to guide EEG-FM selection:}
The availability of multiple EEG-FMs, each purportedly an off-the-shelf tool, poses a practical dilemma for domain experts who must make a principled selection for their research. Comprehensive comparisons of all available options may not be feasible in discovery-based research. The release of model cards \cite{mitchell2019model} that summarize an EEG-FM's functional strengths and weaknesses, inherent or dominant inductive biases, compute requirements, and evaluation scope and rankings could inform the model selection process.

\noindent \textbf{Clinical applications:}
\textcolor{black}{
There is a substantial opportunity for EEG-FMs to augment the clinical EEG review process in both short- and extended-duration monitoring settings. EEG-FMs could speed up review by guiding the reviewer’s attention to noteworthy temporal segments, brain regions, or channels distinct from background activity, especially in longer recordings, such as in epilepsy- or sleep-related applications. Based on these clues, an EEG-FM coupled with a text decoder could summarize the clinically relevant findings as a structured text report. Additionally, EEG-FMs could support case-based clinical reasoning by surfacing historical records with shared EEG phenotypes. Finally, EEG-FMs could improve the performance of spike and seizure detections generated by EEG reviewing software.
}

\noindent \textbf{Translational hurdles:}
\textcolor{black}{
Several hurdles and considerations prevent the translation and adoption of EEG-FMs in clinical practice.
\ul{\textit{Testing and validation.}} Prospective pilot testing for specific use cases using site-specific patient data can provide actionable evidence on the feasibility of model deployment and its benefits to clinical stakeholders. The graceful handling of data noise, corruption, and missing values is required for operational success. 
\ul{\textit{Interpretability.}} Interfaces that relate EEG-FM outputs to clinical domain knowledge and patient-specific physiological factors could increase clinician trust and adoption. 
\ul{\textit{Reproducibility.}} Developers must transparently report and release the data sources, data splits, patient characteristics, and computational pipeline configurations that were utilized to build the EEG-FM to maintain reproducibility.
\ul{\textit{Regulatory considerations.}} 
EEG-FMs and task-specific fine-tuned models will be treated as safety-critical medical artificial intelligence (AI) systems that must adhere to stringent national or international guidelines. Therefore, clinical users should focus on defining clear intended use, rigorous validation, and controlled deployment. 
Attention should be given to documenting data provenance, identifying known limitations and expected failure modes, and establishing human-AI teaming protocols that keep clinicians in the loop. Clinical adoption should follow medical-device regulatory pathways, including risk assessment, post-market monitoring, and predefined change-control procedures to manage model updates as data distributions evolve.
}

%% file: conclusion.tex
\vspace{-0.5em}
\section{Conclusion}
\label{sec:conclusion}
The promise of EEG foundation models, in principle, lies in effective and robust feature learning, feature re-usability, and label efficiency. Our critical analysis of ten early EEG-FMs indicates that these efforts, inspired by their counterparts in the mainstream vision and language domains, have made moderate strides in realizing this promise for the EEG domain. However, to develop universal, robust, and general-purpose EEG feature extractors, future EEG-FMs must prioritize substantial scaling efforts, principled and trustworthy self-supervised representation learning, and practically relevant evaluations. In addition to technical modeling, we believe that future research should also pursue the collaborative development of meaningful EEG benchmarks, \textcolor{black}{including standardized datasets, applications, and holistic model evaluation schemes} that can measurably track the real-world readiness and impact of EEG-FMs. With sustained efforts in these directions, EEG-FMs are poised to advance scientific research, brain-computer interfaces, and clinical decision support systems.

%% file: acknowledgement.tex
\section{Acknowledgments}

We would like to thank Mario Serrafero and Saeid Cheshmi for fruitful discussions on foundation modeling.

%% file: funding.tex
\section{Funding}

This study was supported in part by the Mayo Clinic \& Illinois Alliance Fellowship for Technology-based Healthcare Research, the Edward Heiken Interdisciplinary Health Sciences Institute Fund, and NSF grants IIS-2105233, IIS-2344731, and IIS-2337909.

%% file: data_availability.tex
\vspace{-0.5em}
\section{Data Availability}
\textcolor{black}{
This review did not generate any new data. A visual summary of the EEG-FMs included in this review and links to their original code repositories will be made available via a publicly accessible repository upon acceptance.
}

%% file: conflicts.tex
\section{Conflicts of Interest}

None.

%% file: supplementary.tex
\setcounter{table}{0} 

\section*{Supplement}
\label{sec:supplementary}

\subsection*{Sleep Foundation Models}
\textcolor{black}{
Our literature search yielded two sleep FMs that were excluded from the main review scope. Below, we summarize their salient aspects for interested readers.
}

\noindent\textcolor{black}{\textbf{SleepFM~\cite{thapa2024sleepfm}:}
\textit{Data.} SleepFM is a large-scale multi-modal sleep foundation model trained on over 100,000 hours of polysomnography data from 14,000 participants. The modalities include brain activity (EEG), eye activity (EOG), muscle tone (EMG), cardiac signals (EKG), and respiratory signals. The data are minimally preprocessed and segmented into 30-second clips aligned with expert annotation boundaries. 
\textit{Modeling.} SleepFM uses three modality-specific 1D EfficientNet-based encoders. The model is pretrained with both pairwise contrastive learning and a novel leave-one-out contrastive strategy that aligns each modality with the joint representation of the others. SleepFM supports flexible inference, allowing downstream classifiers to use any subset of the input modalities.
\textit{Evaluation.} The resulting embeddings are evaluated through cross-modal retrieval, demographic prediction, and sleep-related classifications, including sleep staging and sleep disordered breathing detection. Results show that SleepFM consistently outperforms supervised convolutional baselines. Ablations demonstrate that multi-modal pretraining that includes EEG is superior to single-modality pretraining. Additionally, SleepFM demonstrated strong out-of-distribution generalization to an unseen site with a different channel configuration.
}

\noindent\textcolor{black}{\textbf{Self-Supervised Transformer Model Training for a Sleep-EEG Foundation Model \cite{ogg2024self}:}
\textit{Data.} The model is pretrained on 10,897 sleep sessions from 9,013 individuals sourced from multiple public datasets. A single central EEG channel (C3/C4) is resampled to 100 Hz, normalized, clipped, and segmented into 30-second epochs. Sequences of 101 epochs with 25\% overlap are constructed for subsequent modeling.
\textit{Modeling.} Unsupervised k-means clustering (300 clusters) on spectrogram features provides pseudo-labels for self-supervision. The approach follows a HuBERT~\cite{hsu2021hubert}-style masked prediction framework, where the model learns to infer hidden k-means label sequences. The architecture consists of a 7-layer 1D convolutional encoder, a projection layer, positional encoding, and a 4-layer transformer encoder (28.5M parameters). Masking is applied in 10 epoch blocks, and the model is trained for 40 epochs.
\textit{Evaluation.} After pretraining, the model is adapted for the tasks of sleep staging, subject identification, and age prediction. The pretrained model achieves strong representation learning and rapid convergence, outperforming supervised baselines in low-label sleep staging. Experiments with frozen embeddings, i.e., linear probing, suggest that the model can generalize across montages and tasks. Additionally, the masking and k-means-based pseudo-labeling process were shown to support generalization to the external Sleep-EDF dataset. 
}
\newpage

\newcolumntype{L}[1]{>{\raggedright\arraybackslash}p{#1}} 
\newcolumntype{C}[1]{>{\centering\arraybackslash}p{#1}}   
\newcolumntype{R}[1]{>{\raggedleft\arraybackslash}p{#1}}  

\begin{table}[H]
\caption{
\textcolor{black}{\textbf{Pretraining data volume.}
\textit{Channel-hours} were used to represent and compare the data scale at which the EEG foundation models were pretrained. In most cases, channel-hours per dataset were calculated as the product of the number of channels and the recording durations (in hours) obtained from the original studies. Scalp EEG datasets had a fixed channel count across subjects, whereas intracranial EEG datasets used in BrainBERT and Brant had a subject-specific channel count. For Neuro-GPT and EEGFormer, channel-hours were computed from publicly available dataset statistics and were assumed to be applicable to both studies. For BrainBERT, we used the channel-hours reported by the authors. For BrainWave, the total patch count reported was used to calculate the channel-hours by converting the patch size (1 second) to hours. Channel-hours pooled across all pretraining data sources are highlighted in bold. N/A represents cases where channel-hours could not be determined. 
}}
  \centering\scriptsize
  \resizebox{\columnwidth}{!}{
\begin{tabular}{|l|L{4cm}|R{1cm}|R{2.2cm}|R{6cm}|R{2cm}|}
\toprule
\hline

\textbf{EEG-FM}    & 
\textbf{Pretraining Dataset}   & 
\textbf{Number of Channels} &
\textbf{Total Length of Recordings (Hours)}    & 
\textbf{Channel-Hours\newline($=$ Number of Channels $\times$ Recording Length)}   & 
\textbf{Total Channel-Hours}              \\ \hline

BrainBERT & Brain TreeBank~\cite{data:wang2024brain}                     & -         & -       &                            
4,551
   &  \textbf{4,551} \\ \hline

Neuro-GPT  & TUEG~\cite{data:obeid2016temple}                              & 20                            & 27,063                              &  541,260     &  \textbf{541,260}    \\ \hline
Brant     &             Private dataset (unknown source)                       &  -        &  -                                       & (124$\times$235.34)
+ (52$\times$82.39)
+ (120$\times$393.09)
+ (133$\times$137.45)
+ (116$\times$214.22)
+ (101$\times$386.70) 
+ (67$\times$111.55)
+ (47$\times$207.42)
+ (134$\times$759.80)                                                                                  &  \textbf{281,860.10}    \\ \hline
BIOT      & SHHS~\cite{data:quan1997sleep, data:zhang2018national}                               & 2                             & 42,446                                                       & 84,892                                                                                                                                                                                                &                                  \\ \hline
          & PREST                              & 16                            & 14,197                                                       & 227,152                                                                                                                                                                                               &  \textbf{312,044}      \\ \hline
EEGFormer & TUEG                              & 20                            & 27,063                                                       & 541,260                                                                                                                                                                                               &  \textbf{541,260}      \\ \hline
LaBraM    & BCI Competition IV-1~\cite{data:blankertz2007non}               & 59                            & 8.21                                                        & 484.39                                                                                                                                                                                               &                                  \\ \hline
          & Emobrain~\cite{data:savran06_einterface}                           & 64                            & 4.94                                                        & 316.16                                                                                                                                                                                               &                                  \\ \hline
          & Grasp/Lift EEG Challenge~\cite{data:luciw2014multi}       & 32                            & 11.72                                                       & 375.04                                                                                                                                                                                               &                                  \\ \hline
          & Inria BCI Challenge~\cite{data:margaux2012objective}                & 56                            & 29.98                                                       & 1,678.88                                                                                                                                                                                              &                                  \\ \hline
          & EEG Motor Imagery Dataset~\cite{data:schalk2004bci2000} & 64                            & 47.3                                                        & 3,027.2                                                                                                                                                                                               &                                  \\ \hline
          & Visual Categorization EEG Data~\cite{data:trujillo2019mental}                       & 64                            & 34.35                                                       & 2,198.4                                                                                                                                                                                               &                                  \\ \hline
          & Resting State EEG Data ~\cite{data:trujillo2017effect}             & 64                            & 3.04                                                        & 194.56                                                                                                                                                                                               &                                  \\ \hline
          & SEED Series~\cite{data:zheng2015investigating, data:zheng2018emotionmeter, data:liu2022identifying}                        & 62                            & 166.75                                                      & 10,338.5                                                                                                                                                                                              &                                  \\ \hline
          & Siena Scalp EEG Database~\cite{data:detti2020eeg}           & 31                            & 30.47                                                       & 944.57                                                                                                                                                                                               &                                  \\ \hline
          & SPIS Resting State Dataset~\cite{data:torkamani2020prediction}         & 64                            & 0.83                                                        & 53.12                                                                                                                                                                                                &                                  \\ \hline
          & Brain-Invaders~\cite{data:korczowski2019brain}          & 32                            & 16                                                          & 512                                                                                                                                                                                                  &                                  \\ \hline
          & TUAR~\cite{data:obeid2016temple}                               & 23                            & 92.22                                                       & 2,121.06                                                                                                                                                                                              &                                  \\ \hline
          & TUEP~\cite{data:obeid2016temple}                               & 21                            & 591.22                                                      & 12,415.62                                                                                                                                                                                             &                                  \\ \hline
          & TUSZ~\cite{data:obeid2016temple}                               & 21                            & 1,138.53                                                     & 23,909.13                                                                                                                                                                                             &                                  \\ \hline
          & TUSL~\cite{data:obeid2016temple}                               & 23                            & 20.59                                                       & 473.57                                                                                                                                                                                               &                                  \\ \hline
          & Multiple data sources~\cite{data:jiang2021discriminating, data:jiang2023multimodal, data:luo2022multimodal, data:li2021discrimination, data:tao2020emotion}            & 62                            & 342.23                                                      & 21,218.26                                                                                                                                                                                             &  \textbf{80,260.46}    \\ \hline
Mentality & TUSZ                               &  N/A       &  N/A                                    &  N/A                                                                                                                                                                              & \textbf{N/A}                               \\ \hline
NeuroLM   & TUEG                              & 21                            & 24,000                                                       & 504,000                                                                                                                                                                                               &                                  \\ \hline
          & SEED Series                        & 62                            & 170.54                                                      & 10,573.48                                                                                                                                                                                             &                                  \\ \hline
          & BCI Competition IV-1               & 59                            & 8.21                                                        & 484.39                                                                                                                                                                                               &                                  \\ \hline
          & Emobrain                           & 64                            & 4.94                                                        & 316.16                                                                                                                                                                                               &                                  \\ \hline
          & Grasp/Lift EEG Challenge       & 32                            & 11.72                                                       & 375.04                                                                                                                                                                                               &                                  \\ \hline
          & Inria BCI Challenge                & 56                            & 29.98                                                       & 1,678.88                                                                                                                                                                                              &                                  \\ \hline
          & Motor Movement/Imagery Dataset     & 64                            & 47.3                                                        & 3,027.2                                                                                                                                                                                               &                                  \\ \hline
          & Raw EEG Data                       & 64                            & 34.35                                                       & 2,198.4                                                                                                                                                                                               &                                  \\ \hline
          & Resting State EEG Data             & 64                            & 3.04                                                        & 194.56                                                                                                                                                                                               &                                  \\ \hline
          & Siena Scalp EEG Database           & 31                            & 30.47                                                       & 944.57                                                                                                                                                                                               &                                  \\ \hline
          & SPIS Resting State Dataset         & 64                            & 0.83                                                        & 53.12                                                                                                                                                                                                &                                  \\ \hline
          & Brain-Invaders          & 32                            & 16                                                          & 512                                                                                                                                                                                                  &                                  \\ \hline
          & Multiple data sources~\cite{data:jiang2021discriminating, data:jiang2023multimodal, data:luo2022multimodal, data:li2021discrimination, data:tao2020emotion}            & 62                            & 342.23                                                      & 21,218.26                                                                                                                                                                                             &  \textbf{545,576.06}   \\ \hline
FoME      &          Multiple datasets                          &  N/A       &  N/A                                     &  N/A                                                                                                                                                                              & \textbf{N/A}                               \\ \hline
BrainWave & Multiple datasets      & -     & -   &  3,162,233,694$\times$(1$\div$3,600)                                                                            &  \textbf{878,398.25} \\ \hline
\bottomrule
\end{tabular}
 \label{tab:channelhours}
}
\end{table}

\begin{table}[]

\definecolor{clinical}{RGB}{255, 182, 193}  
\definecolor{nonclinical}{RGB}{173, 216, 230}  

\definecolor{pretrained_data}{RGB}{250, 207, 244} 
\definecolor{ood_eval}{RGB}{184, 210, 252} 
\definecolor{few_shot}{RGB}{113, 208, 192} 
\definecolor{heldout_test_ood}{RGB}{250, 246, 207} 

\definecolor{dark_pretrained_data}{RGB}{250, 0, 244} 
\definecolor{dark_ood_eval}{RGB}{80, 146, 252} 
\definecolor{dark_few_shot}{RGB}{0, 77, 64} 
\definecolor{dark_heldout_test_ood}{RGB}{250, 246, 0} 

\caption{
\textcolor{black}{\textbf{EEG foundation model evaluation strategies.} 
Various techniques were used to evaluate EEG foundation models: a) zero-shot transfer learning (ZSTL, in blue) with prior task knowledge but no access to data and labels from the target setting; b) few-shot learning (FSL, in green) with no prior task knowledge but access to a few labels from the target setting; c) linear probing (LP, in pink) where a classifier head is adapted using labels from the target setting; and d) fine-tuning (FT, in yellow) where the backbone and head together are adapted using labels from the target setting. Evaluation scores estimated based on the graphics in the paper are denoted using superscript $^*$.}}

\centering\small
  \resizebox{0.85\textwidth}{!}{
\begin{tabular}{@{}|l|l|>{\columncolor{pretrained_data}}l>{\columncolor{pretrained_data}}l|>{\columncolor{ood_eval}}l>{\columncolor{ood_eval}}l>{\columncolor{heldout_test_ood}}l>{\columncolor{pretrained_data}}l|@{}}

\toprule
                                &                               & \multicolumn{2}{c|}{\textbf{In-Distribution}}                                & \multicolumn{4}{c|}{\textbf{Out-of-Distribution}}                                                         \\ \midrule
	\textbf{EEG-FM}                              & \textbf{Task}                          &  \multicolumn{1}{>{\columncolor{pretrained_data}}l|}{\textbf{LP}} & \multicolumn{1}{>{\columncolor{heldout_test_ood}}l|}{\textbf{FT}} & \multicolumn{1}{>{\columncolor{few_shot}}l|}{\textbf{FSL}} & \multicolumn{1}{>{\columncolor{ood_eval}}l|}{\textbf{ZSTL}} & \multicolumn{1}{>{\columncolor{pretrained_data}}l|}{\textbf{LP}} & \multicolumn{1}{>{\columncolor{heldout_test_ood}}l|}{\textbf{FT}} \\ \midrule
\multicolumn{1}{|c|}{\textbf{BrainBERT}} & Brain TreeBank           & \multicolumn{1}{>{\columncolor{pretrained_data}}l|}{AUC - 0.59}  &  \multicolumn{1}{>{\columncolor{heldout_test_ood}}l|}{AUC - 0.83}  & \multicolumn{1}{>{\columncolor{few_shot}}l|}{}   & \multicolumn{1}{>{\columncolor{ood_eval}}l|}{}   & \multicolumn{1}{>{\columncolor{pretrained_data}}l|}{}   &    \multicolumn{1}{>{\columncolor{heldout_test_ood}}l|}{} \\ \midrule
	\textbf{Neuro-GPT}                       & BCI Competition IV Dataset 2a &  \multicolumn{1}{>{\columncolor{pretrained_data}}l|}{}   &    \multicolumn{1}{>{\columncolor{heldout_test_ood}}l|}{} & \multicolumn{1}{>{\columncolor{few_shot}}l|}{}   & \multicolumn{1}{>{\columncolor{ood_eval}}l|}{}   & \multicolumn{1}{>{\columncolor{pretrained_data}}l|}{Acc. - 44.3}  & \multicolumn{1}{>{\columncolor{heldout_test_ood}}l|}{Acc. - 64.5}  \\ \midrule
\multirow{3}{*}{\textbf{Brant}}          & Private (Pathology Detection)                       & \multicolumn{1}{>{\columncolor{pretrained_data}}l|}{}   & \multicolumn{1}{>{\columncolor{heldout_test_ood}}l|}{}  & \multicolumn{1}{>{\columncolor{few_shot}}l|}{}   & \multicolumn{1}{>{\columncolor{ood_eval}}l|}{}   & \multicolumn{1}{>{\columncolor{pretrained_data}}l|}{}   &    \multicolumn{1}{>{\columncolor{heldout_test_ood}}l|}{Acc. - 91.17} \\ \cmidrule(l){2-8} 
                                & MAYO (Pathology Detection)                           & \multicolumn{1}{>{\columncolor{pretrained_data}}l|}{}   &    \multicolumn{1}{>{\columncolor{heldout_test_ood}}l|}{} & \multicolumn{1}{>{\columncolor{few_shot}}l|}{}   & \multicolumn{1}{>{\columncolor{ood_eval}}l|}{Acc. - 89.40}   & \multicolumn{1}{>{\columncolor{pretrained_data}}l|}{}   & \multicolumn{1}{>{\columncolor{heldout_test_ood}}l|}{}  \\ \cmidrule(l){2-8} 
                                & FNUSA (Pathology Detection)                        & \multicolumn{1}{>{\columncolor{pretrained_data}}l|}{}   &    \multicolumn{1}{>{\columncolor{heldout_test_ood}}l|}{} & \multicolumn{1}{>{\columncolor{few_shot}}l|}{}   & \multicolumn{1}{>{\columncolor{ood_eval}}l|}{Acc. - 83.51}   & \multicolumn{1}{>{\columncolor{pretrained_data}}l|}{}   & \multicolumn{1}{>{\columncolor{heldout_test_ood}}l|}{}  \\ \midrule
\multirow{4}{*}{\textbf{BIOT}}           & TUAB                          & \multicolumn{1}{>{\columncolor{pretrained_data}}l|}{}   & \multicolumn{1}{>{\columncolor{heldout_test_ood}}l|}{AUC - 0.8815}  & \multicolumn{1}{>{\columncolor{few_shot}}l|}{}   & \multicolumn{1}{>{\columncolor{ood_eval}}l|}{}   & \multicolumn{1}{>{\columncolor{pretrained_data}}l|}{}   & \multicolumn{1}{>{\columncolor{heldout_test_ood}}l|}{AUC - 0.8739}  \\ \cmidrule(l){2-8} 
                                & TUEV                          & \multicolumn{1}{>{\columncolor{pretrained_data}}l|}{}   & \multicolumn{1}{>{\columncolor{heldout_test_ood}}l|}{W.F1 - 0.7504}  & \multicolumn{1}{>{\columncolor{few_shot}}l|}{}   & \multicolumn{1}{>{\columncolor{ood_eval}}l|}{}   & \multicolumn{1}{>{\columncolor{pretrained_data}}l|}{}   & \multicolumn{1}{>{\columncolor{heldout_test_ood}}l|}{W. F1 - 0.7322}  \\ \cmidrule(l){2-8} 
                                & CHB-MIT                        & \multicolumn{1}{>{\columncolor{pretrained_data}}l|}{}   & \multicolumn{1}{>{\columncolor{heldout_test_ood}}l|}{AUC - 0.8761}  & \multicolumn{1}{>{\columncolor{few_shot}}l|}{}   & \multicolumn{1}{>{\columncolor{ood_eval}}l|}{}   & \multicolumn{1}{>{\columncolor{pretrained_data}}l|}{}   & \multicolumn{1}{>{\columncolor{heldout_test_ood}}l|}{AUC - 0.8752}  \\ \cmidrule(l){2-8} 
                                & IIIC Seizure                  & \multicolumn{1}{>{\columncolor{pretrained_data}}l|}{}   & \multicolumn{1}{>{\columncolor{heldout_test_ood}}l|}{W. F1 - 0.5737}  & \multicolumn{1}{>{\columncolor{few_shot}}l|}{}   & \multicolumn{1}{>{\columncolor{ood_eval}}l|}{}   & \multicolumn{1}{>{\columncolor{pretrained_data}}l|}{}   & \multicolumn{1}{>{\columncolor{heldout_test_ood}}l|}{W. F1 - 0.5878}  \\ \midrule
\multirow{5}{*}{\textbf{EEGFormer}}      & TUAB                          & \multicolumn{1}{>{\columncolor{pretrained_data}}l|}{}   & \multicolumn{1}{>{\columncolor{heldout_test_ood}}l|}{AUC - 0.876}  & \multicolumn{1}{>{\columncolor{few_shot}}l|}{}   & \multicolumn{1}{>{\columncolor{ood_eval}}l|}{}   & \multicolumn{1}{>{\columncolor{pretrained_data}}l|}{}   &    \multicolumn{1}{>{\columncolor{heldout_test_ood}}l|}{} \\ \cmidrule(l){2-8} 
                                & TUAR                          & \multicolumn{1}{>{\columncolor{pretrained_data}}l|}{AUC - 0.827}  & \multicolumn{1}{>{\columncolor{heldout_test_ood}}l|}{AUC - 0.852}  & \multicolumn{1}{>{\columncolor{few_shot}}l|}{}   & \multicolumn{1}{>{\columncolor{ood_eval}}l|}{}   & \multicolumn{1}{>{\columncolor{pretrained_data}}l|}{}   &    \multicolumn{1}{>{\columncolor{heldout_test_ood}}l|}{} \\ \cmidrule(l){2-8} 
                                & TUSL                           & \multicolumn{1}{>{\columncolor{pretrained_data}}l|}{AUC - 0.657}  & \multicolumn{1}{>{\columncolor{heldout_test_ood}}l|}{AUC - 0.679}  & \multicolumn{1}{>{\columncolor{few_shot}}l|}{}   & \multicolumn{1}{>{\columncolor{ood_eval}}l|}{}   & \multicolumn{1}{>{\columncolor{pretrained_data}}l|}{}   &    \multicolumn{1}{>{\columncolor{heldout_test_ood}}l|}{} \\ \cmidrule(l){2-8} 
                                & TUSZ                           & \multicolumn{1}{>{\columncolor{pretrained_data}}l|}{}   & \multicolumn{1}{>{\columncolor{heldout_test_ood}}l|}{AUC - 0.883}  & \multicolumn{1}{>{\columncolor{few_shot}}l|}{}   & \multicolumn{1}{>{\columncolor{ood_eval}}l|}{}   & \multicolumn{1}{>{\columncolor{pretrained_data}}l|}{}   &    \multicolumn{1}{>{\columncolor{heldout_test_ood}}l|}{} \\ \cmidrule(l){2-8} 
                                & Neonate Dataset              & \multicolumn{1}{>{\columncolor{pretrained_data}}l|}{}   &    \multicolumn{1}{>{\columncolor{heldout_test_ood}}l|}{} & \multicolumn{1}{>{\columncolor{few_shot}}l|}{}   & \multicolumn{1}{>{\columncolor{ood_eval}}l|}{}   & \multicolumn{1}{>{\columncolor{pretrained_data}}l|}{}   & \multicolumn{1}{>{\columncolor{heldout_test_ood}}l|}{AUC - 0.842}  \\ \midrule
\multirow{4}{*}{\textbf{LaBraM}}         & TUAB                           & \multicolumn{1}{>{\columncolor{pretrained_data}}l|}{AUC - 0.8835}  & \multicolumn{1}{>{\columncolor{heldout_test_ood}}l|}{AUC - 0.9162}  & \multicolumn{1}{>{\columncolor{few_shot}}l|}{}   & \multicolumn{1}{>{\columncolor{ood_eval}}l|}{}   & \multicolumn{1}{>{\columncolor{pretrained_data}}l|}{}   &    \multicolumn{1}{>{\columncolor{heldout_test_ood}}l|}{} \\ \cmidrule(l){2-8} 
                                & TUEV                          & \multicolumn{1}{>{\columncolor{pretrained_data}}l|}{B.Acc - 34.61}  & \multicolumn{1}{>{\columncolor{heldout_test_ood}}l|}{B.Acc - 66.16}  & \multicolumn{1}{>{\columncolor{few_shot}}l|}{}   & \multicolumn{1}{>{\columncolor{ood_eval}}l|}{}   & \multicolumn{1}{>{\columncolor{pretrained_data}}l|}{}   &    \multicolumn{1}{>{\columncolor{heldout_test_ood}}l|}{} \\ \cmidrule(l){2-8} 
                                & SEED Series                   & \multicolumn{1}{>{\columncolor{pretrained_data}}l|}{}   & \multicolumn{1}{>{\columncolor{heldout_test_ood}}l|}{Acc. - 41.02}  & \multicolumn{1}{>{\columncolor{few_shot}}l|}{}   & \multicolumn{1}{>{\columncolor{ood_eval}}l|}{}   & \multicolumn{1}{>{\columncolor{pretrained_data}}l|}{}   &    \multicolumn{1}{>{\columncolor{heldout_test_ood}}l|}{} \\ \cmidrule(l){2-8} 
                                & MoBI                          & \multicolumn{1}{>{\columncolor{pretrained_data}}l|}{}   &    \multicolumn{1}{>{\columncolor{heldout_test_ood}}l|}{} & \multicolumn{1}{>{\columncolor{few_shot}}l|}{}   & \multicolumn{1}{>{\columncolor{ood_eval}}l|}{}   & \multicolumn{1}{>{\columncolor{pretrained_data}}l|}{}   & \multicolumn{1}{>{\columncolor{heldout_test_ood}}l|}{RMSE - 0.1196}  \\ \midrule
	\textbf{Mentality}                       & TUSZ                         & \multicolumn{1}{>{\columncolor{pretrained_data}}l|}{}   & \multicolumn{1}{>{\columncolor{heldout_test_ood}}l|}{AUC - 0.72} & \multicolumn{1}{>{\columncolor{few_shot}}l|}{}   & \multicolumn{1}{>{\columncolor{ood_eval}}l|}{}   & \multicolumn{1}{>{\columncolor{pretrained_data}}l|}{}   &    \multicolumn{1}{>{\columncolor{heldout_test_ood}}l|}{} \\ \midrule
\multirow{6}{*}{\textbf{NeuroLM}}        & TUAB                       & \multicolumn{1}{>{\columncolor{pretrained_data}}l|}{}   & \multicolumn{1}{>{\columncolor{heldout_test_ood}}l|}{AUC - 0.7884}  & \multicolumn{1}{>{\columncolor{few_shot}}l|}{}   & \multicolumn{1}{>{\columncolor{ood_eval}}l|}{}   & \multicolumn{1}{>{\columncolor{pretrained_data}}l|}{}   &    \multicolumn{1}{>{\columncolor{heldout_test_ood}}l|}{} \\ \cmidrule(l){2-8} 
                                & TUEV                          & \multicolumn{1}{>{\columncolor{pretrained_data}}l|}{}   & \multicolumn{1}{>{\columncolor{heldout_test_ood}}l|}{B.Acc - 0.4679}  & \multicolumn{1}{>{\columncolor{few_shot}}l|}{}   & \multicolumn{1}{>{\columncolor{ood_eval}}l|}{}   & \multicolumn{1}{>{\columncolor{pretrained_data}}l|}{}   &    \multicolumn{1}{>{\columncolor{heldout_test_ood}}l|}{} \\ \cmidrule(l){2-8} 
                                & TUSL                         & \multicolumn{1}{>{\columncolor{pretrained_data}}l|}{}   & \multicolumn{1}{>{\columncolor{heldout_test_ood}}l|}{B.Acc - 0.6845}  & \multicolumn{1}{>{\columncolor{few_shot}}l|}{}   & \multicolumn{1}{>{\columncolor{ood_eval}}l|}{}   & \multicolumn{1}{>{\columncolor{pretrained_data}}l|}{}   &    \multicolumn{1}{>{\columncolor{heldout_test_ood}}l|}{} \\ \cmidrule(l){2-8} 
                                & HMC                         & \multicolumn{1}{>{\columncolor{pretrained_data}}l|}{}   &    \multicolumn{1}{>{\columncolor{heldout_test_ood}}l|}{} & \multicolumn{1}{>{\columncolor{few_shot}}l|}{}   & \multicolumn{1}{>{\columncolor{ood_eval}}l|}{}   & \multicolumn{1}{>{\columncolor{pretrained_data}}l|}{}   & \multicolumn{1}{>{\columncolor{heldout_test_ood}}l|}{B.Acc - 0.6188}  \\ \cmidrule(l){2-8} 
                                & SEED Series                & \multicolumn{1}{>{\columncolor{pretrained_data}}l|}{}   & \multicolumn{1}{>{\columncolor{heldout_test_ood}}l|}{B.Acc - 0.6034}  & \multicolumn{1}{>{\columncolor{few_shot}}l|}{}   & \multicolumn{1}{>{\columncolor{ood_eval}}l|}{}   & \multicolumn{1}{>{\columncolor{pretrained_data}}l|}{}   &    \multicolumn{1}{>{\columncolor{heldout_test_ood}}l|}{} \\ \cmidrule(l){2-8} 
                                & Workload                     & \multicolumn{1}{>{\columncolor{pretrained_data}}l|}{}   &    \multicolumn{1}{>{\columncolor{heldout_test_ood}}l|}{} & \multicolumn{1}{>{\columncolor{few_shot}}l|}{}   & \multicolumn{1}{>{\columncolor{ood_eval}}l|}{}   & \multicolumn{1}{>{\columncolor{pretrained_data}}l|}{}   & \multicolumn{1}{>{\columncolor{heldout_test_ood}}l|}{B.Acc - 0.6345}  \\ \midrule
\multirow{7}{*}{\textbf{FoME}}           & TUEV                        & \multicolumn{1}{>{\columncolor{pretrained_data}}l|}{F1 - 46.00$^*$}  & \multicolumn{1}{>{\columncolor{heldout_test_ood}}l|}{F1 - 79.00}  & \multicolumn{1}{>{\columncolor{few_shot}}l|}{}   & \multicolumn{1}{>{\columncolor{ood_eval}}l|}{}   & \multicolumn{1}{>{\columncolor{pretrained_data}}l|}{}   &    \multicolumn{1}{>{\columncolor{heldout_test_ood}}l|}{} \\ \cmidrule(l){2-8} 
                                & Sleep-EDFx                     & \multicolumn{1}{>{\columncolor{pretrained_data}}l|}{}   & \multicolumn{1}{>{\columncolor{heldout_test_ood}}l|}{F1 - 99.10}  & \multicolumn{1}{>{\columncolor{few_shot}}l|}{}   & \multicolumn{1}{>{\columncolor{ood_eval}}l|}{}   & \multicolumn{1}{>{\columncolor{pretrained_data}}l|}{}   &    \multicolumn{1}{>{\columncolor{heldout_test_ood}}l|}{} \\ \cmidrule(l){2-8} 
                                & SEED Series                   & \multicolumn{1}{>{\columncolor{pretrained_data}}l|}{}   & \multicolumn{1}{>{\columncolor{heldout_test_ood}}l|}{F1 - 57.89}  & \multicolumn{1}{>{\columncolor{few_shot}}l|}{}   & \multicolumn{1}{>{\columncolor{ood_eval}}l|}{}   & \multicolumn{1}{>{\columncolor{pretrained_data}}l|}{}   &    \multicolumn{1}{>{\columncolor{heldout_test_ood}}l|}{} \\ \cmidrule(l){2-8} 
                                & MAYO (Events Classification)           & \multicolumn{1}{>{\columncolor{pretrained_data}}l|}{F1 - 78.80}   & \multicolumn{1}{>{\columncolor{heldout_test_ood}}l|}{F1 - 87.78}  & \multicolumn{1}{>{\columncolor{few_shot}}l|}{}   & \multicolumn{1}{>{\columncolor{ood_eval}}l|}{}   & \multicolumn{1}{>{\columncolor{pretrained_data}}l|}{}   &    \multicolumn{1}{>{\columncolor{heldout_test_ood}}l|}{} \\ \cmidrule(l){2-8} 
                                & MAYO (Pathology Detection)      & \multicolumn{1}{>{\columncolor{pretrained_data}}l|}{}   & \multicolumn{1}{>{\columncolor{heldout_test_ood}}l|}{F1 - 95.13}  & \multicolumn{1}{>{\columncolor{few_shot}}l|}{}   & \multicolumn{1}{>{\columncolor{ood_eval}}l|}{}   
                                & \multicolumn{1}{>{\columncolor{pretrained_data}}l|}{}   &    \multicolumn{1}{>{\columncolor{heldout_test_ood}}l|}{} \\ \cmidrule(l){2-8} 
                                & FNUSA (Events Classification)            & \multicolumn{1}{>{\columncolor{pretrained_data}}l|}{F1 - 77.44} &  \multicolumn{1}{>{\columncolor{heldout_test_ood}}l|}{F1 - 85.44} & \multicolumn{1}{>{\columncolor{few_shot}}l|}{}   & \multicolumn{1}{>{\columncolor{ood_eval}}l|}{}   & \multicolumn{1}{>{\columncolor{pretrained_data}}l|}{}   &    \multicolumn{1}{>{\columncolor{heldout_test_ood}}l|}{} \\ \cmidrule(l){2-8} 
                                & FNUSA (Pathology Detection)      & \multicolumn{1}{>{\columncolor{pretrained_data}}l|}{}   & \multicolumn{1}{>{\columncolor{heldout_test_ood}}l|}{F1 - 91.81}  & \multicolumn{1}{>{\columncolor{few_shot}}l|}{}   & \multicolumn{1}{>{\columncolor{ood_eval}}l|}{}   & \multicolumn{1}{>{\columncolor{pretrained_data}}l|}{}   &    \multicolumn{1}{>{\columncolor{heldout_test_ood}}l|}{} \\ \midrule
\multirow{14}{*}{\textbf{BrainWave}}     & CHB-MIT                 & \multicolumn{1}{>{\columncolor{pretrained_data}}l|}{}   &    \multicolumn{1}{>{\columncolor{heldout_test_ood}}l|}{} & \multicolumn{1}{>{\columncolor{few_shot}}l|}{}   & \multicolumn{1}{>{\columncolor{ood_eval}}l|}{}  & \multicolumn{1}{>{\columncolor{pretrained_data}}l|}{}   & \multicolumn{1}{>{\columncolor{heldout_test_ood}}l|}{AUC - 0.90$^*$}  \\ \cmidrule(l){2-8} 
                                & Clonic-6                    & \multicolumn{1}{>{\columncolor{pretrained_data}}l|}{}   &    \multicolumn{1}{>{\columncolor{heldout_test_ood}}l|}{} & \multicolumn{1}{>{\columncolor{few_shot}}l|}{}  & \multicolumn{1}{>{\columncolor{ood_eval}}l|}{AUC - 0.79$^*$}   & \multicolumn{1}{>{\columncolor{pretrained_data}}l|}{}   &    \multicolumn{1}{>{\columncolor{heldout_test_ood}}l|}{} \\ \cmidrule(l){2-8} 
                                & Atonic-5                   & \multicolumn{1}{>{\columncolor{pretrained_data}}l|}{}   &    \multicolumn{1}{>{\columncolor{heldout_test_ood}}l|}{} & \multicolumn{1}{>{\columncolor{few_shot}}l|}{}  & \multicolumn{1}{>{\columncolor{ood_eval}}l|}{AUC - 0.77$^*$}   & \multicolumn{1}{>{\columncolor{pretrained_data}}l|}{}   &    \multicolumn{1}{>{\columncolor{heldout_test_ood}}l|}{} \\ \cmidrule(l){2-8}

                                & Absence-16                   & \multicolumn{1}{>{\columncolor{pretrained_data}}l|}{}   &    \multicolumn{1}{>{\columncolor{heldout_test_ood}}l|}{} & \multicolumn{1}{>{\columncolor{few_shot}}l|}{AUC - 0.90$^*$}   & \multicolumn{1}{>{\columncolor{ood_eval}}l|}{}  & \multicolumn{1}{>{\columncolor{pretrained_data}}l|}{}   & \multicolumn{1}{>{\columncolor{heldout_test_ood}}l|}{AUC - 0.96$^*$}  \\ \cmidrule(l){2-8} 

                                & DRE-Clinical              & \multicolumn{1}{>{\columncolor{pretrained_data}}l|}{}   &    \multicolumn{1}{>{\columncolor{heldout_test_ood}}l|}{} & \multicolumn{1}{>{\columncolor{few_shot}}l|}{AUC - 0.75$^*$}   & \multicolumn{1}{>{\columncolor{ood_eval}}l|}{}  & \multicolumn{1}{>{\columncolor{pretrained_data}}l|}{}   & \multicolumn{1}{>{\columncolor{heldout_test_ood}}l|}{AUC - 0.85$^*$}  \\ \cmidrule(l){2-8} 
                                & SD-71                    & \multicolumn{1}{>{\columncolor{pretrained_data}}l|}{}   &    \multicolumn{1}{>{\columncolor{heldout_test_ood}}l|}{} & \multicolumn{1}{>{\columncolor{few_shot}}l|}{AUC - 0.63$^*$}   & \multicolumn{1}{>{\columncolor{ood_eval}}l|}{}  & \multicolumn{1}{>{\columncolor{pretrained_data}}l|}{}   & \multicolumn{1}{>{\columncolor{heldout_test_ood}}l|}{AUC - 0.70$^*$}  \\ \cmidrule(l){2-8} 
                                & ADHD-Adult                & \multicolumn{1}{>{\columncolor{pretrained_data}}l|}{}   &    \multicolumn{1}{>{\columncolor{heldout_test_ood}}l|}{} & \multicolumn{1}{>{\columncolor{few_shot}}l|}{AUC - 0.90$^*$}   & \multicolumn{1}{>{\columncolor{ood_eval}}l|}{}  & \multicolumn{1}{>{\columncolor{pretrained_data}}l|}{}   & \multicolumn{1}{>{\columncolor{heldout_test_ood}}l|}{AUC - 0.96$^*$}  \\ \cmidrule(l){2-8} 
                                & ADHD-Child                 & \multicolumn{1}{>{\columncolor{pretrained_data}}l|}{}   &    \multicolumn{1}{>{\columncolor{heldout_test_ood}}l|}{} & \multicolumn{1}{>{\columncolor{few_shot}}l|}{AUC - 0.77$^*$}   & \multicolumn{1}{>{\columncolor{ood_eval}}l|}{}  & \multicolumn{1}{>{\columncolor{pretrained_data}}l|}{}   & \multicolumn{1}{>{\columncolor{heldout_test_ood}}l|}{AUC - 0.80$^*$}  \\ \cmidrule(l){2-8} 
                                & Schizophrenia-28          & \multicolumn{1}{>{\columncolor{pretrained_data}}l|}{}   &    \multicolumn{1}{>{\columncolor{heldout_test_ood}}l|}{} & \multicolumn{1}{>{\columncolor{few_shot}}l|}{AUC - 0.72$^*$}   & \multicolumn{1}{>{\columncolor{ood_eval}}l|}{}  & \multicolumn{1}{>{\columncolor{pretrained_data}}l|}{}   & \multicolumn{1}{>{\columncolor{heldout_test_ood}}l|}{AUC - 0.88$^*$}  \\ \cmidrule(l){2-8} 
                                & Depression-122           & \multicolumn{1}{>{\columncolor{pretrained_data}}l|}{}   &    \multicolumn{1}{>{\columncolor{heldout_test_ood}}l|}{} & \multicolumn{1}{>{\columncolor{few_shot}}l|}{AUC - 0.75$^*$}   & \multicolumn{1}{>{\columncolor{ood_eval}}l|}{}  & \multicolumn{1}{>{\columncolor{pretrained_data}}l|}{}   & \multicolumn{1}{>{\columncolor{heldout_test_ood}}l|}{AUC - 0.78$^*$}  \\ \cmidrule(l){2-8} 
                                & MDD-64                   & \multicolumn{1}{>{\columncolor{pretrained_data}}l|}{}   &    \multicolumn{1}{>{\columncolor{heldout_test_ood}}l|}{} & \multicolumn{1}{>{\columncolor{few_shot}}l|}{AUC - 0.90$^*$}   & \multicolumn{1}{>{\columncolor{ood_eval}}l|}{}  & \multicolumn{1}{>{\columncolor{pretrained_data}}l|}{}   & \multicolumn{1}{>{\columncolor{heldout_test_ood}}l|}{AUC - 0.92$^*$}  \\ \cmidrule(l){2-8} 
                                & AD-65                     & \multicolumn{1}{>{\columncolor{pretrained_data}}l|}{}   &    \multicolumn{1}{>{\columncolor{heldout_test_ood}}l|}{} & \multicolumn{1}{>{\columncolor{few_shot}}l|}{AUC - 0.63$^*$}   & \multicolumn{1}{>{\columncolor{ood_eval}}l|}{}  & \multicolumn{1}{>{\columncolor{pretrained_data}}l|}{}   & \multicolumn{1}{>{\columncolor{heldout_test_ood}}l|}{AUC - 0.76$^*$}  \\ \cmidrule(l){2-8} 
                                & MAYO (Events Classification)                      & \multicolumn{1}{>{\columncolor{pretrained_data}}l|}{}   &    \multicolumn{1}{>{\columncolor{heldout_test_ood}}l|}{} & \multicolumn{1}{>{\columncolor{few_shot}}l|}{AUC - 0.81$^*$}  & \multicolumn{1}{>{\columncolor{ood_eval}}l|}{AUC - 0.93$^*$}  & \multicolumn{1}{>{\columncolor{pretrained_data}}l|}{}   & \multicolumn{1}{>{\columncolor{heldout_test_ood}}l|}{AUC - 0.97$^*$}  \\ \cmidrule(l){2-8} 
                                & FNUSA (Events Classification)                   & \multicolumn{1}{>{\columncolor{pretrained_data}}l|}{}   &    \multicolumn{1}{>{\columncolor{heldout_test_ood}}l|}{} & \multicolumn{1}{>{\columncolor{few_shot}}l|}{AUC - 0.78$^*$}  & \multicolumn{1}{>{\columncolor{ood_eval}}l|}{AUC - 0.91$^*$}  & \multicolumn{1}{>{\columncolor{pretrained_data}}l|}{}   & \multicolumn{1}{>{\columncolor{heldout_test_ood}}l|}{AUC - 0.91$^*$}  \\ \bottomrule

\end{tabular}
 \label{tab:eval_table_2}
 }
\end{table}

%% file: references.bib
@article{brinkmann2016crowdsourcing,
  title={Crowdsourcing reproducible seizure forecasting in human and canine epilepsy},
  author={Brinkmann, Benjamin H and Wagenaar, Joost and Abbot, Drew and Adkins, Phillip and Bosshard, Simone C and Chen, Min and Tieng, Quang M and He, Jialune and Mu{\~n}oz-Almaraz, FJ and Botella-Rocamora, Paloma and others},
  journal={Brain},
  volume={139},
  number={6},
  pages={1713--1722},
  year={2016},
  publisher={Oxford University Press}
}

@inproceedings{ogg2024self,
  title={Self-supervised transformer model training for a sleep-EEG foundation model},
  author={Ogg, Mattson and Coon, William G},
  booktitle={2024 46th Annual International Conference of the IEEE Engineering in Medicine and Biology Society (EMBC)},
  pages={1--6},
  year={2024},
  organization={IEEE}
}

@article{yuxuan2025foundation,
  title={Foundation models for EEG decoding: current progress and prospective research},
  author={Yuxuan, Yao and Hongbo, Wang and Li, Chen and Yiheng, Peng and Luo, Jingjing},
  journal={Journal of Neural Engineering},
  year={2025},
  publisher={IOP Publishing}
}

@article{weng2025self,
  title={Self-supervised learning for electroencephalogram: A systematic survey},
  author={Weng, Weining and Gu, Yang and Guo, Shuai and Ma, Yuan and Yang, Zhaohua and Liu, Yuchen and Chen, Yiqiang},
  journal={ACM Computing Surveys},
  volume={57},
  number={12},
  pages={1--38},
  year={2025},
  publisher={ACM New York, NY}
}

@article{zhou2025brain,
  title={Brain foundation models: A survey on advancements in neural signal processing and brain discovery},
  author={Zhou, Xinliang and Liu, Chenyu and Chen, Zhisheng and Wang, Kun and Ding, Yi and Jia, Ziyu and Wen, Qingsong},
  journal={arXiv preprint arXiv:2503.00580},
  year={2025}
}

@article{gabor1946theory,
  title={Theory of communication. Part 1: The analysis of information},
  author={Gabor, Dennis},
  journal={Journal of the Institution of Electrical Engineers-part III: radio and communication engineering},
  volume={93},
  number={26},
  pages={429--441},
  year={1946},
  publisher={IET}
}

@article{grossmann1984decomposition,
  title={Decomposition of Hardy functions into square integrable wavelets of constant shape},
  author={Grossmann, Alexander and Morlet, Jean},
  journal={SIAM journal on mathematical analysis},
  volume={15},
  number={4},
  pages={723--736},
  year={1984},
  publisher={SIAM}
}

@article{moca2021time,
  title={Time-frequency super-resolution with superlets},
  author={Moca, Vasile V and B{\^a}rzan, Harald and Nagy-D{\u{a}}b{\^a}can, Adriana and Mureșan, Raul C},
  journal={Nature communications},
  volume={12},
  number={1},
  pages={337},
  year={2021},
  publisher={Nature Publishing Group UK London}
}

@article{gong2014compressing,
  title={Compressing deep convolutional networks using vector quantization},
  author={Gong, Yunchao and Liu, Liu and Yang, Ming and Bourdev, Lubomir},
  journal={arXiv preprint arXiv:1412.6115},
  year={2014}
}

@article{han2015deep,
  title={Deep compression: Compressing deep neural networks with pruning, trained quantization and huffman coding},
  author={Han, Song and Mao, Huizi and Dally, William J},
  journal={arXiv preprint arXiv:1510.00149},
  year={2015}
}

@article{hinton2015distilling,
  title={Distilling the knowledge in a neural network},
  author={Hinton, Geoffrey and Vinyals, Oriol and Dean, Jeff},
  journal={arXiv preprint arXiv:1503.02531},
  year={2015}
}

@article{tatum2016american,
  title={American clinical neurophysiology society guideline 7: guidelines for EEG reporting},
  author={Tatum IV, William O and Selioutski, Olga and Ochoa, Juan G and Clary, Heidi Munger and Cheek, Janna and Drislane, Frank W and Tsuchida, Tammy N},
  journal={The Neurodiagnostic Journal},
  volume={56},
  number={4},
  pages={285--293},
  year={2016},
  publisher={Taylor \& Francis}
}

@inproceedings{mcmahan2017communication,
  title={Communication-efficient learning of deep networks from decentralized data},
  author={McMahan, Brendan and Moore, Eider and Ramage, Daniel and Hampson, Seth and y Arcas, Blaise Aguera},
  booktitle={Artificial intelligence and statistics},
  pages={1273--1282},
  year={2017},
  organization={PMLR}
}

@inproceedings{sun2017revisiting,
  title={Revisiting unreasonable effectiveness of data in deep learning era},
  author={Sun, Chen and Shrivastava, Abhinav and Singh, Saurabh and Gupta, Abhinav},
  booktitle={Proceedings of the IEEE international conference on computer vision},
  pages={843--852},
  year={2017}
}

@article{rosenfeld2019constructive,
  title={A constructive prediction of the generalization error across scales},
  author={Rosenfeld, Jonathan S and Rosenfeld, Amir and Belinkov, Yonatan and Shavit, Nir},
  journal={arXiv preprint arXiv:1909.12673},
  year={2019}
}

@inproceedings{dehghani2023scaling,
  title={Scaling vision transformers to 22 billion parameters},
  author={Dehghani, Mostafa and Djolonga, Josip and Mustafa, Basil and Padlewski, Piotr and Heek, Jonathan and Gilmer, Justin and Steiner, Andreas Peter and Caron, Mathilde and Geirhos, Robert and Alabdulmohsin, Ibrahim and others},
  booktitle={International conference on machine learning},
  pages={7480--7512},
  year={2023},
  organization={PMLR}
}

@article{wenzek2019ccnet,
  title={CCNet: Extracting high quality monolingual datasets from web crawl data},
  author={Wenzek, Guillaume and Lachaux, Marie-Anne and Conneau, Alexis and Chaudhary, Vishrav and Guzm{\'a}n, Francisco and Joulin, Armand and Grave, Edouard},
  journal={arXiv preprint arXiv:1911.00359},
  year={2019}
}

@article{oquab2023dinov2,
  title={Dinov2: Learning robust visual features without supervision},
  author={Oquab, Maxime and Darcet, Timoth{\'e}e and Moutakanni, Th{\'e}o and Vo, Huy and Szafraniec, Marc and Khalidov, Vasil and Fernandez, Pierre and Haziza, Daniel and Massa, Francisco and El-Nouby, Alaaeldin and others},
  journal={arXiv preprint arXiv:2304.07193},
  year={2023}
}

@article{kumar2022fine,
  title={Fine-tuning can distort pretrained features and underperform out-of-distribution},
  author={Kumar, Ananya and Raghunathan, Aditi and Jones, Robbie and Ma, Tengyu and Liang, Percy},
  journal={arXiv preprint arXiv:2202.10054},
  year={2022}
}

@incollection{thrun1998learning,
  title={Learning to learn: Introduction and overview},
  author={Thrun, Sebastian and Pratt, Lorien},
  booktitle={Learning to learn},
  pages={3--17},
  year={1998},
  publisher={Springer}
}

@inproceedings{zhang2023survey,
  title={A Survey on Masked Autoencoder for Visual Self-supervised Learning.},
  author={Zhang, Chaoning and Zhang, Chenshuang and Song, Junha and Yi, John Seon Keun and Kweon, In So},
  booktitle={IJCAI},
  pages={6805--6813},
  year={2023}
}

@article{ericsson2022self,
  title={Self-supervised representation learning: Introduction, advances, and challenges},
  author={Ericsson, Linus and Gouk, Henry and Loy, Chen Change and Hospedales, Timothy M},
  journal={IEEE Signal Processing Magazine},
  volume={39},
  number={3},
  pages={42--62},
  year={2022},
  publisher={IEEE}
}

@inproceedings{wu2021cvt,
  title={Cvt: Introducing convolutions to vision transformers},
  author={Wu, Haiping and Xiao, Bin and Codella, Noel and Liu, Mengchen and Dai, Xiyang and Yuan, Lu and Zhang, Lei},
  booktitle={Proceedings of the IEEE/CVF international conference on computer vision},
  pages={22--31},
  year={2021}
}

@inproceedings{neumann2021smart,
  title={Smart data representations: impact on the accuracy of deep neural networks},
  author={Neumann, Oliver and Ludwig, Nicole and Turowski, Marian and Heidrich, Benedikt and Hagenmeyer, Veit and Mikut, Ralf},
  booktitle={Proceedings 31 workshop computational intelligence},
  pages={113--130},
  year={2021}
}

@article{michel2019eeg,
  title={EEG source imaging: a practical review of the analysis steps},
  author={Michel, Christoph M and Brunet, Denis},
  journal={Frontiers in neurology},
  volume={10},
  pages={325},
  year={2019},
  publisher={Frontiers Media SA}
}

@article{roy2019deep,
  title={Deep learning-based electroencephalography analysis: a systematic review},
  author={Roy, Yannick and Banville, Hubert and Albuquerque, Isabela and Gramfort, Alexandre and Falk, Tiago H and Faubert, Jocelyn},
  journal={Journal of neural engineering},
  volume={16},
  number={5},
  pages={051001},
  year={2019},
  publisher={IOP Publishing}
}

@book{tatum2021handbook,
  title={Handbook of EEG interpretation},
  author={Tatum IV, William O},
  year={2021},
  publisher={Springer Publishing Company}
}

@article{popa2020role,
  title={The role of quantitative EEG in the diagnosis of neuropsychiatric disorders},
  author={Popa, Livia Livint and Dragos, Hanna and Pantelemon, Cristina and Rosu, Olivia Verisezan and Strilciuc, Stefan},
  journal={Journal of medicine and life},
  volume={13},
  number={1},
  pages={8},
  year={2020}
}

@article{gao2017inferring,
  title={Inferring synaptic excitation/inhibition balance from field potentials},
  author={Gao, Richard and Peterson, Erik J and Voytek, Bradley},
  journal={Neuroimage},
  volume={158},
  pages={70--78},
  year={2017},
  publisher={Elsevier}
}

@article{cespedes2020influence,
  title={Influence of Patient-Specific Head Modeling on EEG Source Imaging},
  author={C{\'e}spedes-Villar, Yohan and Martinez-Vargas, Juan David and Castellanos-Dominguez, Germ{\'a}n},
  journal={Computational and Mathematical Methods in Medicine},
  volume={2020},
  number={1},
  pages={5076865},
  year={2020},
  publisher={Wiley Online Library}
}

@article{smit2012individual,
  title={Individual differences in EEG spectral power reflect genetic variance in gray and white matter volumes},
  author={Smit, Dirk JA and Boomsma, Dorret I and Schnack, Hugo G and Pol, Hilleke E Hulshoff and de Geus, Eco JC},
  journal={Twin research and human genetics},
  volume={15},
  number={3},
  pages={384--392},
  year={2012},
  publisher={Cambridge University Press}
}

@inproceedings{li2021review,
  title={A review of EEG acquisition, processing and application},
  author={Li, Bohao and Cheng, Tianshuo and Guo, Zexuan},
  booktitle={Journal of Physics: Conference Series},
  volume={1907},
  number={1},
  pages={012045},
  year={2021},
  organization={IOP Publishing}
}

@article{blume2006drug,
  title={Drug effects on EEG},
  author={Blume, Warren T},
  journal={Journal of Clinical Neurophysiology},
  volume={23},
  number={4},
  pages={306--311},
  year={2006},
  publisher={LWW}
}

@article{amin2023normal,
  title={Normal variants and artifacts: importance in EEG interpretation},
  author={Amin, Ushtar and Nascimento, F{\'a}bio A and Karakis, Ioannis and Schomer, Donald and Benbadis, Selim R},
  journal={Epileptic disorders},
  volume={25},
  number={5},
  pages={591--648},
  year={2023},
  publisher={Wiley Online Library}
}

@article{perinelli2022power,
  title={Power shift and connectivity changes in healthy aging during resting-state EEG},
  author={Perinelli, Alessio and Assecondi, Sara and Tagliabue, Chiara F and Mazza, Veronica},
  journal={NeuroImage},
  volume={256},
  pages={119247},
  year={2022},
  publisher={Elsevier}
}

@article{oken2006vigilance,
  title={Vigilance, alertness, or sustained attention: physiological basis and measurement},
  author={Oken, Barry S and Salinsky, Martin C and Elsas, Siegward-M},
  journal={Clinical neurophysiology},
  volume={117},
  number={9},
  pages={1885--1901},
  year={2006},
  publisher={Elsevier}
}

@article{klimesch1996memory,
  title={Memory processes, brain oscillations and EEG synchronization},
  author={Klimesch, Wolfgang},
  journal={International journal of psychophysiology},
  volume={24},
  number={1-2},
  pages={61--100},
  year={1996},
  publisher={Elsevier}
}

@article{da2013eeg,
  title={EEG and MEG: relevance to neuroscience},
  author={da Silva, Fernando Lopes},
  journal={Neuron},
  volume={80},
  number={5},
  pages={1112--1128},
  year={2013},
  publisher={Elsevier}
}

@article{tatum2018clinical,
  title={Clinical utility of EEG in diagnosing and monitoring epilepsy in adults},
  author={Tatum, William O and Rubboli, G and Kaplan, Peter W and Mirsatari, SM and Radhakrishnan, Kuruppath and Gloss, D and Caboclo, LO and Drislane, FW and Koutroumanidis, M and Schomer, DL and others},
  journal={Clinical Neurophysiology},
  volume={129},
  number={5},
  pages={1056--1082},
  year={2018},
  publisher={Elsevier}
}

@article{varbu2022past,
  title={Past, present, and future of EEG-based BCI applications},
  author={V{\"a}rbu, Kaido and Muhammad, Naveed and Muhammad, Yar},
  journal={Sensors},
  volume={22},
  number={9},
  pages={3331},
  year={2022},
  publisher={MDPI}
}

@article{petroff2016comparison,
  title={A comparison of the power spectral density of scalp EEG and subjacent electrocorticograms},
  author={Petroff, Ognen A and Spencer, Dennis D and Goncharova, Irina I and Zaveri, Hitten P},
  journal={Clinical Neurophysiology},
  volume={127},
  number={2},
  pages={1108--1112},
  year={2016},
  publisher={Elsevier}
}

@article{brunner2016volume,
  title={Volume conduction influences scalp-based connectivity estimates},
  author={Brunner, Clemens and Billinger, Martin and Seeber, Martin and Mullen, Timothy R and Makeig, Scott},
  journal={Frontiers in computational neuroscience},
  volume={10},
  pages={121},
  year={2016},
  publisher={Frontiers Media SA}
}

@article{buzsaki2012origin,
  title={The origin of extracellular fields and currents—EEG, ECoG, LFP and spikes},
  author={Buzs{\'a}ki, Gy{\"o}rgy and Anastassiou, Costas A and Koch, Christof},
  journal={Nature reviews neuroscience},
  volume={13},
  number={6},
  pages={407--420},
  year={2012},
  publisher={Nature Publishing Group UK London}
}

@article{donoghue2020parameterizing,
  title={Parameterizing neural power spectra into periodic and aperiodic components},
  author={Donoghue, Thomas and Haller, Matar and Peterson, Erik J and Varma, Paroma and Sebastian, Priyadarshini and Gao, Richard and Noto, Torben and Lara, Antonio H and Wallis, Joni D and Knight, Robert T and others},
  journal={Nature neuroscience},
  volume={23},
  number={12},
  pages={1655--1665},
  year={2020},
  publisher={Nature Publishing Group US New York}
}

@article{klonowski2009everything,
  title={Everything you wanted to ask about EEG but were afraid to get the right answer},
  author={Klonowski, Wlodzimierz},
  journal={Nonlinear biomedical physics},
  volume={3},
  pages={1--5},
  year={2009},
  publisher={Springer}
}

@article{kiessner2024reaching,
  title={Reaching the ceiling? Empirical scaling behaviour for deep EEG pathology classification},
  author={Kiessner, Ann-Kathrin and Schirrmeister, Robin T and Boedecker, Joschka and Ball, Tonio},
  journal={Computers in Biology and Medicine},
  volume={178},
  pages={108681},
  year={2024},
  publisher={Elsevier}
}

@article{edwards2024scaling,
  title={Scaling-laws for large time-series models},
  author={Edwards, Thomas DP and Alvey, James and Alsing, Justin and Nguyen, Nam H and Wandelt, Benjamin D},
  journal={arXiv preprint arXiv:2405.13867},
  year={2024}
}

@inproceedings{isik2024scaling,
  title={Scaling laws for downstream task performance of large language models},
  author={Isik, Berivan and Ponomareva, Natalia and Hazimeh, Hussein and Paparas, Dimitris and Vassilvitskii, Sergei and Koyejo, Sanmi},
  booktitle={ICLR 2024 Workshop on Mathematical and Empirical Understanding of Foundation Models},
  year={2024}
}

@article{yao2024towards,
  title={Towards Neural Scaling Laws for Time Series Foundation Models},
  author={Yao, Qingren and Yang, Chao-Han Huck and Jiang, Renhe and Liang, Yuxuan and Jin, Ming and Pan, Shirui},
  journal={arXiv preprint arXiv:2410.12360},
  year={2024}
}

@article{kaplan2020scaling,
  title={Scaling laws for neural language models},
  author={Kaplan, Jared and McCandlish, Sam and Henighan, Tom and Brown, Tom B and Chess, Benjamin and Child, Rewon and Gray, Scott and Radford, Alec and Wu, Jeffrey and Amodei, Dario},
  journal={arXiv preprint arXiv:2001.08361},
  year={2020}
}

@inproceedings{mitchell2019model,
  title={Model cards for model reporting},
  author={Mitchell, Margaret and Wu, Simone and Zaldivar, Andrew and Barnes, Parker and Vasserman, Lucy and Hutchinson, Ben and Spitzer, Elena and Raji, Inioluwa Deborah and Gebru, Timnit},
  booktitle={Proceedings of the conference on fairness, accountability, and transparency},
  pages={220--229},
  year={2019}
}

@inproceedings{schiratti2018ensemble,
  title={An ensemble learning approach to detect epileptic seizures from long intracranial EEG recordings},
  author={Schiratti, J-B and Le Douget, Jean-Eudes and Le Van Quyen, Michel and Essid, Slim and Gramfort, Alexandre},
  booktitle={2018 IEEE International Conference on Acoustics, Speech and Signal Processing (ICASSP)},
  pages={856--860},
  year={2018},
  organization={IEEE}
}

@article {HBM:HBM23730,
author = {Schirrmeister, Robin Tibor and Springenberg, Jost Tobias and Fiederer,
  Lukas Dominique Josef and Glasstetter, Martin and Eggensperger, Katharina and Tangermann, Michael and
  Hutter, Frank and Burgard, Wolfram and Ball, Tonio},
title = {Deep learning with convolutional neural networks for EEG decoding and visualization},
journal = {Human Brain Mapping},
issn = {1097-0193},
url = {http://dx.doi.org/10.1002/hbm.23730},
doi = {10.1002/hbm.23730},
month = {aug},
year = {2017},
}

@article{gramfort2013meg,
  title={MEG and EEG data analysis with MNE-Python},
  author={Gramfort, Alexandre and Luessi, Martin and Larson, Eric and Engemann, Denis A and Strohmeier, Daniel and Brodbeck, Christian and Goj, Roman and Jas, Mainak and Brooks, Teon and Parkkonen, Lauri and others},
  journal={Frontiers in Neuroinformatics},
  volume={7},
  pages={267},
  year={2013},
  publisher={Frontiers Media SA}
}

@article{dan2024szcore,
  title={SzCORE: Seizure Community Open-Source Research Evaluation framework for the validation of electroencephalography-based automated seizure detection algorithms},
  author={Dan, Jonathan and Pale, Una and Amirshahi, Alireza and Cappelletti, William and Ingolfsson, Thorir Mar and Wang, Xiaying and Cossettini, Andrea and Bernini, Adriano and Benini, Luca and Beniczky, S{\'a}ndor and others},
  journal={Epilepsia},
  year={2024},
  publisher={Wiley Online Library}
}

@inproceedings{reyna2023predicting,
  title={Predicting neurological recovery from coma after cardiac arrest: The George B. Moody PhysioNet Challenge 2023},
  author={Reyna, Matthew A and Amorim, Edilberto and Sameni, Reza and Weigle, James and Elola, Andoni and Rad, Ali Bahrami and Seyedi, Salman and Kwon, Hyeokhyen and Zheng, Wei-Long and Ghassemi, Mohammad M and others},
  booktitle={2023 Computing in Cardiology (CinC)},
  volume={50},
  pages={1--4},
  year={2023},
  organization={IEEE}
}

@inproceedings{wei20222021,
  title={2021 BEETL competition: Advancing transfer learning for subject independence and heterogenous EEG data sets},
  author={Wei, Xiaoxi and Faisal, A Aldo and Grosse-Wentrup, Moritz and Gramfort, Alexandre and Chevallier, Sylvain and Jayaram, Vinay and Jeunet, Camille and Bakas, Stylianos and Ludwig, Siegfried and Barmpas, Konstantinos and others},
  booktitle={NeurIPS 2021 Competitions and Demonstrations Track},
  pages={205--219},
  year={2022},
  organization={PMLR}
}

@article{craikDeepLearningElectroencephalogram2019,
  title = {Deep Learning for Electroencephalogram ({{EEG}}) Classification Tasks: A Review},
  shorttitle = {Deep Learning for Electroencephalogram ({{EEG}}) Classification Tasks},
  author = {Craik, Alexander and He, Yongtian and {Contreras-Vidal}, Jose L},
  year = {2019},
  month = apr,
  journal = {Journal of Neural Engineering},
  volume = {16},
  number = {3},
  pages = {031001},
  publisher = {IOP Publishing},
  issn = {1741-2552},
  doi = {10.1088/1741-2552/ab0ab5},
  urldate = {2025-02-19},
  langid = {english}
}

@article{royDeepLearningbasedElectroencephalography2019a,
  title = {Deep Learning-Based Electroencephalography Analysis: A Systematic Review},
  shorttitle = {Deep Learning-Based Electroencephalography Analysis},
  author = {Roy, Yannick and Banville, Hubert and Albuquerque, Isabela and Gramfort, Alexandre and Falk, Tiago H and Faubert, Jocelyn},
  year = {2019},
  month = aug,
  journal = {Journal of Neural Engineering},
  volume = {16},
  number = {5},
  pages = {051001},
  publisher = {IOP Publishing},
  issn = {1741-2552},
  doi = {10.1088/1741-2552/ab260c},
  urldate = {2025-02-19},
  langid = {english}
}

@article{chen2024eegformer,
      title={EEGFormer: Towards transferable and interpretable large-scale EEG foundation model},
      author={Chen, Yuqi and Ren, Kan and Song, Kaitao and Wang, Yansen and Wang, Yifan and Li, Dongsheng and Qiu, Lili},
      journal={arXiv preprint arXiv:2401.10278},
      year={2024}
}

@article{zhang2024brant,
      title={Brant: Foundation model for intracranial neural signal},
      author={Zhang, Daoze and Yuan, Zhizhang and Yang, Yang and Chen, Junru and Wang, Jingjing and Li, Yafeng},
      journal={Advances in Neural Information Processing Systems},
      volume={36},
      year={2024}
}

@misc{yuan2024brainwavebrainsignalfoundation,
      title={BrainWave: A Brain Signal Foundation Model for Clinical Applications}, 
      author={Zhizhang Yuan and Fanqi Shen and Meng Li and Yuguo Yu and Chenhao Tan and Yang Yang},
      year={2024},
      eprint={2402.10251},
      archivePrefix={arXiv},
      primaryClass={q-bio.NC},
      url={https://arxiv.org/abs/2402.10251}, 
}

@article{nejedly_multicenter_2020,
	title = {Multicenter intracranial {EEG} dataset for classification of graphoelements and artifactual signals},
	volume = {7},
	copyright = {2020 The Author(s)},
	issn = {2052-4463},
	url = {https://www.nature.com/articles/s41597-020-0532-5},
	doi = {10.1038/s41597-020-0532-5},
	language = {en},
	number = {1},
	urldate = {2024-04-12},
	journal = {Scientific Data},
	author = {Nejedly, Petr and Kremen, Vaclav and Sladky, Vladimir and Cimbalnik, Jan and Klimes, Petr and Plesinger, Filip and Mivalt, Filip and Travnicek, Vojtech and Viscor, Ivo and Pail, Martin and Halamek, Josef and Brinkmann, Benjamin H. and Brazdil, Milan and Jurak, Pavel and Worrell, Gregory},
	month = jun,
	year = {2020},
	note = {Publisher: Nature Publishing Group},
	pages = {179},
}

@misc{jiang_neurolm_2024,
	title = {{NeuroLM}: {A} {Universal} {Multi}-task {Foundation} {Model} for {Bridging} the {Gap} between {Language} and {EEG} {Signals}},
	shorttitle = {{NeuroLM}},
	url = {http://arxiv.org/abs/2409.00101},
	doi = {10.48550/arXiv.2409.00101},
	urldate = {2024-09-25},
	publisher = {arXiv},
	author = {Jiang, Wei-Bang and Wang, Yansen and Lu, Bao-Liang and Li, Dongsheng},
	month = aug,
	year = {2024},
	note = {arXiv:2409.00101 [cs, eess]},
}

@misc{jiang_large_2024,
	title = {Large Brain Model for Learning Generic Representations with Tremendous EEG Data in BCI},
	url = {http://arxiv.org/abs/2405.18765},
	language = {en},
	urldate = {2024-09-25},
	publisher = {arXiv},
	author = {Jiang, Wei-Bang and Zhao, Li-Ming and Lu, Bao-Liang},
	month = may,
	year = {2024},
	note = {arXiv:2405.18765 [cs]},
}

@misc{shi_fome_2024,
	title = {{FoME}: {A} {Foundation} {Model} for {EEG} using {Adaptive} {Temporal}-{Lateral} {Attention} {Scaling}},
	shorttitle = {{FoME}},
	url = {http://arxiv.org/abs/2409.12454},
	language = {en},
	urldate = {2024-09-25},
	publisher = {arXiv},
	author = {Shi, Enze and Zhao, Kui and Yuan, Qilong and Wang, Jiaqi and Hu, Huawen and Yu, Sigang and Zhang, Shu},
	month = sep,
	year = {2024},
	note = {arXiv:2409.12454 [cs, eess]},
}

@article{wang2023brainbert,
  title={BrainBERT: Self-supervised representation learning for intracranial recordings},
  author={Wang, Christopher and Subramaniam, Vighnesh and Yaari, Adam Uri and Kreiman, Gabriel and Katz, Boris and Cases, Ignacio and Barbu, Andrei},
  journal={arXiv preprint arXiv:2302.14367},
  year={2023}
}

@article{ansuini2019intrinsic,
  title={Intrinsic dimension of data representations in deep neural networks},
  author={Ansuini, Alessio and Laio, Alessandro and Macke, Jakob H and Zoccolan, Davide},
  journal={Advances in Neural Information Processing Systems},
  volume={32},
  year={2019}
}

@inproceedings{cui2024neuro,
  title={Neuro-gpt: Towards a foundation model for eeg},
  author={Cui, Wenhui and Jeong, Woojae and Th{\"o}lke, Philipp and Medani, Takfarinas and Jerbi, Karim and Joshi, Anand A and Leahy, Richard M},
  booktitle={2024 IEEE International Symposium on Biomedical Imaging (ISBI)},
  pages={1--5},
  year={2024},
  organization={IEEE}
}

@article{bommasani2021opportunities,
  title={On the opportunities and risks of foundation models},
  author={Bommasani, Rishi and Hudson, Drew A and Adeli, Ehsan and Altman, Russ and Arora, Simran and von Arx, Sydney and Bernstein, Michael S and Bohg, Jeannette and Bosselut, Antoine and Brunskill, Emma and others},
  journal={arXiv preprint arXiv:2108.07258},
  year={2021}
}

@inproceedings{
        panchavati2024mentality,
        title={Mentality},
        author={Saarang Panchavati and William Speier},
        booktitle={ICLR 2024 Workshop on Learning from Time Series For Health},
        year={2024},
        url={https://openreview.net/forum?id=O6T38rRiFp}
}

@article{gu2023mamba,
  title={Mamba: Linear-time sequence modeling with selective state spaces},
  author={Gu, Albert and Dao, Tri},
  journal={arXiv preprint arXiv:2312.00752},
  year={2023}
}

@inproceedings{thapa2024sleepfm,
  title={SleepFM: Multi-modal Representation Learning for Sleep across ECG, EEG and Respiratory Signals},
  author={Thapa, Rahul and He, Bryan and Kjaer, Magnus Ruud and Moore IV, Hyatt and Ganjoo, Gauri and Mignot, Emmanuel and Zou, James Y},
  booktitle={AAAI 2024 Spring Symposium on Clinical Foundation Models},
  year={2024}
}

@article{yang2024biot,
  title={Biot: Biosignal transformer for cross-data learning in the wild},
  author={Yang, Chaoqi and Westover, M and Sun, Jimeng},
  journal={Advances in Neural Information Processing Systems},
  volume={36},
  year={2024}
}

@inproceedings{wagh2021domain,
  title={Domain-guided self-supervision of eeg data improves downstream classification performance and generalizability},
  author={Wagh, Neeraj and Wei, Jionghao and Rawal, Samarth and Berry, Brent and Barnard, Leland and Brinkmann, Benjamin and Worrell, Gregory and Jones, David and Varatharajah, Yogatheesan},
  booktitle={Machine Learning for Health},
  pages={130--142},
  year={2021},
  organization={PMLR}
}

@misc{hms-harmful-brain-activity-classification,
    author = {Jin Jing and Zhen Lin and Chaoqi Yang and Ashley Chow and Sohier Dane and Jimeng Sun and M. Brandon Westover},
    title = {HMS - Harmful Brain Activity Classification },
    year = {2024},
    howpublished = {\url{https://kaggle.com/competitions/hms-harmful-brain-activity-classification}},
    note = {Kaggle}
}

@article{van2017neural,
  title={Neural discrete representation learning},
  author={Van Den Oord, Aaron and Vinyals, Oriol and others},
  journal={Advances in neural information processing systems},
  volume={30},
  year={2017}
}

@article{peng2022beit,
  title={Beit v2: Masked image modeling with vector-quantized visual tokenizers},
  author={Peng, Zhiliang and Dong, Li and Bao, Hangbo and Ye, Qixiang and Wei, Furu},
  journal={arXiv preprint arXiv:2208.06366},
  year={2022}
}

@article{wu2022timesnet,
  title={Timesnet: Temporal 2d-variation modeling for general time series analysis},
  author={Wu, Haixu and Hu, Tengge and Liu, Yong and Zhou, Hang and Wang, Jianmin and Long, Mingsheng},
  journal={arXiv preprint arXiv:2210.02186},
  year={2022},
  publisher={arXivpreprint}
}

@article{nie2022time,
  title={A time series is worth 64 words: Long-term forecasting with transformers},
  author={Nie, Yuqi and Nguyen, Nam H and Sinthong, Phanwadee and Kalagnanam, Jayant},
  journal={arXiv preprint arXiv:2211.14730},
  year={2022}
}

@article{goswami2024moment,
  title={Moment: A family of open time-series foundation models},
  author={Goswami, Mononito and Szafer, Konrad and Choudhry, Arjun and Cai, Yifu and Li, Shuo and Dubrawski, Artur},
  journal={arXiv preprint arXiv:2402.03885},
  year={2024}
}

@article{woo2022cost,
  title={Cost: Contrastive learning of disentangled seasonal-trend representations for time series forecasting},
  author={Woo, Gerald and Liu, Chenghao and Sahoo, Doyen and Kumar, Akshat and Hoi, Steven},
  journal={arXiv preprint arXiv:2202.01575},
  year={2022}
}

@article{mushtaqOneHundredYears2024,
  title = {One Hundred Years of {{EEG}} for Brain and Behaviour Research},
  author = {Mushtaq, Faisal and Welke, Dominik and Gallagher, Anne and Pavlov, Yuri G. and Kouara, Layla and {Bosch-Bayard}, Jorge and {van den Bosch}, Jasper J. F. and Arvaneh, Mahnaz and Bland, Amy R. and Chaumon, Maximilien and Borck, Cornelius and He, Xun and Luck, Steven J. and Machizawa, Maro G. and Pernet, Cyril and Puce, Aina and Segalowitz, Sidney J. and Rogers, Christine and Awais, Muhammad and Babiloni, Claudio and Bailey, Neil W. and Baillet, Sylvain and Bendall, Robert C. A. and Brady, Daniel and {Bringas-Vega}, Maria L. and Busch, Niko A. and {Calzada-Reyes}, Ana and Chatard, Armand and Clayson, Peter E. and Cohen, Michael X. and Cole, Jonathan and Constant, Martin and Corneyllie, Alexandra and Coyle, Damien and Cruse, Damian and Delis, Ioannis and Delorme, Arnaud and Fair, Damien and Falk, Tiago H. and Gamer, Matthias and Ganis, Giorgio and Gloy, Kilian and Gregory, Samantha and Hassall, Cameron D. and Hiley, Katherine E. and Ivry, Richard B. and Jerbi, Karim and Jenkins, Michael and Kaiser, Jakob and Keil, Andreas and Knight, Robert T. and Kochen, Silvia and Kotchoubey, Boris and Krigolson, Olave E. and Langer, Nicolas and Liesefeld, Heinrich R. and Lipp{\'e}, Sarah and London, Raquel E. and MacNamara, Annmarie and Makeig, Scott and Marinovic, Welber and {Mart{\'i}nez-Montes}, Eduardo and Marzuki, Aleya A. and Mathew, Ryan K. and Michel, Christoph and Mill{\'a}n, Jos{\'e} d R. and {Mon-Williams}, Mark and {Morales-Chac{\'o}n}, Lilia and Naar, Richard and Nilsonne, Gustav and Niso, Guiomar and Nyhus, Erika and Oostenveld, Robert and Paul, Katharina and Paulus, Walter and Pfabigan, Daniela M. and Pourtois, Gilles and Rampp, Stefan and Rausch, Manuel and Robbins, Kay and Rossini, Paolo M. and Ruzzoli, Manuela and Schmidt, Barbara and Senderecka, Magdalena and Srinivasan, Narayanan and Stegmann, Yannik and Thompson, Paul M. and {Valdes-Sosa}, Mitchell and {van der Molen}, Melle J. W. and Veniero, Domenica and Verona, Edelyn and Voytek, Bradley and Yao, Dezhong and Evans, Alan C. and {Valdes-Sosa}, Pedro},
  year = {2024},
  month = aug,
  journal = {Nature Human Behaviour},
  volume = {8},
  number = {8},
  pages = {1437--1443},
  publisher = {Nature Publishing Group},
  issn = {2397-3374},
  doi = {10.1038/s41562-024-01941-5},
  urldate = {2025-02-17},
  copyright = {2024 Springer Nature Limited},
  langid = {english}
}

@misc{nieTimeSeriesWorth2023,
  title = {A {{Time Series}} Is {{Worth}} 64 {{Words}}: {{Long-term Forecasting}} with {{Transformers}}},
  shorttitle = {A {{Time Series}} Is {{Worth}} 64 {{Words}}},
  author = {Nie, Yuqi and Nguyen, Nam H. and Sinthong, Phanwadee and Kalagnanam, Jayant},
  year = {2023},
  month = mar,
  number = {arXiv:2211.14730},
  eprint = {2211.14730},
  primaryclass = {cs},
  publisher = {arXiv},
  doi = {10.48550/arXiv.2211.14730},
  urldate = {2025-02-17},
  archiveprefix = {arXiv}
}

@article{data:obeid2016temple,
  title={The temple university hospital EEG data corpus},
  author={Obeid, Iyad and Picone, Joseph},
  journal={Frontiers in neuroscience},
  volume={10},
  pages={196},
  year={2016},
  publisher={Frontiers Media SA}
}

@article{data:guttag2010chb,
  title={CHB-MIT scalp EEG database (version 1.0. 0)},
  author={Guttag, John},
  journal={PhysioNet},
  year={2010}
}

@article{data:kemp2000analysis,
  title={Analysis of a sleep-dependent neuronal feedback loop: the slow-wave microcontinuity of the EEG},
  author={Kemp, Bob and Zwinderman, Aeilko H and Tuk, Bert and Kamphuisen, Hilbert AC and Oberye, Josefien JL},
  journal={IEEE Transactions on Biomedical Engineering},
  volume={47},
  number={9},
  pages={1185--1194},
  year={2000},
  publisher={IEEE}
}

@article{data:detti2020eeg,
  title={EEG synchronization analysis for seizure prediction: A study on data of noninvasive recordings},
  author={Detti, Paolo and Vatti, Giampaolo and Zabalo Manrique de Lara, Garazi},
  journal={Processes},
  volume={8},
  number={7},
  pages={846},
  year={2020},
  publisher={MDPI}
}

@article{data:quan1997sleep,
  title={The sleep heart health study: design, rationale, and methods},
  author={Quan, Stuart F and Howard, Barbara V and Iber, Conrad and Kiley, James P and Nieto, F Javier and O'Connor, George T and Rapoport, David M and Redline, Susan and Robbins, John and Samet, Jonathan M and others},
  journal={Sleep},
  volume={20},
  number={12},
  pages={1077--1085},
  year={1997},
  publisher={Oxford University Press}
}

@article{data:zhang2018national,
  title={The National Sleep Research Resource: towards a sleep data commons},
  author={Zhang, Guo-Qiang and Cui, Licong and Mueller, Remo and Tao, Shiqiang and Kim, Matthew and Rueschman, Michael and Mariani, Sara and Mobley, Daniel and Redline, Susan},
  journal={Journal of the American Medical Informatics Association},
  volume={25},
  number={10},
  pages={1351--1358},
  year={2018},
  publisher={Oxford University Press}
}

@article{data:terzano2001atlas,
  title={Atlas, rules, and recording techniques for the scoring of cyclic alternating pattern (CAP) in human sleep},
  author={Terzano, Mario Giovanni and Parrino, Liborio and Sherieri, Adriano and Chervin, Ronald and Chokroverty, Sudhansu and Guilleminault, Christian and Hirshkowitz, Max and Mahowald, Mark and Moldofsky, Harvey and Rosa, Agostino and others},
  journal={Sleep medicine},
  volume={2},
  number={6},
  pages={537--554},
  year={2001},
  publisher={Amsterdam; New York: Elsevier, c2000-}
}

@article{data:alvarez2021inter,
  title={Inter-database validation of a deep learning approach for automatic sleep scoring},
  author={Alvarez-Estevez, Diego and Rijsman, Roselyne M},
  journal={PloS one},
  volume={16},
  number={8},
  pages={e0256111},
  year={2021},
  publisher={Public Library of Science San Francisco, CA USA}
}

@article{data:hatlestad2022bids,
  title={BIDS-structured resting-state electroencephalography (EEG) data extracted from an experimental paradigm},
  author={Hatlestad-Hall, Christoffer and Rygvold, Trine Waage and Andersson, Stein},
  journal={Data in Brief},
  volume={45},
  pages={108647},
  year={2022},
  publisher={Elsevier}
}

@article{data:olejarczyk2017graph,
  title={Graph-based analysis of brain connectivity in schizophrenia},
  author={Olejarczyk, Elzbieta and Jernajczyk, Wojciech},
  journal={PloS one},
  volume={12},
  number={11},
  pages={e0188629},
  year={2017},
  publisher={Public Library of Science San Francisco, CA USA}
}

@article{data:liu2024eeg,
  title={An EEG motor imagery dataset for brain computer interface in acute stroke patients},
  author={Liu, Haijie and Wei, Penghu and Wang, Haochong and Lv, Xiaodong and Duan, Wei and Li, Meijie and Zhao, Yan and Wang, Qingmei and Chen, Xinyuan and Shi, Gaige and others},
  journal={Scientific Data},
  volume={11},
  number={1},
  pages={131},
  year={2024},
  publisher={Nature Publishing Group UK London}
}

@dataset{data:ds002778:1.0.5,
  author = {Alexander P. Rockhill and Nicko Jackson and Jobi George and Adam Aron and Nicole C. Swann},
  title = {"UC San Diego Resting State EEG Data from Patients with Parkinson's Disease"},
  year = {2021},
  doi = {doi:10.18112/openneuro.ds002778.v1.0.5},
  publisher = {OpenNeuro}
}

@article{data:cavanagh2018diminished,
  title={Diminished EEG habituation to novel events effectively classifies Parkinson’s patients},
  author={Cavanagh, James F and Kumar, Praveen and Mueller, Andrea A and Richardson, Sarah Pirio and Mueen, Abdullah},
  journal={Clinical Neurophysiology},
  volume={129},
  number={2},
  pages={409--418},
  year={2018},
  publisher={Elsevier}
}

@article{data:anjum2020linear,
  title={Linear predictive coding distinguishes spectral EEG features of Parkinson's disease},
  author={Anjum, Md Fahim and Dasgupta, Soura and Mudumbai, Raghuraman and Singh, Arun and Cavanagh, James F and Narayanan, Nandakumar S},
  journal={Parkinsonism \& related disorders},
  volume={79},
  pages={79--85},
  year={2020},
  publisher={Elsevier}
}

@article{data:vicchietti2023computational,
  title={Computational methods of EEG signals analysis for Alzheimer’s disease classification},
  author={Vicchietti, M{\'a}rio L and Ramos, Fernando M and Betting, Luiz E and Campanharo, Andriana SLO},
  journal={Scientific Reports},
  volume={13},
  number={1},
  pages={8184},
  year={2023},
  publisher={Nature Publishing Group UK London}
}

@article{data:stevenson2019dataset,
  title={A dataset of neonatal EEG recordings with seizure annotations},
  author={Stevenson, Nathan J and Tapani, Karoliina and Lauronen, Leena and Vanhatalo, Sampsa},
  journal={Scientific data},
  volume={6},
  number={1},
  pages={1--8},
  year={2019},
  publisher={Nature Publishing Group}
}

@article{data:ge2021deep,
  title={Deep active learning for interictal ictal injury continuum EEG patterns},
  author={Ge, Wendong and Jing, Jin and An, Sungtae and Herlopian, Aline and Ng, Marcus and Struck, Aaron F and Appavu, Brian and Johnson, Emily L and Osman, Gamaleldin and Haider, Hiba A and others},
  journal={Journal of neuroscience methods},
  volume={351},
  pages={108966},
  year={2021},
  publisher={Elsevier}
}

@article{data:xiang2024resting,
  title={A resting-state EEG dataset for sleep deprivation},
  author={Xiang, Chuqin and Fan, Xinrui and Bai, Duo and Lv, Ke and Lei, Xu},
  journal={Scientific Data},
  volume={11},
  number={1},
  pages={427},
  year={2024},
  publisher={Nature Publishing Group UK London}
}

@inproceedings{data:trinh2023task,
  title={Task-related and resting-state EEG classification of adult patients with ADHD using machine learning},
  author={Trinh, Nam and Whelan, Robert and Ward, Tomas and Derosiere, Gerard},
  booktitle={2023 IEEE 19th International Conference on Body Sensor Networks (BSN)},
  pages={1--4},
  year={2023},
  organization={IEEE}
}

@article{data:motie2020eeg,
  title={EEG data for ADHD/Control children},
  author={Motie Nasrabadi, Ali and Allahverdy, Armin and Samavati, Mehdi and Mohammadi, Mohammad Reza},
  journal={(No Title)},
  year={2020},
  publisher={IEEE DataPort}
}

@dataset{data:ds003478:1.1.0,
  author = {James F Cavanagh  jcavanagh@unm.edu},
  title = {"EEG: Depression rest"},
  year = {2021},
  doi = {10.18112/openneuro.ds003478.v1.1.0},
  publisher = {OpenNeuro}
}

@article{data:mumtaz2016mdd,
  title={MDD patients and healthy controls EEG data (new)},
  author={Mumtaz, Wajid},
  journal={figshare, Dataset},
  year={2016}
}

@article{data:miltiadous2023dataset,
  title={A dataset of scalp EEG recordings of Alzheimer’s disease, frontotemporal dementia and healthy subjects from routine EEG},
  author={Miltiadous, Andreas and Tzimourta, Katerina D and Afrantou, Theodora and Ioannidis, Panagiotis and Grigoriadis, Nikolaos and Tsalikakis, Dimitrios G and Angelidis, Pantelis and Tsipouras, Markos G and Glavas, Euripidis and Giannakeas, Nikolaos and others},
  journal={Data},
  volume={8},
  number={6},
  pages={95},
  year={2023},
  publisher={MDPI}
}

@article{data:schalk2004bci2000,
  title={BCI2000: a general-purpose brain-computer interface (BCI) system},
  author={Schalk, Gerwin and McFarland, Dennis J and Hinterberger, Thilo and Birbaumer, Niels and Wolpaw, Jonathan R},
  journal={IEEE Transactions on biomedical engineering},
  volume={51},
  number={6},
  pages={1034--1043},
  year={2004},
  publisher={IEEE}
}

@article{data:blankertz2007non,
  title={The non-invasive Berlin brain--computer interface: fast acquisition of effective performance in untrained subjects},
  author={Blankertz, Benjamin and Dornhege, Guido and Krauledat, Matthias and M{\"u}ller, Klaus-Robert and Curio, Gabriel},
  journal={NeuroImage},
  volume={37},
  number={2},
  pages={539--550},
  year={2007},
  publisher={Elsevier}
}

@inproceedings{data:savran06_einterface,
  title     = {Emotion detection in the loop from brain signals and facial images},
  author    = {Arman Savran and Koray Ciftci and Guillame Chanel and Javier {Cruz Mota} and Luong Hong Viet and Bülent Sankur and Lale Akarun and Alice Caplier and Michele Rombaut},
  year      = {2006},
  booktitle = {Summer Workshop on Multimodal Interfaces (eINTERFACE 2006)},
  pages     = {69--80},
}

@article{data:luciw2014multi,
  title={Multi-channel EEG recordings during 3,936 grasp and lift trials with varying weight and friction},
  author={Luciw, Matthew D and Jarocka, Ewa and Edin, Benoni B},
  journal={Scientific data},
  volume={1},
  number={1},
  pages={1--11},
  year={2014},
  publisher={Nature Publishing Group}
}

@article{data:margaux2012objective,
  title={Objective and Subjective Evaluation of Online Error Correction during P300-Based Spelling},
  author={Margaux, Perrin and Emmanuel, Maby and S{\'e}bastien, Daligault and Olivier, Bertrand and J{\'e}r{\'e}mie, Mattout},
  journal={Advances in Human-Computer Interaction},
  volume={2012},
  number={1},
  pages={578295},
  year={2012},
  publisher={Wiley Online Library}
}

@article{data:trujillo2019mental,
  title={Mental effort and information-processing costs are inversely related to global brain free energy during visual categorization},
  author={Trujillo, Logan T},
  journal={Frontiers in neuroscience},
  volume={13},
  pages={1292},
  year={2019},
  publisher={Frontiers Media SA}
}

@article{data:trujillo2017effect,
  title={The effect of electroencephalogram (EEG) reference choice on information-theoretic measures of the complexity and integration of EEG signals},
  author={Trujillo, Logan T and Stanfield, Candice T and Vela, Ruben D},
  journal={Frontiers in neuroscience},
  volume={11},
  pages={425},
  year={2017},
  publisher={Frontiers Media SA}
}

@article{data:zheng2015investigating,
  title={Investigating critical frequency bands and channels for EEG-based emotion recognition with deep neural networks},
  author={Zheng, Wei-Long and Lu, Bao-Liang},
  journal={IEEE Transactions on autonomous mental development},
  volume={7},
  number={3},
  pages={162--175},
  year={2015},
  publisher={IEEE}
}

@article{data:zheng2018emotionmeter,
  title={Emotionmeter: A multimodal framework for recognizing human emotions},
  author={Zheng, Wei-Long and Liu, Wei and Lu, Yifei and Lu, Bao-Liang and Cichocki, Andrzej},
  journal={IEEE transactions on cybernetics},
  volume={49},
  number={3},
  pages={1110--1122},
  year={2018},
  publisher={IEEE}
}

@article{data:liu2022identifying,
  title={Identifying similarities and differences in emotion recognition with EEG and eye movements among Chinese, German, and French People},
  author={Liu, Wei and Zheng, Wei-Long and Li, Ziyi and Wu, Si-Yuan and Gan, Lu and Lu, Bao-Liang},
  journal={Journal of Neural Engineering},
  volume={19},
  number={2},
  pages={026012},
  year={2022},
  publisher={IOP Publishing}
}

@article{data:torkamani2020prediction,
  title={Prediction of reaction time and vigilance variability from spatio-spectral features of resting-state EEG in a long sustained attention task},
  author={Torkamani-Azar, Mastaneh and Kanik, Sumeyra Demir and Aydin, Serap and Cetin, Mujdat},
  journal={IEEE journal of biomedical and health informatics},
  volume={24},
  number={9},
  pages={2550--2558},
  year={2020},
  publisher={IEEE}
}

@phdthesis{data:korczowski2019brain,
  title={Brain Invaders calibration-less P300-based BCI with modulation of flash duration Dataset (bi2015a)},
  author={Korczowski, Louis and Cederhout, Martine and Andreev, Anton and Cattan, Gr{\'e}goire and Rodrigues, Pedro Luiz Coelho and Gautheret, Violette and Congedo, Marco},
  year={2019},
  school={GIPSA-lab}
}

@inproceedings{data:jiang2021discriminating,
  title={Discriminating surprise and anger from EEG and eye movements with a graph network},
  author={Jiang, Wei-Bang and Zhao, Li-Ming and Guo, Ping and Lu, Bao-Liang},
  booktitle={2021 IEEE International Conference on Bioinformatics and Biomedicine (BIBM)},
  pages={1353--1357},
  year={2021},
  organization={IEEE}
}

@inproceedings{data:jiang2023multimodal,
  title={Multimodal adaptive emotion transformer with flexible modality inputs on a novel dataset with continuous labels},
  author={Jiang, Wei-Bang and Liu, Xuan-Hao and Zheng, Wei-Long and Lu, Bao-Liang},
  booktitle={proceedings of the 31st ACM international conference on multimedia},
  pages={5975--5984},
  year={2023}
}

@inproceedings{data:luo2022multimodal,
  title={Multimodal emotion recognition in response to oil paintings},
  author={Luo, Shuai and Lan, Yu-Ting and Peng, Dan and Li, Ziyi and Zheng, Wei-Long and Lu, Bao-Liang},
  booktitle={2022 44th Annual International Conference of the IEEE Engineering in Medicine \& Biology Society (EMBC)},
  pages={4167--4170},
  year={2022},
  organization={IEEE}
}

@inproceedings{data:tao2020emotion,
  title={Emotion recognition under sleep deprivation using a multimodal residual LSTM network},
  author={Tao, Le-Yan and Lu, Bao-Liang},
  booktitle={2020 International Joint Conference on Neural Networks (IJCNN)},
  pages={1--8},
  year={2020},
  organization={IEEE}
}

@article{data:brunner2008bci,
  title={BCI Competition 2008--Graz data set A},
  author={Brunner, Clemens and Leeb, Robert and M{\"u}ller-Putz, Gernot and Schl{\"o}gl, Alois and Pfurtscheller, Gert},
  journal={Institute for knowledge discovery (laboratory of brain-computer interfaces), Graz University of Technology},
  volume={16},
  number={1-6},
  pages={34},
  year={2008},
  publisher={Citeseer}
}

@article{data:zyma2019electroencephalograms,
  title={Electroencephalograms during mental arithmetic task performance},
  author={Zyma, Igor and Tukaev, Sergii and Seleznov, Ivan and Kiyono, Ken and Popov, Anton and Chernykh, Mariia and Shpenkov, Oleksii},
  journal={Data},
  volume={4},
  number={1},
  pages={14},
  year={2019},
  publisher={MDPI}
}

@article{data:wang2024brain,
  title={Brain treebank: Large-scale intracranial recordings from naturalistic language stimuli},
  author={Wang, Christopher and Yaari, Adam and Singh, Aaditya and Subramaniam, Vighnesh and Rosenfarb, Dana and DeWitt, Jan and Misra, Pranav and Madsen, Joseph and Stone, Scellig and Kreiman, Gabriel and others},
  journal={Advances in Neural Information Processing Systems},
  volume={37},
  pages={96505--96540},
  year={2024}
}

@article{data:nejedly2020multicenter,
  title={Multicenter intracranial EEG dataset for classification of graphoelements and artifactual signals},
  author={Nejedly, Petr and Kremen, Vaclav and Sladky, Vladimir and Cimbalnik, Jan and Klimes, Petr and Plesinger, Filip and Mivalt, Filip and Travnicek, Vojtech and Viscor, Ivo and Pail, Martin and others},
  journal={Scientific data},
  volume={7},
  number={1},
  pages={179},
  year={2020},
  publisher={Nature Publishing Group UK London}
}

@dataset{data:ds004080:1.2.4,
  author = {D. van Blooijs and M.A. van den Boom and J.F. van der Aar and G.J.M. Huiskamp and G. Castegnaro and M. Demuru and W.J.E.M. Zweiphenning and P. van Eijsden and K. J. Miller and F.S.S. Leijten and D. Hermes},
  title = {"CCEP ECoG dataset across age 4-51"},
  year = {2023},
  doi = {doi:10.18112/openneuro.ds004080.v1.2.4},
  publisher = {OpenNeuro}
}

@inproceedings{data:li2021discrimination,
  title={Discrimination of decision confidence levels from EEG signals},
  author={Li, Rui and Liu, Le-Dian and Lu, Bao-Liang},
  booktitle={2021 10th International IEEE/EMBS Conference on Neural Engineering (NER)},
  pages={946--949},
  year={2021},
  organization={IEEE}
}

@article{data:he2018mobile,
  title={A mobile brain-body imaging dataset recorded during treadmill walking with a brain-computer interface},
  author={He, Yongtian and Luu, Trieu Phat and Nathan, Kevin and Nakagome, Sho and Contreras-Vidal, Jose L},
  journal={Scientific data},
  volume={5},
  number={1},
  pages={1--10},
  year={2018},
  publisher={Nature Publishing Group}
}

@article{wagh2022evaluating,
  title={Evaluating latent space robustness and uncertainty of EEG-ML models under realistic distribution shifts},
  author={Wagh, Neeraj and Wei, Jionghao and Rawal, Samarth and Berry, Brent M and Varatharajah, Yogatheesan},
  journal={Advances in Neural Information Processing Systems},
  volume={35},
  pages={21142--21156},
  year={2022}
}

@article{banville2021uncovering,
  title={Uncovering the structure of clinical EEG signals with self-supervised learning},
  author={Banville, Hubert and Chehab, Omar and Hyv{\"a}rinen, Aapo and Engemann, Denis-Alexander and Gramfort, Alexandre},
  journal={Journal of Neural Engineering},
  volume={18},
  number={4},
  pages={046020},
  year={2021},
  publisher={IOP Publishing}
}

@article{obeid2016temple,
  title={The temple university hospital EEG data corpus},
  author={Obeid, Iyad and Picone, Joseph},
  journal={Frontiers in neuroscience},
  volume={10},
  pages={196},
  year={2016},
  publisher={Frontiers Media SA}
}

@inproceedings{devlin2019bert,
  title={Bert: Pre-training of deep bidirectional transformers for language understanding},
  author={Devlin, Jacob and Chang, Ming-Wei and Lee, Kenton and Toutanova, Kristina},
  booktitle={Proceedings of the 2019 conference of the North American chapter of the association for computational linguistics: human language technologies, volume 1 (long and short papers)},
  pages={4171--4186},
  year={2019}
}

@article{mosher2002eeg,
  title={EEG and MEG: forward solutions for inverse methods},
  author={Mosher, John C and Leahy, Richard M and Lewis, Paul S},
  journal={IEEE Transactions on biomedical engineering},
  volume={46},
  number={3},
  pages={245--259},
  year={2002},
  publisher={IEEE}
}

@inproceedings{goel2022s,
  title={It’s raw! audio generation with state-space models},
  author={Goel, Karan and Gu, Albert and Donahue, Chris and R{\'e}, Christopher},
  booktitle={International conference on machine learning},
  pages={7616--7633},
  year={2022},
  organization={PMLR}
}

@inproceedings{ronneberger2015u,
  title={U-net: Convolutional networks for biomedical image segmentation},
  author={Ronneberger, Olaf and Fischer, Philipp and Brox, Thomas},
  booktitle={Medical image computing and computer-assisted intervention--MICCAI 2015: 18th international conference, Munich, Germany, October 5-9, 2015, proceedings, part III 18},
  pages={234--241},
  year={2015},
  organization={Springer}
}

@article{lawhern2018eegnet,
  title={EEGNet: a compact convolutional neural network for EEG-based brain--computer interfaces},
  author={Lawhern, Vernon J and Solon, Amelia J and Waytowich, Nicholas R and Gordon, Stephen M and Hung, Chou P and Lance, Brent J},
  journal={Journal of neural engineering},
  volume={15},
  number={5},
  pages={056013},
  year={2018},
  publisher={iOP Publishing}
}

@inproceedings{chen2020simple,
  title={A simple framework for contrastive learning of visual representations},
  author={Chen, Ting and Kornblith, Simon and Norouzi, Mohammad and Hinton, Geoffrey},
  booktitle={International conference on machine learning},
  pages={1597--1607},
  year={2020},
  organization={PmLR}
}

@inproceedings{he2022masked,
  title={Masked autoencoders are scalable vision learners},
  author={He, Kaiming and Chen, Xinlei and Xie, Saining and Li, Yanghao and Doll{\'a}r, Piotr and Girshick, Ross},
  booktitle={Proceedings of the IEEE/CVF conference on computer vision and pattern recognition},
  pages={16000--16009},
  year={2022}
}

@article{kostas2021bendr,
  title={BENDR: Using transformers and a contrastive self-supervised learning task to learn from massive amounts of EEG data},
  author={Kostas, Demetres and Aroca-Ouellette, Stephane and Rudzicz, Frank},
  journal={Frontiers in Human Neuroscience},
  volume={15},
  pages={653659},
  year={2021},
  publisher={Frontiers Media SA}
}

@inproceedings{oikonomou2017comparison,
  title={A comparison study on EEG signal processing techniques using motor imagery EEG data},
  author={Oikonomou, Vangelis P and Georgiadis, Kostas and Liaros, George and Nikolopoulos, Spiros and Kompatsiaris, Ioannis},
  booktitle={2017 IEEE 30th international symposium on computer-based medical systems (CBMS)},
  pages={781--786},
  year={2017},
  organization={IEEE}
}

@article{schirrmeister2017deep,
  title={Deep learning with convolutional neural networks for EEG decoding and visualization},
  author={Schirrmeister, Robin Tibor and Springenberg, Jost Tobias and Fiederer, Lukas Dominique Josef and Glasstetter, Martin and Eggensperger, Katharina and Tangermann, Michael and Hutter, Frank and Burgard, Wolfram and Ball, Tonio},
  journal={Human brain mapping},
  volume={38},
  number={11},
  pages={5391--5420},
  year={2017},
  publisher={Wiley Online Library}
}

@article{amin2019deep,
  title={Deep Learning for EEG motor imagery classification based on multi-layer CNNs feature fusion},
  author={Amin, Syed Umar and Alsulaiman, Mansour and Muhammad, Ghulam and Mekhtiche, Mohamed Amine and Hossain, M Shamim},
  journal={Future Generation computer systems},
  volume={101},
  pages={542--554},
  year={2019},
  publisher={Elsevier}
}

@article{zhang2020motor,
  title={Motor imagery classification via temporal attention cues of graph embedded EEG signals},
  author={Zhang, Dalin and Chen, Kaixuan and Jian, Debao and Yao, Lina},
  journal={IEEE journal of biomedical and health informatics},
  volume={24},
  number={9},
  pages={2570--2579},
  year={2020},
  publisher={IEEE}
}

@article{radford2019language,
  title={Language models are unsupervised multitask learners},
  author={Radford, Alec and Wu, Jeffrey and Child, Rewon and Luan, David and Amodei, Dario and Sutskever, Ilya and others},
  journal={OpenAI blog},
  volume={1},
  number={8},
  pages={9},
  year={2019}
}

@inproceedings{potter2022unsupervised,
  title={Unsupervised multivariate time-series transformers for seizure identification on eeg},
  author={Potter, Ilkay Y{\i}ld{\i}z and Zerveas, George and Eickhoff, Carsten and Duncan, Dominique},
  booktitle={2022 21st IEEE International Conference on Machine Learning and Applications (ICMLA)},
  pages={1304--1311},
  year={2022},
  organization={IEEE}
}

@article{zhang2022self,
  title={Self-supervised contrastive pre-training for time series via time-frequency consistency},
  author={Zhang, Xiang and Zhao, Ziyuan and Tsiligkaridis, Theodoros and Zitnik, Marinka},
  journal={Advances in neural information processing systems},
  volume={35},
  pages={3988--4003},
  year={2022}
}

@article{eldele2021time,
  title={Time-series representation learning via temporal and contextual contrasting},
  author={Eldele, Emadeldeen and Ragab, Mohamed and Chen, Zhenghua and Wu, Min and Kwoh, Chee Keong and Li, Xiaoli and Guan, Cuntai},
  journal={arXiv preprint arXiv:2106.14112},
  year={2021}
}

@article{zhang2015low,
  title={Low-complexity seizure prediction from iEEG/sEEG using spectral power and ratios of spectral power},
  author={Zhang, Zisheng and Parhi, Keshab K},
  journal={IEEE transactions on biomedical circuits and systems},
  volume={10},
  number={3},
  pages={693--706},
  year={2015},
  publisher={IEEE}
}

@inproceedings{handa2021epileptic,
  title={Epileptic seizure detection using rhythmicity spectrogram and cross-patient test set},
  author={Handa, Palak and Goel, Nidhi},
  booktitle={2021 8th international conference on signal processing and integrated networks (SPIN)},
  pages={898--902},
  year={2021},
  organization={IEEE}
}

@article{kadivar2019comparison,
  title={A comparison of conventional electroencephalography with amplitude-integrated EEG in detection of neonatal seizures},
  author={Kadivar, Maliheh and Moghadam, Elahe Movahedi and Shervin Badv, Reza and Sangsari, Raziye and Saeedy, Maryam},
  journal={Medical Devices: Evidence and Research},
  pages={489--496},
  year={2019},
  publisher={Taylor \& Francis}
}

@article{wang2022seeg,
  title={SEEG-Net: An explainable and deep learning-based cross-subject pathological activity detection method for drug-resistant epilepsy},
  author={Wang, Yiping and Yang, Yanfeng and Cao, Gongpeng and Guo, Jinjie and Wei, Penghu and Feng, Tao and Dai, Yang and Huang, Jinguo and Kang, Guixia and Zhao, Guoguang},
  journal={Computers in Biology and Medicine},
  volume={148},
  pages={105703},
  year={2022},
  publisher={Elsevier}
}

@article{jing2023development,
  title={Development of expert-level classification of seizures and rhythmic and periodic patterns during EEG interpretation},
  author={Jing, Jin and Ge, Wendong and Hong, Shenda and Fernandes, Marta Bento and Lin, Zhen and Yang, Chaoqi and An, Sungtae and Struck, Aaron F and Herlopian, Aline and Karakis, Ioannis and others},
  journal={Neurology},
  volume={100},
  number={17},
  pages={e1750--e1762},
  year={2023},
  publisher={Lippincott Williams \& Wilkins Hagerstown, MD}
}

@article{yang2021self,
  title={Self-supervised EEG representation learning for automatic sleep staging},
  author={Yang, Chaoqi and Xiao, Danica and Westover, M Brandon and Sun, Jimeng},
  journal={arXiv preprint arXiv:2110.15278},
  year={2021}
}

@inproceedings{peh2022transformer,
  title={Transformer convolutional neural networks for automated artifact detection in scalp EEG},
  author={Peh, Wei Yan and Yao, Yuanyuan and Dauwels, Justin},
  booktitle={2022 44th Annual International Conference of the IEEE Engineering in Medicine \& Biology Society (EMBC)},
  pages={3599--3602},
  year={2022},
  organization={IEEE}
}

@article{li2022motor,
  title={Motor imagery EEG classification algorithm based on CNN-LSTM feature fusion network},
  author={Li, Hongli and Ding, Man and Zhang, Ronghua and Xiu, Chunbo},
  journal={Biomedical signal processing and control},
  volume={72},
  pages={103342},
  year={2022},
  publisher={Elsevier}
}

@article{song2021transformer,
  title={Transformer-based spatial-temporal feature learning for EEG decoding},
  author={Song, Yonghao and Jia, Xueyu and Yang, Lie and Xie, Longhan},
  journal={arXiv preprint arXiv:2106.11170},
  year={2021}
}

@article{bai2018empirical,
  title={An empirical evaluation of generic convolutional and recurrent networks for sequence modeling},
  author={Bai, Shaojie and Kolter, J Zico and Koltun, Vladlen},
  journal={arXiv preprint arXiv:1803.01271},
  year={2018}
}

@article{tang2021self,
  title={Self-supervised graph neural networks for improved electroencephalographic seizure analysis},
  author={Tang, Siyi and Dunnmon, Jared A and Saab, Khaled and Zhang, Xuan and Huang, Qianying and Dubost, Florian and Rubin, Daniel L and Lee-Messer, Christopher},
  journal={arXiv preprint arXiv:2104.08336},
  year={2021}
}

@inproceedings{tang2023modeling,
  title={Modeling multivariate biosignals with graph neural networks and structured state space models},
  author={Tang, Siyi and Dunnmon, Jared A and Liangqiong, Qu and Saab, Khaled K and Baykaner, Tina and Lee-Messer, Christopher and Rubin, Daniel L},
  booktitle={Conference on health, inference, and learning},
  pages={50--71},
  year={2023},
  organization={PMLR}
}

@article{hochreiter1997long,
  title={Long short-term memory},
  author={Hochreiter, Sepp and Schmidhuber, J{\"u}rgen},
  journal={Neural computation},
  volume={9},
  number={8},
  pages={1735--1780},
  year={1997},
  publisher={MIT press}
}

@inproceedings{liu2022convnet,
  title={A convnet for the 2020s},
  author={Liu, Zhuang and Mao, Hanzi and Wu, Chao-Yuan and Feichtenhofer, Christoph and Darrell, Trevor and Xie, Saining},
  booktitle={Proceedings of the IEEE/CVF conference on computer vision and pattern recognition},
  pages={11976--11986},
  year={2022}
}

@article{mivalt2023impedance,
  title={Impedance rhythms in human limbic system},
  author={Mivalt, Filip and Kremen, Vaclav and Sladky, Vladimir and Cui, Jie and Gregg, Nicholas M and Balzekas, Irena and Marks, Victoria and St Louis, Erik K and Croarkin, Paul and Lundstrom, Brian Nils and others},
  journal={Journal of Neuroscience},
  volume={43},
  number={39},
  pages={6653--6666},
  year={2023},
  publisher={Society for Neuroscience}
}

@article{yuan2021florence,
  title={Florence: A new foundation model for computer vision},
  author={Yuan, Lu and Chen, Dongdong and Chen, Yi-Ling and Codella, Noel and Dai, Xiyang and Gao, Jianfeng and Hu, Houdong and Huang, Xuedong and Li, Boxin and Li, Chunyuan and others},
  journal={arXiv preprint arXiv:2111.11432},
  year={2021}
}

@inproceedings{kirillov2023segment,
  title={Segment anything},
  author={Kirillov, Alexander and Mintun, Eric and Ravi, Nikhila and Mao, Hanzi and Rolland, Chloe and Gustafson, Laura and Xiao, Tete and Whitehead, Spencer and Berg, Alexander C and Lo, Wan-Yen and others},
  booktitle={Proceedings of the IEEE/CVF international conference on computer vision},
  pages={4015--4026},
  year={2023}
}

@inproceedings{radford2021learning,
  title={Learning transferable visual models from natural language supervision},
  author={Radford, Alec and Kim, Jong Wook and Hallacy, Chris and Ramesh, Aditya and Goh, Gabriel and Agarwal, Sandhini and Sastry, Girish and Askell, Amanda and Mishkin, Pamela and Clark, Jack and others},
  booktitle={International conference on machine learning},
  pages={8748--8763},
  year={2021},
  organization={PmLR}
}

@article{achiam2023gpt,
  title={Gpt-4 technical report},
  author={Achiam, Josh and Adler, Steven and Agarwal, Sandhini and Ahmad, Lama and Akkaya, Ilge and Aleman, Florencia Leoni and Almeida, Diogo and Altenschmidt, Janko and Altman, Sam and Anadkat, Shyamal and others},
  journal={arXiv preprint arXiv:2303.08774},
  year={2023}
}

@article{touvron2023llama,
  title={Llama: Open and efficient foundation language models},
  author={Touvron, Hugo and Lavril, Thibaut and Izacard, Gautier and Martinet, Xavier and Lachaux, Marie-Anne and Lacroix, Timoth{\'e}e and Rozi{\`e}re, Baptiste and Goyal, Naman and Hambro, Eric and Azhar, Faisal and others},
  journal={arXiv preprint arXiv:2302.13971},
  year={2023}
}

@article{chowdhery2023palm,
  title={Palm: Scaling language modeling with pathways},
  author={Chowdhery, Aakanksha and Narang, Sharan and Devlin, Jacob and Bosma, Maarten and Mishra, Gaurav and Roberts, Adam and Barham, Paul and Chung, Hyung Won and Sutton, Charles and Gehrmann, Sebastian and others},
  journal={Journal of Machine Learning Research},
  volume={24},
  number={240},
  pages={1--113},
  year={2023}
}

@article{vaswani2017attention,
  title={Attention is all you need},
  author={Vaswani, Ashish and Shazeer, Noam and Parmar, Niki and Uszkoreit, Jakob and Jones, Llion and Gomez, Aidan N and Kaiser, {\L}ukasz and Polosukhin, Illia},
  journal={Advances in neural information processing systems},
  volume={30},
  year={2017}
}

@article{dosovitskiy2020image,
  title={An image is worth 16x16 words: Transformers for image recognition at scale},
  author={Dosovitskiy, Alexey and Beyer, Lucas and Kolesnikov, Alexander and Weissenborn, Dirk and Zhai, Xiaohua and Unterthiner, Thomas and Dehghani, Mostafa and Minderer, Matthias and Heigold, Georg and Gelly, Sylvain and others},
  journal={arXiv preprint arXiv:2010.11929},
  year={2020}
}

@article{wan2025openecg,
  title={Openecg: Benchmarking ecg foundation models with public 1.2 million records},
  author={Wan, Zhijiang and Yu, Qianhao and Mao, Jia and Duan, Wenfeng and Ding, Cheng},
  journal={arXiv preprint arXiv:2503.00711},
  year={2025}
}

@article{arora2024evaluation,
  title={On the evaluation of speech foundation models for spoken language understanding},
  author={Arora, Siddhant and Pasad, Ankita and Chien, Chung-Ming and Han, Jionghao and Sharma, Roshan and Jung, Jee-weon and Dhamyal, Hira and Chen, William and Shon, Suwon and Lee, Hung-yi and others},
  journal={arXiv preprint arXiv:2406.10083},
  year={2024}
}

@article{wang2024deep,
  title={Deep time series models: A comprehensive survey and benchmark},
  author={Wang, Yuxuan and Wu, Haixu and Dong, Jiaxiang and Liu, Yong and Long, Mingsheng and Wang, Jianmin},
  journal={arXiv preprint arXiv:2407.13278},
  year={2024}
}

@article{hsu2021hubert,
  title={Hubert: Self-supervised speech representation learning by masked prediction of hidden units},
  author={Hsu, Wei-Ning and Bolte, Benjamin and Tsai, Yao-Hung Hubert and Lakhotia, Kushal and Salakhutdinov, Ruslan and Mohamed, Abdelrahman},
  journal={IEEE/ACM transactions on audio, speech, and language processing},
  volume={29},
  pages={3451--3460},
  year={2021},
  publisher={IEEE}
}
